\documentclass[preprint, 12pt, proof, authoryear]{elsarticle}

\makeatletter
\def\ps@pprintTitle{%
 \let\@oddhead\@empty
 \let\@evenhead\@empty
 \def\@oddfoot{}%
 \let\@evenfoot\@oddfoot}
\makeatother

\usepackage{amsmath, mathtools}
\usepackage[authoryear]{natbib}
\usepackage{amsthm,amssymb,centernot, amsfonts}
\usepackage{graphics}
\usepackage{graphicx}
\usepackage{caption}
\usepackage{enumitem}
\usepackage{color}
\usepackage[table]{xcolor}
\usepackage[calc]{picture}
\usepackage[top=1.15in, bottom=1.15in, outer=1in, inner=1in, lines=22]{geometry}
\usepackage{secdot}
\usepackage{float}
\usepackage{tikz}
\usepackage[toc, header]{appendix}
\usetikzlibrary{arrows,calc,shapes.geometric, positioning,trees,shadows, fit, arrows.meta, decorations}
\usepackage{setspace, comment}
\usepackage{titlesec}
\usepackage{subfig}
\usepackage{tcolorbox}
\usepackage{centernot}
\usepackage{ulem}
\usepackage{comment}
\usepackage{booktabs}
\usepackage{mathbbol}
\usepackage{multirow}
\usepackage{pdflscape}
\usepackage{rotating}
\usepackage{footnote}
\usepackage{tablefootnote}
\makesavenoteenv{tabular}
\usepackage[flushleft]{threeparttable}
\usepackage{adjustbox}
\usepackage{changepage}
\usepackage{multibib}
\newcites{Main}{References}
\newcites{Meth}{Articles included in methodological review}
\newcites{App}{Articles included in applied review for $\gamma_{RDE}$}
\newcites{SM}{Appendix References}
\usepackage{afterpage}
\usepackage{mwe}

\usepackage{titletoc}

\newcommand\DoToC{%
  \startcontents
  \printcontents{}{1}{\textbf{ }\vskip3pt\hrule\vskip5pt}
  \vskip3pt\hrule\vskip5pt
}

\tikzset{
    -Latex,auto,node distance =1 cm and 1 cm,semithick,
    state/.style ={ellipse, draw, minimum width = 0.7 cm},
    point/.style = {circle, draw, inner sep=0.04cm,fill,node contents={}},
    bidirected/.style={Latex-Latex,dashed},
    el/.style = {inner sep=2pt, align=left, sloped}
}

\newcommand{\CI}{\mathrel{\perp\mspace{-10mu}\perp}}

\newtheorem{ppst}{Proposition}
\newtheorem{theorem}{Theorem}
\newtheorem{definition}{Definition}
\newtheorem{corollary}{Corollary}

 \usepackage[hyperindex,breaklinks]{hyperref}

\begin{document}

\begin{frontmatter}
\title{Interpretational errors in statistical causal inference}

\author[1]{Aaron L. Sarvet  \corref{cor1}}
\author[1]{Mats J. Stensrud}
\author[2]{Lan Wen}

\address[1]{Department of Mathematics, École Polytechnique Fédérale de Lausanne, Switzerland}
\address[2]{Department of Statistics and Actuarial Science, University of Waterloo, Waterloo, Ontario, Canada}
\cortext[cor1]{\textbf{Contact information for corresponding author:}\\
Aaron L. Sarvet, Department of Mathematics, École Polytechnique Fédérale de Lausanne, Switzerland. 
}


\begin{abstract}

We formalize an interpretational error that is common in statistical causal inference, termed \textit{identity slippage}. This formalism is used to describe historically-recognized fallacies, and analyse a fast-growing literature in statistics and applied fields. We conducted a systematic review of natural language claims in the literature on stochastic mediation parameters, and documented extensive evidence of \textit{identity slippage} in applications. This framework for error detection is applicable whenever policy decisions depend on the accurate interpretation of statistical results, which is nearly always the case. Therefore, broad awareness of \textit{identity slippage} will aid statisticians in the successful translation of data into public good.

\end{abstract}
\end{frontmatter}

\date{October 2022}

\newpage

\doublespacing
\section{Introduction}\label{sec: intro}

The validity of a statistical analysis hinges on human decisions. Thus, it is important to identify and characterize sources of human error in this process. Errors in interpretation are especially impactful because they occur at the interface between technical procedures and practical consequences.

The formal integration of causal models in statistics directly intervenes at this interface by providing ``formal mathematical definitions for the English sentences expressing the investigator's causal inferences'' \citepMain{robins1987addendum}. This integration, considered by some a ``causal revolution'', is a critical guard-rail for efforts to translate data into public good.  However, the contemporary proliferation of causal inference in statistics has presented new challenges in interpretation. 

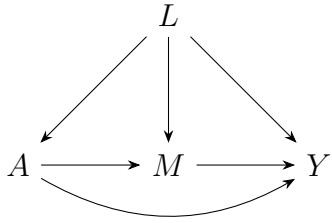
\begin{figure}[!h]
\centering
\begin{tikzpicture}
\begin{scope}[every node/.style={thick,draw=none}]

    \node (L) at (2,2) {$L$};
    \node (A) at (0,0) {$A$};
    \node (M) at (2,0) {$M$};
    \node (Y) at (4,0){$Y$};

\end{scope}

\begin{scope}[>={Stealth},
              every node/.style={fill=white,circle},
              every edge/.style={draw=black}]

\path [->] (L) edge (A);
\path [->] (L) edge (M);
\path [->] (L) edge (Y);
\path [->] (A) edge (M);
\path [->] (M) edge (Y);
\path [->] (A) edge[bend right] (Y);

\end{scope}
\end{tikzpicture}

\caption{Directed acyclic graph (DAG) representing a nonparametric structural equation with independent errors. We suppose that there exists an organic intervention variable $I$ that meets the conditions in \citet{lok2016defining}, and that the dismissable component conditions in \citet{stensrud2022separable} hold with respect to modified treatments $A_M$ and $A_Y$.}
\label{fig: IntroDAG}
\end{figure}

Consider a simple observed data structure consisting of baseline covariates ($L$), treatment ($A$), causal intermediate ($M$), and outcome ($Y$). This is the classical mediation structure considered in canonical works by \citetMain{baron1986moderator} and later by \citetMain{robins1992identifiability} and \citetMain{pearl2001direct} that currently enjoy over 115000 citations. Let the statistical parameter $\Psi$ be the functional of the observed data termed the ``mediation formula'' \citepMain{pearl2009causality},
$$\Psi(P) \equiv \int_{m}\int_l \Big\{\mathbb{E}_P[Y \mid m, a=1, l] - \mathbb{E}_P[Y \mid m, a=0, l]\Big\} f_P(m \mid a=0, l)f_P(l)d\mu(m, l),$$
where $P$ is the probability distribution of $(L,A,M,Y)$, and $\mu$ is a counting measure if a variable is discrete or Lebesgue measure if the variable is continuous. Suppose also that the directed acyclic graph (DAG) in Figure \ref{fig: IntroDAG} represents a non-parametric structural equation model $\mathcal{M}$ with independent errors (NPSEM-IE, \citealp{pearl2009causality}), and that additional causal conditions hold (see Figure \ref{fig: IntroDAG}). 

Then $\Psi(P)$ may simultaneously be interpreted as at least four distinct causal parameters: the natural (pure) direct effect of treatment $A$, denoted by $\gamma_{NDE}$ \citepMain{robins1992identifiability, pearl2001direct}; the separable effect of the component $A_Y$ of treatment $A$ that does not cause the mediator $M$, denoted by $\gamma_{Sep}$ \citepMain{robins2010alternative,robins2022interventionist, stensrud2022separable}; the randomized interventional analogue to the natural direct effect, denoted by $\gamma_{RDE}$ \citepMain{ vanderweele2014effect}; and the organic direct effect, denoted by $\gamma_{Org}$ \citepMain{lok2016defining}. Equivalently, $\Psi$, $\gamma_{NDE}$,$\gamma_{Sep}$, $\gamma_{RDE}$, and $\gamma_{Org}$ all map to the same value at all distributions $P$ in the causal model $\mathcal{M}$. Despite their equality, the distinctions between these causal parameters are not merely nominal; their values at a given distribution may have very different practical implication for decision making and scientific theory. Furthermore, their concordance in this case is merely a coincidence of $\mathcal{M}$; for joint distributions outside of $\mathcal{M}$, their values may diverge substantially.

This example illustrates the subtle model-dependence in the interpretation of statistical parameters of observed data, like $\Psi$. The rapid growth of literature articulating well-defined causal parameters and developing statistical theory for their efficient estimation undoubtedly enriches the methodological toolbox of applied investigators. However, this toolbox also creates opportunities for errors in interpretation, where estimates for one well-defined causal parameter are mistakenly interpreted as another. Furthermore, it potentially encourages an economy of statistical research wherein well-defined causal parameters are \textit{designed}, not for their precise correspondence with particular scientific tasks, but rather for their desirable statistical properties. The existence of such parameters may support a phenomenon conceptually related to, but distinct from, ``p-hacking'' \citepMain{simmons2016false} and the ``garden of forking paths'' \citepMain{gelman2013garden}.  The formal definition and investigation of this phenomenon, which we term \textit{identity slippage}, is the focus of this article. The resulting framework for interpretational error detection may not only provide a crucial aid to applied statisticians and consumers of quantitative research; it may guide methodological researchers to develop and enhance tools that remain faithful to stakeholder's real-world scientific questions.   


The simplest case of \textit{identity slippage} is merely a reformulation of the interpretive error associated with the maxim ``association is not causation''. To fix ideas, suppose $Y$ indicates a long-term functional outcome for a patient with acute ischemic stroke (AIS), and $A=1$ indicates off-label use of tenecteplase, an alternative thrombolytic agent, versus alteplase ($A=0$), a current standard of care \citepMain{katsanos2022off}. We let $Y^a$ denote the potential outcome under an intervention that sets $A$ to $a$. Suppose an investigator then considers two models: a fully nonparametric causal model ($\mathcal{M}_1$) that places no restrictions on the law of the counterfactual and observed data; and a second model ($\mathcal{M}_2$) that is defined by augmenting $\mathcal{M}_1$ with the exchangeability assumptions that $Y^a \CI A \mid L$ for all values $a$. Under this second model, suppose the average treatment effect  $\mathbb{E}[Y^{a=1} - Y^{a=0}]$ is identified by an associational parameter: a difference in covariate-adjusted conditional outcome means, $ \int_{l}\big\{\mathbb{E}[Y \mid A=1, l] - \mathbb{E}[Y \mid A=0,l]\big\}f(l)d\mu(l) $. However, some investigators are not willing to accept the stricter model $\mathcal{M}_2$, perhaps in accordance with The American Medical Association's Manual of Style from 2020 \citepMain{ama2020ama}, which asserts ``residual confounding is \textit{unavoidable} in observational studies''.  Thus, the investigators instead explicitly target the associational parameter, which yields a positive value after meta-analysis of several studies with large sample sizes. They are careful to avoid explicit causal language such as ``effect'' and ``efficacy'', but instead make several other claims as if the value they computed corresponded not to  the associational parameter, but to the (unidentified) average treatment effect, for example writing: ``tenecteplase is as safe as alteplase for the treatment of AIS'' \citepMain{katsanos2022off}. This is a case of \textit{identity slippage}. 

We anchor \textit{identity slippage} in this example in order to unify seemingly disparate cases as structurally isomorphic to ``association is not causation''. We demonstrate the analytic capacity of \textit{identity slippage} through consideration of historical examples, and contemporary examples that are currently obscured. Particularly, we systematically review \textit{identity slippage} between canonical mediation parameters and novel stochastic mediation parameters that are currently the subject of substantial attention in the statistical literature.  

Furthermore, the formal conceptualization of human errors and other fallacies facilitates objective examination of the sociological structures that promote them. For example, the conceptualization of ``p-hacking'' and ``publication bias'' has shone a light on sociological pressures to publish statistically significant results, and led to debates and policy changes to counteract them, for example the systematic registration of protocols for clinical trials \citepMain{easterbrook1991publication}. Analogously, we leverage the conceptualization of \textit{identity slippage} to identify contrasting \textit{pragmatic} and \textit{ideal} frameworks for causal inference that create different sets of conditions for the occurrence of \textit{identity slippage} in practice. Recognition and characterization of these operative frameworks will help illuminate both technical and sociological vulnerabilities in current statistical practice.  

The remainder of the manuscript is organized as follows: Section \ref{sec: IDslip} proposes a formal definition of \textit{identity slippage}, and locates it within a process for causal analysis that distinguishes other previously well-described human errors in statistics. Section \ref{sec: contemp motiv} introduces a contemporary motivating example for \textit{identity slippage} involving the canonical stochastic mediation parameter $\gamma_{RDE}$ popularized by \citet{vanderweele2014effect}. Section \ref{sec: pragideal} describes two frameworks for causal inference, and analyses an investigator's potential risk of \textit{identity slippage} within each. Section \ref{sec: case1}  presents a case study of \textit{identity slippage} for stochastic mediation parameters. Therein, we present a comprehensive review of its methodological literature and also a systematic review of an applied literature that this methodological literature produced. We emphasize that this systematic review represents a novel methodology for analysing the use and misuse of novel causal and statistical methodologies in practice. A second case study of \textit{identity slippage} for parameters defined by dynamic treatment regimes and stochastic interventions in a sequential treatment setting is presented in Web Appendix A. 

\section{Defining \textit{identity slippage}} \label{sec: IDslip}

\subsection{Preliminaries}
Let $P^F$ denote a joint distribution (full law) over observed variables and their counterfactual counterparts under all possible regimes, and let $P_0^{F}$ denote the true distribution. Let $\mathcal{M}$ denote a causal model, where a model is understood as a set of candidate laws $P^F$ satisfying causal and other statistical conditions, for example, exchangeability conditions, or parametric conditions. We let $P\equiv P(P^F)$ denote the joint distribution of the observed variables that is implied by $P^F$, and let  $\mathcal{M}^{obs} \equiv \{P \mid P^{F} \in \mathcal{M}\}$ denote the statistical model for the observed data induced by $\mathcal{M}$.
We let $\gamma$ denote a causal parameter of the full law defined as a mapping such that  $\gamma: \mathcal{M}\rightarrow \mathbb{R}$. For example, we may write the average treatment effect (ATE) at the true law as $\gamma_{\text{ATE}}(P^F_0) \coloneqq \mathbb{E}_{P^F_0}(Y^{a=1} -Y^{a=0})$. Let $\Psi$ denote a statistical parameter of the observed data defined as a mapping such that $\Psi: \mathcal{M}^{obs}\rightarrow \mathbb{R}$. For example, we may write the difference in covariate-adjusted conditional outcome means at the true law as $\Psi_{\text{CMN}}(P_0) \coloneqq \int_{l}\big\{\mathbb{E}_{P_0}[Y \mid A=1, L=l] - \mathbb{E}_{P_0}[Y \mid A=0,L=l]\big\}f_{P_0}(l)d\mu(l)$. Here, $(Y,A, L)$ are also variables described by the full law $P^F$, so we might equivalently define a $\gamma_{\text{CMN}}$ such that $\gamma_{\text{CMN}}(P^F_0) \coloneqq \int_{l}\big\{\mathbb{E}_{P^F_0}[Y \mid A=1, L=l] - \mathbb{E}_{P^F_0}[Y \mid A=0,L=l]\big\}f_{P^F_0}(l)d\mu(l)$. However, we cannot define a statistical parameter of the observed data equivalent to $\gamma_{\text{ATE}}$ because $(Y^{a=1},Y^{a=0})$ are not variables of the observed data law $P$.


Because the interpretive error we describe involves identification, we provide the following definition of identifiability.
 
\begin{definition}[Identifiability]\label{def: ID}
A causal parameter $\gamma$ is identified by $\Psi$ under $\mathcal{M}$ whenever $\gamma(P^F)$ equals $\Psi(P) \equiv \Psi(P(P^F))$, for all $P^F\in\mathcal{M}$. 
\end{definition}

Any interpretive error involves \textit{interpretation}. Thus, it follows that a formal definition of such an error would require a formal definition of a valid interpretation. We propose the following such definition. 

\begin{definition}[Interpretive map for $\gamma$]
Let $\Gamma$ denote the codomain of $\gamma$. Let $\mathfrak{I}$ be the set of all possible claims and scientific implications, expressed in natural language. Let $\mathcal{P}(\mathfrak{I})$ denote the power set of $\mathfrak{I}$. Let $\mathcal{I}_{\gamma}: \Gamma \rightarrow \mathcal{P}(\mathfrak{I})$ be a function that maps each value $x\in \Gamma$ to the largest element in $\mathcal{P}(\mathfrak{I})$ that only includes the interpretive claims and scientific implications that would be appropriate to make if $\gamma(P_0^F)=x$. Then, $\mathcal{I}_{\gamma}$ is an interpretive map for $\gamma$, and a claim $\mathfrak{i}$ for $x\in \Gamma$ is valid for $\gamma(P_0^F)$ if $\mathfrak{i}\in \mathcal{I}_{\gamma}(x)$.
\end{definition}

While the notion of an interpretive map $\mathcal{I}_{\gamma}$ may seem abstract, the implicit use of such functions is ubiquitous in applied research, for example in medical articles; they are used when the values of parameters, such as an ATE or an average natural indirect effect, are translated into an action, e.g., the approval of a COVID-19 vaccine for emergency use among children \citepMain{usfoodanddrugadministration_2022} or the refutation of a theory that the effect of prescribing an extended release formulation for a medication is mediated by increased adherence \citepMain{rudolph2021explaining}. Values in the codomain of the interpretive map appear in nearly every discussion section of applied research articles. As such, interpretive maps are critical components for the successful translation of statistical science into social good. Furthermore, they are of special concern to statisticians because of their unique expertise in the interpretation of statistical objects \citepMain{gardner1986use, altman1998statistical}.


As an example of an interpretive map, let $\gamma$ denote the average effect of a novel immunotherapeutic drug $(a=1)$ vs.\ standard of care $(a=0)$  on 5-year survival $(Y)$ among the study population of patients with stage IV small-cell carcinoma enrolled in a clinical trial: $\gamma(P^F) = E_{P^F}[Y^{a=1} - Y^{a=0}]$.  The function $\mathcal{I}_{\gamma}$ maps values in $[-1,1]$ to sets of permissible statements. For example, elements in $\mathcal{I}_{\gamma}(0.2)$ might include 
\begin{itemize}
    \item $\mathfrak{i}_1 \coloneqq \{$``The probability of 5-year survival among those taking the new therapeutic drug is 20 percentage points greater than among those following current standards of care.''$\}$, or 
    \item $\mathfrak{i}_2 \coloneqq \{$``Evidence suggests the novel immunotherapeutic drug can improve survival vs.\ current standards of care.''$\}$. 
\end{itemize}

Alternatively, elements of the set $\mathcal{I}_{\gamma}(-0.1)$ may include the claim 
\begin{itemize}
    \item $\mathfrak{i}_3 \coloneqq  \{$``Evidence suggests the novel immunotherapeutic drug is harmful with respect to survival vs.\ current standards of care.''$\}$. 
\end{itemize}

Of course $\mathfrak{i}_3 \not\in \mathcal{I}_{\gamma}(0.2)$, because it would be invalid to claim that the drug is harmful when the ATE $\gamma(P^F_0)$ is positive. Alternatively, elements in the image of $\mathcal{I}_{\gamma}$ may have non-empty intersections - that is, the same claim $\mathfrak{i}$ may be valid for multiple values in $\Gamma$. For example $\mathfrak{i}_3$ may be an element of $\mathcal{I}_{\gamma}(x)$ for all $x<0$. 

We now provide a definition of \textit{identity slippage}, formalizing a frequent error in interpretative claims. 



\begin{definition}[Identity slippage] \label{def: IDslip}
Suppose the following conditions hold:
\begin{align*}
   \textbf{I1. }&  \mathcal{M}_2  \subset \mathcal{M}_1 ,\\
   \textbf{I2. }& \gamma_1 \text{ is identified by } \Psi_1 \text{ under } \mathcal{M}_1 ,\\
   \textbf{I3. }& \gamma_2 \text{ is identified by } \Psi_2 \text{ under } \mathcal{M}_2 ,\\
   \textbf{I4. }& \gamma_2 \text{ is not identified by either } \Psi_2 \text{ or } \Psi_1 \text{ under } \mathcal{M}_1.
\end{align*} 
Let $\hat{\psi}_1$ be an estimate of $\Psi_1(P_0)$. If an investigator selects $\gamma_1$ as their target of inference and assumes $\mathcal{M}_1$, but makes the claim $\mathfrak{i} \in \mathcal{I}_{\gamma_2}(\hat{\psi}_1)$ and $\mathfrak{i} \notin \mathcal{I}_{\gamma_1}(\hat{\psi}_1)$, then \textit{identity slippage} has occurred.
\end{definition}

In words, we suppose that an investigator considers the nested models $\mathcal{M}_2 \subset \mathcal{M}_1$, and two parameters $\gamma_2$ and $\gamma_1$. Here, $\gamma_1$ is identified by $\Psi_1$ under $\mathcal{M}_1$, and $\gamma_2$ is identified by $\Psi_2$ under $\mathcal{M}_2$ but $\gamma_2$ is not identified by either $\Psi_2$ or $\Psi_1$ under $\mathcal{M}_1$. The investigator assumes the larger model $\mathcal{M}_1$ and then estimates $\gamma_1(P_0^{F})$ with $\hat{\psi}_1$.  However, the investigator makes an interpretive claim $\mathfrak{i}$ in the set of claims about the parameter $\gamma_2$ \textit{that would be valid had $\gamma_2$ equaled $\hat{\psi}_1$}, i.e. $\mathfrak{i}\in \mathcal{I}_{\gamma_2}(\hat{\psi}_1)$, and further this claim did not happen to be in the set of such permissible claims had $\gamma_1$ equaled $\hat{\psi}_1$, i.e. $\mathfrak{i}\notin \mathcal{I}_{\gamma_1}(\hat{\psi}_1)$. In other words, the investigator makes an interpretive claim corresponding to the unidentified parameter $\gamma_2$ but using the estimated value of the identified $\gamma_1$. Then we say \textit{identity slippage} has occurred. When an analysis uses a consistent estimator for $\gamma_1$ under $\mathcal{M}_1$, an error would be harmful at increasing sample sizes almost surely if $\gamma_1(P_0^F) \neq \gamma_2(P_0^F)$, such that interpretive claims in $\mathcal{I}_{\gamma_2}(\gamma_2(P_0^F))$ qualitatively depart from those of $\mathcal{I}_{\gamma_2}(\gamma_1(P_0^F))$. We will consider several concrete examples beginning in Section \ref{subsec: hist1}.

\subsection{Distinguishing \textit{identity slippage} from errors of other kinds}

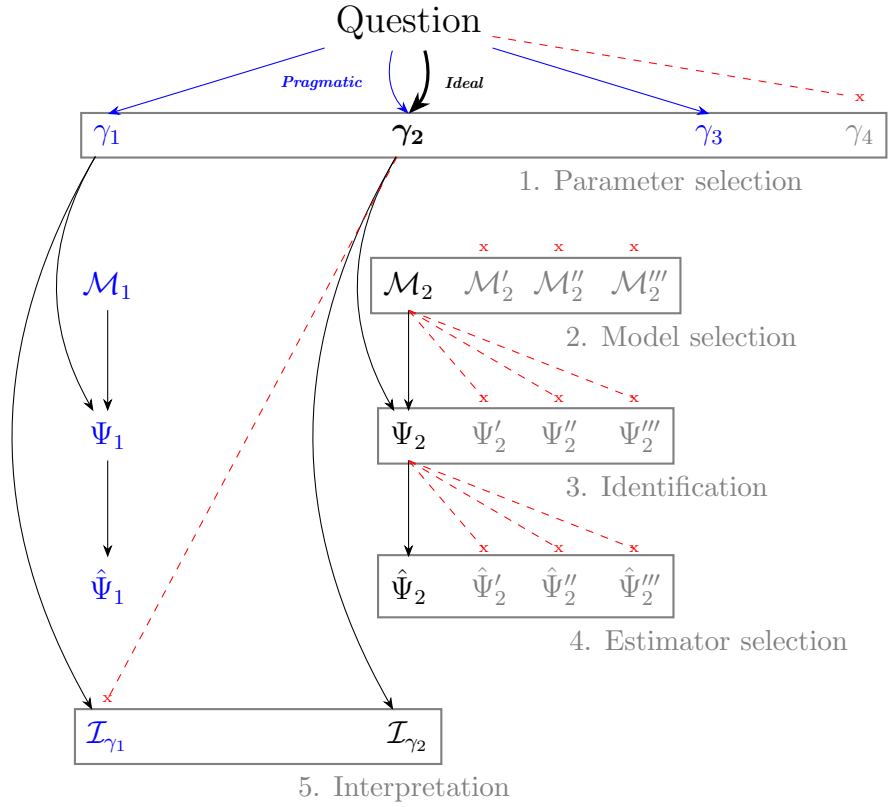
\begin{figure}
\centering
\begin{tikzpicture}
\begin{scope}[every node/.style={thick,draw=none}]


    \node (sci) at (4,5.5) {\text{\large Question}};
    \node (gamma1) at (0,4) {\color{blue}{$\gamma_1$}};
    \node (gamma2) at (4,4) {\color{black}{$\boldsymbol{\gamma_2}$}};
    \node (gamma3) at (8,4) {\color{blue}{$\gamma_3$}};
    \node (gamma4) at (10,4) {\color{gray}{$\gamma_4$}};
    \node[draw=gray, fit={(gamma1) (gamma2)(gamma3) (gamma4)} , inner sep=.01cm, color=gray, label=below right:\text{\small \textcolor{gray}{1. Parameter selection}}]   (paramselect) {};
    \node (xg4) at   (10,4.5)  {\text{\tiny \textcolor{red}{x}}};

    \node (model2) at    (4,2)    {\color{black}{$\mathcal{M}_2$}};
    \node (model2') at   (5,2)    {\color{gray}{ $\mathcal{M}_2'$}};
    \node (model2'') at  (6,2)    {\color{gray}{$\mathcal{M}_2''$}};
    \node (model2''') at (7,2)    {\color{gray}{ $\mathcal{M}_2'''$}};
    \node (model1) at    (0,2)    {\color{blue}{$\mathcal{M}_1$}};
    \node[draw=gray, fit={(model2) (model2') (model2'') (model2''')} , inner sep=.01cm, color=gray, label=below right:\text{\small \textcolor{gray}{2. Model selection}}]   (modelselect) {};  
    \node (xm2') at   (5,2.5)  {\text{\tiny \textcolor{red}{x}}};
    \node (xm2'') at   (6,2.5)  {\text{\tiny \textcolor{red}{x}}};
    \node (xm2''') at   (7,2.5)  {\text{\tiny \textcolor{red}{x}}};

    \node (psi2) at    (4,0)     {\color{black}{$\Psi_2$}};
    \node (psi2') at   (5,0)     {\color{gray}{ $\Psi_2'$}};
    \node (psi2'') at  (6,0)     {\color{gray}{$\Psi_2''$}};
    \node (psi2''') at (7,0)     {\color{gray}{ $\Psi_2'''$}};
    \node (psi1) at    (0,0)     {\color{blue}{$\Psi_1$}};
    \node[draw=gray, fit={(psi2) (psi2') (psi2'') (psi2''')} , inner sep=.01cm, color=gray, label=below right:\text{\small \textcolor{gray}{3. Identification}}]   (identification) {}; 
    \node (xp2') at    (5,0.5)  {\text{\tiny \textcolor{red}{x}}};
    \node (xp2'') at   (6,0.5)  {\text{\tiny \textcolor{red}{x}}};
    \node (xp2''') at  (7,0.5)  {\text{\tiny \textcolor{red}{x}}};

    \node (psihat2) at   (4,-2)    {\color{black}{$\hat{\Psi}_2$}};
    \node (psihat2') at  (5,-2)    {\color{gray}{ $\hat{\Psi}_2'$}};
    \node (psihat2'') at (6,-2)    {\color{gray}{$\hat{\Psi}_2''$}};
    \node (psihat2''') at(7,-2)    {\color{gray}{ $\hat{\Psi}_2'''$}};
    \node (psihat1) at   (0,-2)    {\color{blue}{$\hat{\Psi}_1$}};
    \node[draw=gray, fit={(psihat2) (psihat2') (psihat2'') (psihat2''')} , inner sep=.01cm, color=gray, label=below right:\text{\small \textcolor{gray}{4. Estimator selection}}]   (estselect) {};
    \node (xph2') at    (5,-1.5)  {\text{\tiny \textcolor{red}{x}}};
    \node (xph2'') at   (6,-1.5)  {\text{\tiny \textcolor{red}{x}}};
    \node (xph2''') at  (7,-1.5)  {\text{\tiny \textcolor{red}{x}}};

    \node (inter2) at (4,-4) {\color{black}{$\mathcal{I}_{\gamma_2}$}};
    \node (inter1) at (0,-4) {\color{blue}{$\mathcal{I}_{\gamma_1}$}};
    \node[draw=gray, fit={(inter1) (inter2)} , inner sep=.01cm, color=gray, label=below right:\text{\small \textcolor{gray}{5. Interpretation}}]   (interpret) {};
    \node (xi1) at  (0,-3.5)  {\text{\tiny \textcolor{red}{x}}};

    \node (prag) at (2.8, 4.7) {\tiny \text{ \textcolor{blue}{\textbf{\textit{Pragmatic}}}}};
    \node (ideal) at (4.7, 4.7) {\tiny \text{ \textcolor{black}{\textbf{\textit{Ideal}}}}};

\end{scope}

\begin{scope}[>={Stealth},
              every node/.style={fill=white,circle},
              every edge/.style={draw=black}]

    \path [->] (gamma2) edge[bend right] (psi2);
    \path [->] (model2) edge (psi2);
    \path [-] (model2.south) edge[color=red, dashed] (xp2'.center);
    \path [-] (model2.south) edge[color=red, dashed] (xp2''.center);
    \path [-] (model2.south) edge[color=red, dashed] (xp2'''.center);

    \path [->] (psi2) edge (psihat2);
    \path [-] (psi2.south) edge[red, dashed] (xph2'.center);
    \path [-] (psi2.south) edge[red, dashed] (xph2''.center);
    \path [-] (psi2.south) edge[red, dashed] (xph2'''.center);

    \path [->] (gamma2) edge[bend right] (inter2);

    \path [->] (gamma1) edge[bend right] (psi1);
    \path [->] (model1) edge (psi1);
    \path [->] (psi1) edge (psihat1);
    \path [->] (gamma1) edge[bend right] (inter1);

\path [-] (gamma2) edge[red, dashed] (xi1.center);

\path [->] (sci) edge[color=blue] (gamma1.north);
\path [->] (sci) edge[color=blue, bend right] (gamma2.north);
\path [->] (sci) edge[color=black, very thick, bend left] (gamma2.north);
\path [->] (sci) edge[color=blue] (gamma3.north);
\path [-] (sci) edge[red, dashed] (xg4);

\end{scope}
\end{tikzpicture}

\caption{Conceptual diagram illustrating a 5-step process under \textit{pragmatic} and \textit{ideal} frameworks for causal inference. Arrows between objects in the figure represent dependencies in the process of causal analysis. For example, selection of an observed data functional $\Psi$ is a deductive task that depends on the causal parameter $\gamma$ and a selected model $\mathcal{M}$. Valid analyses will involve valid decisions at each step. Red arrows (x's) represent errors, with identity slippage located as a type of error in step 5 (interpretation). Pragmatic and ideal approaches are distinguished in the diagram by differences in correspondences between a scientific question and causal parameters; in an ideal approach there is a one-to-one correspondence between scientific question and causal parameter ($\gamma_2$), in a pragmatic approach there is a one-to-many correspondence between scientific question and causal parameters ($\{\gamma_1, \gamma_2, \gamma_3\}$). 
}
\label{fig: concept}
\end{figure}

To clarify understanding of \textit{identity slippage}, Figure \ref{fig: concept} depicts a process for a causal analysis. We partition this analysis into five steps: 1) \textit{Parameter selection}; 2) \textit{Model selection}; 3) \textit{Identification}; 4) \textit{Estimator selection}; and 5) \textit{Interpretation}. A valid causal analysis is understood as a set of valid decisions such that interpretive claims are made using the appropriate interpretive map applied to the computed value of an estimator consistent for the causal parameter at $P_0^F$. Conversely, an invalid analysis may arise from an error in any one of these steps. We have located \textit{identity slippage} as an error in the \textit{Step 5. Interpretation}. For example, the red arrow between $\gamma_2$ and $\mathcal{I}_{\gamma_1}$ in Figure \ref{fig: concept} indicates \textit{identity slippage}. We review errors in the remaining steps in reverse order, in the context of the terms in Definition \ref{def: IDslip}. 

\subsubsection*{Step 4. Estimator selection.}
Suppose the conditions of Definition \ref{def: IDslip} hold, and the investigator both correctly specifies $\mathcal{M}_1$, and correctly identifies $\gamma_1$ with $\Psi_1$ under $\mathcal{M}_1$. However, suppose they select an estimator $\hat{\Psi}_1$ that is consistent for some value not equal to $\Psi_1$ at $P_0$. Nevertheless the investigator makes a claim $\mathfrak{i} \in \mathcal{I}_{\gamma_1}(\hat{\psi}_1)$.  This will almost surely lead to an error in an investigator's conclusions about $\gamma_1(P_0^F)$, but this is not a case of \textit{identity slippage}. This error occurred in selecting an inconsistent estimator of $\gamma_1$. This error is depicted in Figure \ref{fig: concept} with the red-dashed lines from the statistical parameter $\Psi_2$ to the estimators $\hat{\Psi}'_2$, $\hat{\Psi}^{''}_2$, and $\hat{\Psi}^{'''}_2$.

\subsubsection*{Step 3. Identification.}
Suppose the conditions of Definition \ref{def: IDslip} hold and an investigator follows the analysis described in that definition, except that the investigator explicitly selects $\gamma_2$ as their target of inference. Then the investigator has made an error in identification because  $\gamma_2(P^F) \neq  \Psi_1(P)$ for at least one law $P^F\in \mathcal{M}_1$, which may be the true law $P^F_0$. The investigator might have chosen an estimator $\hat{\Psi}_1$ that is consistent for $\Psi_1$ under $\mathcal{M}_1$, and furthermore may have restricted themselves to claims in $\mathcal{I}_{\gamma_2}(\hat{\psi}_1)$, thus using an interpretive map that corresponds to the causal parameter $\gamma_2$ that they explicitly selected. Further, if $P^F_0 \in \mathcal{M}_1$, then the investigator has specified a valid model. However, if $\gamma_2(P^F_0) \neq  \Psi_1(P_0)$, then this will almost surely lead to an error in an investigator's conclusions because $\hat{\Psi}_1$ will not be a consistent estimator for $\gamma_2$ at $P^F_0$. This error is depicted in Figure \ref{fig: concept} with the red-dashed lines from the causal model $\mathcal{M}_2$ to the statistical parameters ${\Psi}'_2$, ${\Psi}^{''}_2$, and ${\Psi}^{'''}_2$.

\subsubsection*{Step 2. Model selection.}
Suppose the conditions of Definition \ref{def: IDslip} hold, except that the investigator both selects $\gamma_2$ as their target of inference and selects $\mathcal{M}_2$. Let $\hat{\psi}_2$ be an estimate of $\Psi_2(P_0)$, calculated by a consistent estimator for $\Psi_2$ under $\mathcal{M}_2$. Suppose then the investigator makes a claim $\mathfrak{i} \in \mathcal{I}_{\gamma_2}(\hat{\psi}_2)$, and thus this is not a case of \textit{identity slippage}. Indeed, $\Psi_2$ identifies $\gamma_2$ under $\mathcal{M}_2$, so it is also not an error in identification. However, if $P_0^F \notin \mathcal{M}_2$, then the investigator will not have selected a valid model for $P_0^F$, and $\Psi_2(P_0)$ may not equal $\gamma_2(P^F_0)$. This error is depicted in Figure \ref{fig: concept} with the red x's above the causal models $\mathcal{M}'_2$, $\mathcal{M}^{''}_2$, and $\mathcal{M}^{'''}_2$. 

\subsubsection*{Step 1. Parameter selection.}

Suppose the conditions of Definition \ref{def: IDslip} hold, except that the investigator makes a claim $\mathfrak{i} \in \mathcal{I}_{\gamma_1}(\hat{\psi}_1)$. This would not be a case of \textit{identity slippage} because the investigator had successfully interpreted their estimate using the interpretive map corresponding to their explicitly selected causal parameter  ($\gamma_1$). However, if $\gamma_1$ does not correspond to their scientific question of interest, then an investigator will not have produced a useful analysis for their scientific task. This has been called a Type III error by \citetMain{schwartz1999right}, under which an investigator might find the ``right answer for the wrong question''. This error is depicted in Figure \ref{fig: concept} with the red-dashed lines from the ''Question'' to the causal parameter $\gamma_4$. We defer exposition of the blue arrows in the figure until Section \ref{sec: pragideal}.

\subsection{Historical examples} \label{subsec: hist1}
\subsubsection*{Historical Example 1: Causation vs.\ association. (revisited)} 

The canonical example of \textit{identity slippage} is the interpretation of adjusted associations as causal effects under a model without causal assumptions. Using the terms of  Definition \ref{def: IDslip}, we consider two parameters, $\gamma_2 \coloneqq \gamma_{ATE}$ and $\gamma_1\coloneqq\gamma_{CMN} = \Psi_{CMN}$, and two models, $\mathcal{M}_1 \coloneqq \{\text{all } P^F\}$ and $\mathcal{M}_2 \coloneqq \{P^F \text{ s.t. } Y^a \CI A \mid L \ \forall \ a\}$. An investigator selects $\gamma_{CMN}$ and $\mathcal{M}_1$, and computes $\hat{\psi}_1$ using a consistent estimator for $\Psi_{CMN}$, but then makes an interpretive claim using the interpretive map for $\gamma_{ATE}$. For example, the claim ``tenecteplase is as safe as alteplase for the treatment of AIS'' \citepMain{katsanos2022off} would be an element in the set $\mathcal{I}_{\gamma_{ATE}}(x)$ for all values $x > 0$. This would be a case of \textit{identity slippage}.

\subsubsection*{Historical Example 2: Local vs.\ population average treatment effects in instrumental variable (IV) settings.}

Let $\gamma_2$ be an ATE in a population of interest, and let $\gamma_1$ be that ATE among the unknown subset of the population who would take treatment regardless of their assignment to some value of an instrumental variable. Across various literatures, $\gamma_1$ is known as a complier average causal effect (CACE), a local average treatment effect (LATE), or a principle stratum effect. Let $\mathcal{M}_1$ be defined by a set of assumptions sufficient for identifying $\gamma_1$ by some $\Psi_1$, for example the classical assumptions in \citetMain{imbens1994identification}, and let $\mathcal{M}_2$ be a model augmented by a homogeneity assumption, for example on the conditional ATEs over levels of the unmeasured confounders, such that $\gamma_2$ is just identified by $\Psi_1$. Suppose an investigator is primarily interested in an ATE, but the investigator instead selects $\mathcal{M}_1$ because they find the augmenting assumptions of $\mathcal{M}_2$ too strong, and explicitly targets the LATE, $\gamma_1$. The investigator continues their data analysis and computes an estimate $\hat{\psi}_1$ using a consistent estimator of the LATE, $\gamma_1$, under $\mathcal{M}_1$. However, in the discussion and abstract of their paper, the investigator interprets their estimate as an ATE, that is, they make an interpretive claim $\mathfrak{i} \in \mathcal{I}_{\gamma_2}(\hat{\psi}_1)$. 

This is a case of \textit{identity slippage} that is possibly quite common.  The authors of an extensive review of IV methods and assumptions \citepMain{swanson2018partial} observe: 

{\small{\textit{``because the effect homogeneity assumptions necessary for point identification for the ATE are often unrealistic, the LATE has often been favored in IV applications''.} }}

Taking one prominent application, re-analyzed by \citetMain{swanson2018partial}, investigators target a LATE: the effect of Medicaid expansion on emergency department use \textit{among low-income adults who would only enroll in Medicaid if selected by a state-run lottery program} \citepMain{taubman2014medicaid}. However, the possibly eccentric, unobservable subgroup, i.e. the principal stratum that defines their LATE, $\gamma_1$, is not mentioned a single time outside of the Statistical Analysis section of \citetMain{taubman2014medicaid}. In other words, all interpretive claims are drawn from the set $\mathcal{I}_{\gamma_2}(\hat{\psi}_1)$. 

We present a third historical example in Web Appendix B.

\section{Contemporary motivating example: randomized interventional analogues of natural (in)direct effects}
\label{sec: contemp motiv}

We now present a contemporary example of \textit{identity slippage}, involving classical mediation parameters and their ``analogue'' parameters defined by stochastic interventions. We briefly review their definitions and the canonical models for their identification, and refer readers to Web Appendix C for an elaborated exposition. We also summarize the historical motivations for the development of these parameters because such motivations may often impact the risk of \textit{identity slippage} in practice. We elaborate on this argument in Section \ref{sec: pragideal}. To fix ideas, we consider a single randomized binary treatment $A$, post-treatment covariate $W$, binary potential mediator $M$, and  outcome $Y$. 

\subsection{A brief history of mediation in statistics}
Since Sewall Wright's influential work, mediation has become nearly synonymous with path-analysis: a partitioning of a total effect of an exposure $A$ on an outcome $Y$ into a direct effect (path) unmediated by a third variable $M$, and an indirect effect (path) under natural conditions.  While \citetMain{baron1986moderator}, hereafter referred to as BK, arguably popularized this statistical formulation, it was not until \citetMain{robins1992identifiability} and later \citetMain{pearl2001direct}, hereafter RGP, that these path-specific parameters were given well-defined ``model-free'' formulations. They named these parameters the average natural (or pure) direct and indirect effects, $\gamma_{NDE}$ and $\gamma_{NIE}$, where
\begin{align}  
    \gamma_{NDE}(P^F) \coloneqq \mathbb{E}_{P^F}[Y^{a=1, M^{a=0}} - Y^{a=0, M^{a=0}}], \label{NDE}\\
    \gamma_{NIE}(P^F) \coloneqq \mathbb{E}_{P^F}[Y^{a=1, M^{a=1}}- Y^{a=1, M^{a=0}}]. \label{NIE}
\end{align}

In BK the mediation parameters were symbolically synonymous with coefficient terms in linear models, and as such, the definitions of these parameters were effectively conflated with the parametric assumptions of the models in which they were defined. In contrast, the definition of the individual natural direct effect $Y^{a=1, M^{a=0}} - Y^{a=0, M^{a=0}}$ is agnostic to modelling and identification assumptions. RGP's clear separation between \textit{parameter selection} and \textit{model selection} made it possible to scrutinize the strategies and assumptions employed to identify $\gamma_{NDE}$ and $\gamma_{NIE}$ that had been previously neglected.

In understanding the method of BK as a particular identification strategy for the natural direct and indirect effects, RGP highlighted the unnecessary strength, and often implausibility, of the assumptions involved in BK's approach. More broadly, RGP's articulation of the natural direct and indirect effects revealed features of their definition with important practical and philosophical implications: principally, these parameters could not possibly be observed in any real world trial without additional assumptions, and so any causal model under which \eqref{NDE} and \eqref{NIE} are identified must depend on at least some (in-principle) untestable assumptions. In particular, investigators often impose so-called cross-world counterfactual independence conditions when they do inference on $\gamma_{NDE}$ (see \citetMain{pearl2009causality}). However, these same investigators will often conclude that these conditions are contradicted by other obvious characteristics of the data generating mechanism. Famously, the presence of a common cause of the mediator and the outcome that is itself affected by the exposure \citepMain{avin2005identifiability}, termed a ``recanting witness'', will contradict the particular cross-world independence that is assumed in canonical identification strategies for $\gamma_{NDE}$. These variables commonly arise for investigators who construct causal DAGs as a routine part of their model specification process. As such, point-identification of $\gamma_{NDE}$ is precluded for investigators in many settings. In this context, a class of novel parameters involving stochastic regimes were introduced as alternatives to $\gamma_{NDE}$. In this motivating example, we review one of these parameters, termed the ``randomized interventional analogue'' of the natural direct effect, deferring full consideration of this class of parameters until Section \ref{sec: case1}.

\subsection{Defining randomized interventional analogues}

The randomized interventional analogues of the natural direct and indirect effects   $\gamma_{RDE}$ and $\gamma_{RIE}$ are defined as
\begin{align}  
    \gamma_{RDE}(P^F) = \mathbb{E}_{P^F}[Y^{a=1, G_{a=0}} - Y^{a=0, G_{a=0}}], \label{RDE}\\
    \gamma_{RIE}(P^F) = \mathbb{E}_{P^F}[Y^{a=1, G_{a=1}}- Y^{a=1, G_{a=0}}]. \label{RIE}
\end{align}

These effects are defined by counterfactuals of the type $Y^{a, G_{a^*}}$, where $G_{a^*}$ indicates a variable following the counterfactual distribution of the mediator $M$ under an intervention that sets $A$ to $a^*$. We let $\pi_{P^F}^{a*}$ denote this distribution. Thus, $Y^{a, G_{a^*}}$ is the potential outcome that would occur under an intervention that sets treatment $A$ to $a$ and sets $M$ to a value randomly drawn from $\pi_{P^F}^{a*}$. Web Appendix C provides additional details on the definition of  $\gamma_{RDE}$ that clarifies the features that distinguish  $\gamma_{RDE}$ and $\gamma_{NDE}$ in terms of experimental identifiability. For simplicity, we consider a setting without baseline covariates, but $\gamma_{RDE}$ may also be defined such that $M$ is assigned according to the counterfactual distribution of the mediator $M$ conditional on some pre-treatment confounders $L$.

We focus on $\gamma_{RDE}$ for historical reasons (see Section \ref{sec: case1}), but $\gamma_{RDE}$ is certainly not the only parameter defined by stochastic interventions on the mediator using counterfactual distributions. We also introduce an alternatively defined randomized interventional analogue $\gamma_{RDE-W}$, because they have been used to heuristically construct a conceptual bridge between $\gamma_{RDE}$ and $\gamma_{NDE}$: under different models $\gamma_{RDE-W}$ may either coincide with $\gamma_{NDE}$ or be understood as a special case of  $\gamma_{RDE}$ that conditions on pre-treatment covariates. Specifically, we define $\gamma_{RDE-W}$ as the expected outcome under a stochastic intervention using the \textit{conditional} distribution of $M^{a^*}$ given $W^{a^*}$,
\begin{align}  
    \gamma_{RDE-W}(P^F) = \mathbb{E}_{P^F}[Y^{a=1, G_{a=0 \mid W}} - Y^{a=0, G_{a=0\mid W}}]. \label{RDE-W}
\end{align}

Similarly, $\gamma_{RDE-W}$ is defined by counterfactuals of the type $Y^{a, G_{a^*\mid W}}$. Here, $G_{a^*\mid W}$ indicates a variable following the counterfactual distribution of the mediator conditional on the intermediate covariate $W$ under an intervention that sets $A$ to $a^*$. We can let $\pi_{W, P^F}^{a*}(w)$ denote that particular distribution conditional on $W^{a^*}=w$.  Then, $Y^{a, G_{a^*\mid W}}$ is the potential outcome that would occur under the following regime: first, treatment $A$ is set to $a$ and subsequently the value of $W^a$ is observed (say, $w$); second, $M$ is set to a value randomly drawn from the distribution $\pi_{W, P^F}^{a*}(w)$. Note that $W$ is in general distinguished crucially from a pre-treatment confounder $L$ because $W$ may possibly be caused by treatment $A$. However, under models in which $W$ is \textit{a priori} assumed a non-descendent of $A$, $\gamma_{RDE-W}$ has simply been treated as a particular randomized interventional analogue $\gamma_{RDE}$ where $W$ is subsumed into the vector of baseline ``pre-treatment'' covariates $L$. 

\subsection{Identifying randomized interventional analogues}

We now review some classical identification results that have historically been used to justify conceptual connections between $\gamma_{RDE}$ and $\gamma_{NDE}$. Suppose that an investigator defines $\mathcal{M}_1$ solely via conditional independencies between treatment, mediator, and outcome counterfactuals that could directly be tested in ideal experiments (see Web Appendix C for details). Investigators might justify the assumptions of $\mathcal{M}_1$ through beliefs in no-unmeasured confounding between $\{A, M, Y\}$. It is well known that $\gamma_{NDE}$ is not identified under $\mathcal{M}_1$. In contrast, both $\gamma_{RDE}$ and $\gamma_{RDE-W}$ are identified by $\Psi_{RDE}$ and $\Psi_{RDE-W}$, respectively,
\begin{align}
    \Psi_{RDE}(P) & = \int_{m}\int_{w}\mathbb{E}_P[Y \mid m, w, a] f_P(m \mid a^*)f_P(w \mid a)d\mu(w,m), \label{eq: psi RDE} \\
    \Psi_{RDE-W}(P) & = \int_{m}\int_{w}\mathbb{E}_P[Y \mid m, w, a] f_P(m \mid w, a^*)f_P(w \mid a)d\mu(w,m). \label{eq: psi RDE-W}
\end{align}

An investigator might further consider augmenting $\mathcal{M}_1$ with additional cross-world independence assumptions such that $\gamma_{NDE}$ is classicly identified by $\Psi_{RDE-W}$ under the resulting submodel $\mathcal{M}_2$ (see Web Appendix C for details). Investigators might justify the additional assumptions of $\mathcal{M}_2$ because cross-world independencies commonly follow from $NPSEM-IE$ models represented by DAGs without unmeasured confounders \citepMain{pearl2009causality}. 

Since, $\mathcal{M}_2$ is a submodel of $\mathcal{M}_1$ and $\Psi_{RDE-W}$ identifies $\gamma_{RDE-W}$ under $\mathcal{M}_1$, then it certainly also identifies $\gamma_{RDE-W}$ under $\mathcal{M}_2$. Thus, $\gamma_{NDE}\equiv\gamma_{RDE-W}$ under $\mathcal{M}_2$: the natural direct effect $\gamma_{NDE}$ and the particular randomized interventional analogue $\gamma_{RDE-W}$ are conceptually linked under $\mathcal{M}_2$, because their values coincide at all laws in this model. This strict concordance between $\gamma_{NDE}$ and $\gamma_{RDE-W}$ under models like $\mathcal{M}_2$ (in which $\gamma_{NDE}$ is identified) has been emphasized in many articles focusing on methodological development for $\gamma_{RDE}$ (see Section \ref{sec: case1}). Under such models, $W$ will surely be a non-descendent of $A$, since a causal structure to the contrary would contradict the cross-world independences that defines $\mathcal{M}_2$. Thus, $\gamma_{RDE-W}$ is also conceptually linked to $\gamma_{RDE}$  under $\mathcal{M}_2$, because $W$ may be considered in an analysis as a pre-treatment covariate and thus $\gamma_{RDE-W}$ may be understood as a particular $\gamma_{RDE}$.

\subsection{An example of \textit{identity slippage}}

We emphasize that the concordances and definitional equivalencies between  $\gamma_{RDE}$, $\gamma_{RDE-W}$, and $\gamma_{NDE}$ only hold under laws in models like $\mathcal{M}_2$ in which $\gamma_{RDE-W}$, and $\gamma_{NDE}$ coincide and $W$ is not possibly a recanting witness \citepMain{avin2005identifiability}. Ironically, it is this latter feature of these laws that will often lead investigators originally interested in $\gamma_{NDE}$ to abandon $\mathcal{M}_2$ and revert to $\mathcal{M}_1$: investigators will often struggle to justify the absence of a recanting witness. These investigators will often shift attention to $\gamma_{RDE}$ (identified under $\mathcal{M}_1$) as their target of inference, even though the conceptual relations between $\gamma_{NDE}$ and $\gamma_{RDE}$ may not hold for some laws included in their selected model $\mathcal{M}_1$. 

Suppose such an investigator deduces the correct functional of observed data parameters, $\Psi_{RDE}$, that identifies $\gamma_{RDE}$ under $\mathcal{M}_1$, and suppose further that they select a consistent estimator of $\Psi_{RDE}$ under $\mathcal{M}_1$ and compute $\hat{\psi}_{RDE}$ from their observed data. However, in remarks, the investigator interprets $\hat{\psi}_{RDE}$ \textit{as if} it were an estimate for $\gamma_{NDE}$. For example, they may make the claim $\mathfrak{i}$ that $\{$``the mediator $M$ explains $X\%$ of the total effect of $A$ on $Y$''$\}$: they have made a claim $\mathfrak{i} \in \mathcal{I}_{\gamma_{NDE}}(\hat{\psi}_{RDE})$, and further $\mathfrak{i} \notin \mathcal{I}_{\gamma_{RDE}}(\hat{\psi}_{RDE})$, because $\gamma_{RDE}$ and $\gamma_{RIE}$ do not generally decompose the total effect. This would be a case of \textit{identity slippage}.

\section{Pragmatic and ideal frameworks for causal inference} \label{sec: pragideal}

The motivating example in Section \ref{sec: contemp motiv} illustrated one way investigators may make an error of \textit{identity slippage}, by interpreting an estimate of an identified randomized interventional analogue $\gamma_{RDE}$ in \eqref{RDE} as an unidentified natural direct effect $\gamma_{NDE}$ in \eqref{NDE}. In particular, the example illustrated a process of model and causal parameter selection which created the cognitive conditions for this type of error: a parameter of primary interest, here $\gamma_{NDE}$, is first considered with respect to a canonical model in which it is identified. Under this model, it is noted that the $\gamma_{NDE}$ concords with $\gamma_{RDE-W}$, a member of a class of parameters defined by stochastic interventions on the mediator. This canonical model, however, is rejected due to the implausibility of a subset of its defining assumptions, e.g. the absence of recanting witnesses. Under the resulting super-model, a member of this class of parameters defined by stochastic interventions, $\gamma_{RDE}$, is identified, and, while the original scientific task is retained, this parameter becomes the nominal target of inference. Cognitively, the investigator has \textit{shifted} focus from a parameter of original interest to what they believe is an semantically proximal parameter: i.e. there is a belief that the images of $\mathcal{I}_{\gamma_{NDE}}$ and $\mathcal{I}_{\gamma_{RDE}}$ substantially overlap. But then there is an interpretational \textit{slip} back to the parameter of original interest. In Web Appendix A, we present a second elaborated example of \textit{identity slippage} between expected potential outcomes under so-called longitudinal modified treatment plans \citepMain{diaz2021mtpJASA}, and parameters under dynamic regimes defined by stochastic interventions. Therein, we illustrate an isomorphic process of model and parameter selection. 

\subsection{A pragmatic framework}

We conjecture that the cognitive conditions in these examples arise from a particular framework for parameter and model selection. In this process, an investigator with a particular scientific question is permitted to select from a menu of parameters with different identifiability conditions. Under this process, different parameters could sensibly be considered different ``alternatives'' to answering a fixed scientific question, each with their own properties, including those of identification. We refer to this process as a \textit{pragmatic} framework for causal inference because it views parameters as practical strategies whose selection could in principle be driven by modeling beliefs and the data on hand. Figure \ref{fig: concept} illustrates this feature with the set of blue arrows between the scientific question and the \textit{set} of parameters $\{\gamma_1, \gamma_2, \gamma_3\}$, representing a one-to-many correspondence between scientific question and parameters. 

Apparently, \textit{pragmatic frameworks} operate both implicitly and explicitly in different scientific literatures. For example a \textit{pragmatic framework} seems to operate implicitly for investigators who routinely implement instrumental variable analyses and commonly select the LATE as their parameter of interest. Indeed, \citetMain{imbens2014instrumental} states ``\textit{The local average treatment effect is an unusual estimand}'' and later writes that, in a ``\textit{typical approach}'' to a LATE analysis, \textit{``it appears difficult to argue that [the principal stratum] is a substantially interesting group, and in fact no attempt [is] made to do so}''. Similarly, \citetMain{pearl2011principal} observes of the proponents of LATEs that: \textit{``Realizing that the [ATE] is not identifiable in experiments marred by noncompliance, they have shifted attention to a specific response type (i.e., compliers) for which the causal effect was identifiable, and presented the latter as an approximation for the ATE.}''

Alternatively, an explicit formulation of a \textit{pragmatic} framework can be found in an emerging literature on ``assumption-lean inference'' \citepMain{vansteelandt2020assumption}. Therein, investigators are explicitly invited to consider a \emph{pair} of parameters $\{\gamma_1, \gamma_2\}$ and a \emph{pair} of nested models $\{\mathcal{M}_2 \subset \mathcal{M}_1\}$ and do inference on $\Psi(P_0)$ with guarantees that $\Psi(P) = \gamma_1(P^F) =\gamma_2(P^F)$ under $\mathcal{M}_2$ and $\Psi(P) = \gamma_1(P^F)$ under $\mathcal{M}_1$. An investigator committed to a particular scientific question need not have any commitment to a particular parameter: the two parameters represent two approaches and the specific interpretation of $\Psi$ depends on whether $P^F_0\in \mathcal{M}_2$ or $P^F_0\in \mathcal{M}_1\setminus\mathcal{M}_2$, which of course will remain unknown. For another example of an explicitly pragmatic approach in an applied setting, see also \citetMain{rudolph2022effects}, titled ``\textit{When effects cannot be estimated: redefining estimands to understand the effects of naloxone access laws}''.

\subsection{An ideal framework}
In contrast to the \textit{pragmatic} framework, an investigator might adopt a process in which a scientific question is synonymous with a single causal parameter. In this sense, parameter selection would be a primitive step in the formulation of a scientific inquiry, as it is, for example, in the `Causal Roadmap' promoted by \citetMain{van2011targeted}. We call this the \textit{ideal} framework, which has also been referred to by \citetMain{vansteelandt2020assumption} as the ``hygienic causal inference approach.'' Figure \ref{fig: concept} illustrates this feature with the single bolded black arrow between the scientific question and parameter $\gamma_2$, representing a one-to-one correspondence between between the scientific question and parameter.  \citetMain{pearl2011principal} alludes to such an approach when he wonders whether investigators targeting the LATE are \textit{``motivated by mathematical convenience, mathematical necessity (to achieve identification) or \emph{a genuine interest in the stratum under analysis}.''} (emphasis added.)

In the \textit{ideal} framework, an understanding of a causal parameter as an ``approach'' to a scientific question, or an understanding of one parameter as an ``alternative'' to another parameter, is nonsensical. If causal parameters \textit{are} questions, then they cannot possibly be considered \textit{approaches} to questions (for which several ``alternatives'' could exist).  

Although, some, e.g. \citetMain{vansteelandt2020assumption}, cite advantages of a \textit{pragmatic} framework, we argue that, at the very least, a \textit{pragmatic} framework is more conducive to \textit{identity slippage} than an \textit{ideal} one. For example, under an \textit{ideal} framework an investigator could only correctly select $\gamma_{RDE}$ as a target of interest if, say, an investigator was for some reason truly considering implementing a policy that randomly assigned a mediator according to some distribution, which happened to be that of the mediator under a specific treatment assignment. We would expect interpretive slippage to the NDE would be very unlikely for this investigator because the relations between the NDE and the RDE would not likely be in the cognitive foreground.


\section{Systematic review: stochastic mediation parameters}\label{sec: case1}

We hypothesize that the stochastic mediation literature is primarily developed in the context of a \textit{pragmatic} framework for causal inference. We thus provide a comprehensive evaluation of the relative prominence of a \textit{pragmatic} vs.\ an \textit{ideal} framework to causal inference in this methodological literature. We then describe and present the results of a systematic review of the \textit{applied} stochastic mediation literature, which we conduct in order to evaluate the relative prominence of \textit{pragmatic} vs.\ an \textit{ideal} frameworks in this applied literature and to measure the prevalence of \textit{identity slippage} therein. 

\subsection{Stochastic mediation parameters under a pragmatic framework}\label{subsec: case1 met}

We identified 22 articles published in leading statistics and epidemiology journals that introduce or develop statistical methods for causal parameters defined by stochastic interventions involving a variable $M$ that hypothetically mediates a causal effect of treatment $A$ on an outcome $Y$. The references to these articles are listed in Web Appendix D. These articles are grouped in Tables \ref{tab: RIAstoch} and \ref{tab: altstoch} according to the specific causal parameter they would map to if expressed in potential outcomes notation for a single time-point, single-mediator setting. Specifically, Table \ref{tab: RIAstoch} lists articles developing $\gamma_{RDE}$, popularly termed the ``randomized interventional analogue to the natural direct effect'', but also referred to by various authors as the ``generated'', ``stochastic'', ''interventional'', or even simply the ``natural'' direct effect.  Table \ref{tab: altstoch}  lists definitionally distinct stochastic mediation parameters.

To the best of our knowledge, this set of articles comprises the complete methodological literature for stochastic mediation parameters. We systematically reviewed the natural language remarks in this literature concerning their motivations, their definitions, their interpretations, and their use of stochastic mediation parameters in practice. We identified two recurring motifs consistent with a \textit{pragmatic} framework for causal inference. We let $\gamma_1$ and $\gamma_2$ represent the stochastic mediation parameter and a natural direct effect parameter, respectively, and let $\mathcal{M}_2 \subset \mathcal{M}_1$ be two nested models. Then, these two motifs are
\begin{enumerate}
    \item \textbf{Identifiability concerns:} the statistical formulation of the novel parameter ($\gamma_1$) being explicitly motivated, not entirely by concordance with real subject-matter scientific and policy questions, but also by properties of identifiability, namely the non-necessity of assumptions (in the set defining $\mathcal{M}_2$) that were necessary for the identification of an index parameter ($\gamma_2$, under $\mathcal{M}_2$), and 
    \item \textbf{Interpretability concerns:}  a nomenclature and/or description for the novel parameter ($\gamma_1$) that makes explicit reference to the a corresponding index parameter ($\gamma_2$), thereby implying a strong conceptual proximity between them. 
\end{enumerate}

Evidence of these motifs is summarized in Tables \ref{tab: RIAstoch} and \ref{tab: altstoch}, and is extensively documented in Web Appendix D; for each article we abstract prominent excerpts (when present) of \textit{identifiability concerns} and \textit{interpretability concerns}. Under the motif of \textit{interpretability concerns}, 15 of the 22 articles refer to the stochastic mediation parameter as an ``approach'' to an unelaborated mediation question  and 14 of 22 as an ``alternative'' to other definitions of direct and indirect effects. Eleven of 22 articles explicitly name or describe the stochastic mediation parameter as \textit{analogues} to the natural direct effect, $\gamma_{NDE}$, and 5 articles even refer to the stochastic parameter as a ``natural direct effect'' itself. Finally, 15 of the 22 articles emphasized the concordance of the stochastic mediation parameter and $\gamma_{NDE}$ when $\gamma_{NDE}$ is identified, a fact originally noted by \citetMain{robins2003semantics} in the earliest known articulation of the parameter $\mathbb{E}_{P^F}[Y^{a=1, G_{a=0}}]$.

Under the motif of \textit{identifiability concerns}, nearly every single article in this literature emphasizes the identifiability of the stochastic mediation parameter in settings where the natural direct effect is not identified. The lone exceptions is the article by \citetMeth{didelez2012direct}, perhaps because they develop causal parameters in a decision theoretic framework without counterfactuals in which $\gamma_{NDE}$ is not sensibly defined.

Several more recent articles, however, use stronger language, framing the stochastic mediation parameter as ''solutions'' to identifiability issues, or as actors that themselves ``relax'', ``avoid'', ``overcome'', ``resolve'', or ``allow identification'' in the absence of cross-world assumptions  (see Table \ref{tab: rationales} and Web Appendix D). As perhaps the most extreme example, \citetMain[pg. ~7]{van2008direct} invite an investigator to directly base their choice of causal parameter on its identifiability, writing: 

{\small{\textit{``Therefore, if one is not comfortable with the identifiability [assumptions of the NDE], then one can view the [proposed parameter] as the parameter of interest$[\dots]$''}}}

\newgeometry{margin=1cm} 
\begin{landscape}
\thispagestyle{empty}

{\renewcommand{\arraystretch}{0.9}
\begin{table}[!htb]
\begin{threeparttable}
\caption{Randomized-interventional analogues and equivalent parameters. \label{tab: RIAstoch}}
\centering
\begin{tabular}{|m{2.2cm}|m{2cm}|m{3cm}|m{8cm}|m{9cm}|}
\toprule
\multirow{2}{*}{\textbf{\emph{Parameter}} }& \multirow{2}{*}{\textbf{\emph{Author}}} & \multirow{2}{*}{\textbf{\emph{Name}}} &  \multicolumn{2}{|c|}{\small{\textbf{\emph{Excerpts}}}} \\ \cline{4-5} 
   & & &\multicolumn{1}{|c|}{\small{\textbf{{Identifiability}}}} & \multicolumn{1}{|c|}{{\textbf{Interpretability}} }     \\  \hline
 \multirow{8}[75]{*}{$\mathbb{E}_{P^F}[Y^{a, G_{a*\mid L}}]$}   & \tiny{\citeMeth[pg. ~204, 205]{geneletti2007identifying}}&  {Generated DE} & \tiny{{``Thus, in the DT framework[$\dots$], neither the assumption of \citetSM{pearl2001direct} nor of \citetMeth{petersen2006estimation} is necessary; nor are any additional counterfactual assumptions (see \citetSM{robins2003semantics}) necessary for the identification of direct and indirect effects[$\dots$].''}} & \tiny{{``This \textbf{approach} is similar to Pearl's [NDE].''} }      \\  \cline{2-5}  
 & \tiny{\citeMeth[pg. ~6]{didelez2012direct}} & {Natural DE} & \centering * &   \tiny{{``The NDE can be formulated as a special case of a standardized DE[$\dots$].''}}    \\  \cline{2-5}  
 & \tiny{\citeMeth[pg. ~303, 300, 304]{vanderweele2014effect}}&  \multirow{5}{3cm}{{Randomized interventional \textbf{analogue} of the NDE}} & \tiny{{``Although [NDEs] with $M$ alone as the mediator of interest are not identified, \textbf{alternative} effects that randomly set $M$ to a value chosen from the distribution of a particular exposure level can be identified.''}} & \tiny{{``[$\dots$]we describe three \textbf{alternative} \textbf{approaches} to effect decomposition[$\dots$].''; ``These expressions reduce to the mediation formulae[$\dots$]when [$W$] does not confound the association between $M$ and $Y$[$\dots$].''} }      \\  \cline{2-2}\cline{4-5}  
 & \tiny{\citeMeth[pg. ~126, 183]{vanderweele2015explanation}}& & \tiny{{``[$\dots$]we will discuss \textbf{alternative approaches} to [NDEs] that can be used in the presence of exposure-induced $[M-Y]$ confounders even though [NDEs] themselves are not identified.''}} & \tiny{{``One way to \textbf{approach} the interpretation of the estimators for [NDEs$\dots]$is that, without further assumptions, we can still interpret them as the interventional analogues of [NDEs]. If we are further willing to make the cross-world independence assumption, then we can also interpret these estimates as estimates of the [NDEs] themselves.''
 }}      \\ \cline{2-2}\cline{4-5} 
 & \tiny{\citeMeth[pg. ~157]{chen2016mediation}} & & \multicolumn{2}{|c|}{\tiny{{``[$\dots$]in settings [with a recanting witness, NDEs] are not identified. Instead we consider randomized interventional \textbf{analogs} of these effects[$\dots$].''}}}        \\ \cline{2-2}\cline{4-5}    
 & \tiny{\citeMeth[pg. ~923, 927]{vanderweele2017mediation}} & & \tiny{{``Although these interventional[$\dots$] effects defined here are not identical to [the NDE], they are in some sense the best we may be able to do as the [NDEs] themselves will not be identified when a [$M-Y$] confounder is affected by the exposure;''
 ''}} & \tiny{{``As yet another \textbf{alternative}, though one that we argue is not suitable for mediation analysis, we could have[$\dots$].''}}      \\ \cline{2-2}\cline{4-5} 
  & \tiny{\citeMeth[pg. ~6, 1]{lin2017interventional}} & & \tiny{{``When the time-varying confounders are not affected by exposure[$\dots$ the RIA ]reduces to[$\dots$]the expression of the standard mediation parameter[$\dots$].''}} & \tiny{{``Mediation analysis is a \textbf{technique} to decompose the total effect of an exposure on an outcome into a direct effect (the effect not through a mediator) and an indirect effect (the effect through a mediator).''} }      \\ \cline{2-5} 
 & \tiny{\citeMeth[pg. ~3, 2]{rudolph2018robust}}& \multirow{2}{3cm}{{Stochastic DE}} &\tiny{{``There has been recent work to \textbf{relax} the assumption of no intermediate confounder, $M_{a^*} \CI Y^{a,m} \mid W$, by using a stochastic intervention on $M$.''}} & \tiny{{``The SDE and SIE coincide with the NDE and NIE in the absence of intermediate confounders.''}}      \\ \cline{2-2}\cline{4-5}  
 & \tiny{\citeMeth[pg. ~199, 199]{rudolph2021transporting}}& &\tiny{{``Stochastic [DEs] are similar to [NDEs] but do not require the[$\dots$]assumption of no[$\dots$]post-treatment confounder[$\dots$].''}} & \tiny{{``Stochastic [DEs$\dots$], also called randomized interventional [DEs], represent the[$\dots$]direct effect of $A$ on $Y$ not through $M$[$\dots$].''}}      \\ \cline{2-5} 
  & \tiny{\citeMeth[pg. ~258, 260]{vansteelandt2017interventional}}& \multirow{5}{3cm}{{Interventional DE}} & \tiny{{``These concerns all originate from the fact that [NDEs] are defined in terms of so-called cross-world counterfactuals that are unobservable, even from experimental data; they call for \textbf{alternative} effect measures that are less remote from the observed data.''}} & \tiny{{``Under these assumptions, these effects reduce to[$\dots$]average direct and indirect interventional effects, but with $L$ empty. It thus follows that in single mediator models without post-treatment confounding, [NDEs] obtained under assumption (iv) can also be interpreted as interventional [DEs] (even when that assumption is violated).}}      \\ \cline{2-2}\cline{4-5}
   & \tiny{\citeMeth[pg. ~5086, 5095]{mittinty2019effect}}&  &  \tiny{{``These can be identified under much weaker conditions than [NDEs], but sum to a total interventional causal effect, not the TCE.''}} &   \tiny{{``These expressions reduce to mediation formulae[$\dots$]when [$W$] does not confound the association between $M_1,M_2$, and $Y$[$\dots$].''}}   \\ \cline{2-2}\cline{4-5}
   & \tiny{\citeMeth[pg. ~172, 172]{benkeser2021nonparametric}}&  & \tiny{{``A debate in this literature has emerged pertaining to the reliance[$\dots$]on cross-world independence assumptions that are fundamentally untestable even in randomized controlled experiments.''}} & \tiny{{``A second \textbf{approach} considers seeking \textbf{alternative} definitions of mediation parameters that do not require such cross-world assumptions[$\dots$].''}}      \\ \cline{2-2}\cline{4-5} 
    & \tiny{\citeMeth[pg. ~2, 2]{loh2020heterogeneous}}&  & \tiny{{``This distinction \textbf{allows} identification of interventional indirect effects in the context of multiple mediators under empirically verifiable assumptions[$\dots$].''}}  & \tiny{ {``[$\dots$]the natural and interventional effects may coincide empirically.''}}    \\ \cline{2-2}\cline{4-5}
    & \tiny{\citeMeth[pg. ~627, 631]{diaz2021nonparametric}}&  & \tiny{{``Interventional effects for mediation analysis were proposed as a solution to the lack of identifiability of natural (in)direct effects in the presence of a mediator-outcome confounder affected by exposure.''}} & \tiny{ {``To \textbf{overcome} this problem while retaining the path decomposition employed by the [NDEs], we adopt an \textbf{approach}[$\dots$]defining [DEs] using stochastic interventions on the mediator.''}}      \\ \hline 

\bottomrule
\end{tabular}
 \begin{tablenotes}
      \tiny
      \item N(I)DE: Natural (in)direct effect; DE: direct effect; S(I)DE: stochastic (in)direct effect; M: mediator; Y: outcome; L: pre-treatment covariate; W: post-treatment covariate; DT: decision theoretic; TCE: total causal effect.
      \item We use [brackets] to indicate when we have modified excerpts, for brevity, and refer readers to Tables \ref{tab: Intconcern1} to \ref{tab: IDconcern3}, which provide excerpts in their entirety. 
      \item $Y^{a, G_{a*\mid L}}$ is defined by a regime in which patients are assigned treatment level $a$ and a patient with baseline covariate $L=l$ is assigned a value of the mediator according to the conditional distribution of $M^{a*}$ given  $L=l$. 
      \item The parameters presented by \citetMeth{geneletti2007identifying} and \citetMeth{didelez2012direct} are not defined in terms of counterfactuals - we consider them equivalent to $\mathbb{E}_{P^F}[Y^{a, G_{a*\mid L}}]$ because the regime defining the potential outcome $Y^{a, G_{a*\mid L}}$ corresponds parameter considered by these authors, which they interpret as the expected outcome in a trial that implements that regime.
      \item \citet{vanderweele2017mediation} and \citet{chen2016mediation} describe extensions for sequential treatments and mediators, and \citet{lin2017interventional},  \citet{mittinty2019effect}, \citet{benkeser2021nonparametric}, and \citet{loh2020heterogeneous} describe extensions for multiple mediators or pathways; each describe parameters that reduce to $\mathbb{E}_{P^F}[Y^{a, G_{a*\mid L}}]$ for the special case of a single time-point, single mediator setting.
    \end{tablenotes}
\end{threeparttable}
\end{table}
}

\end{landscape}
\restoregeometry

\newgeometry{margin=1cm} 
\begin{landscape}
\thispagestyle{empty}

{\renewcommand{\arraystretch}{1}
\begin{table}[!htb]
\begin{threeparttable}

\caption{Additional stochastic mediation parameters \label{tab: altstoch}}


\begin{tabular}{|p{5.8cm}|p{1,5cm}|p{2.5cm}|p{7cm}|p{7cm}|}
\toprule
\multirow{2}{*}{\textbf{\emph{Parameter}} }& \multirow{2}{*}{\textbf{\emph{Author}}} & \multirow{2}{*}{\textbf{\emph{Name}}} &  \multicolumn{2}{|c|}{\textbf{\emph{Excerpts}}} \\ \cline{4-5} 
   & & &\multicolumn{1}{|c|}{\textbf{{Identifiability}}} & \multicolumn{1}{|c|}{{\textbf{Interpretability}} }     \\  \hline
 \multirow{2}[15]{*}{$\mathbb{E}_{P^F}\Big[\sum\limits_{m}Y^{a,m}P(M^{a^*}=m \mid L)\Big]$}   & \tiny{\citeMeth[pg. ~280, 281]{petersen2006estimation}}&  \multirow{4}{2cm}{{Natural DE}} & \tiny{{{``We note that, even when our [DE] assumption (7) fails to hold, equation (3) still estimates an interesting causal parameter: a summary of the [DE] of the exposure in the population with [$M$] controlled at its mean counterfactual level in the absence of exposure.''}}} & \tiny{{``Similarly[$\dots$], the [NDE] is simply an average of the controlled [DEs] at each level of the [$M$] weighted with respect to the distribution of [$M$] in the unexposed.''} }      \\  \cline{2-2}\cline{4-5} 
 & \tiny{\citeMeth[pg. ~24, 3]{van2008direct}} & & \tiny{ {``We note that these \textbf{alternative} definitions no longer depend on untestable assumptions regarding unmeasured confounding, nor are any additional assumptions necessary to make the [NDE] identifiable.''}} &   \tiny{{``The article focuses on modelling and estimation of this [NDE] parameter, which happens to agree with the conventional [NDE] parameter[$\dots$]under any of the identifying assumptions of \citetSM{robins1992identifiability}, \citetSM{van2004estimation} or \citetSM{pearl2001direct}, but does not rely on any of these additional assumptions.''}}   \\  \cline{1-2}\cline{4-5}
 \multirow{2}[15]{*}{$\mathbb{E}_{P^F}[Y^{a, G_{a*\mid L, W}}]$}   & \tiny{\citeMeth[pg. ~3, 30]{zheng2012causal}}&    & \tiny{{``Consequently, the identifiability conditions of the resulting parameters would impose restrictions on the event indicators — conditions that we find too strong for the purpose of a survival study[$\dots$].''}} &  \tiny{{``As an \textbf{alternative}, we proposed to adopt the stochastic interventions (SI) \textbf{approach} of \citetMeth{didelez2012direct}, where the mediator is considered an intervention variable, onto which a given distribution is enforced.''}}     \\  \cline{2-2}\cline{4-5}
& \tiny{\citeMeth[pg. ~2, 2]{zheng2017longitudinal}} & & \tiny{ {``The corresponding [NDE] parameters have different interpretation than those under the formulations in \citetSM{robins1992identifiability} and \citetSM{pearl2001direct}, but their identifiability is at hand even in the presence of [a recanting witness].''}} &   \tiny{{``Under the proposed formulation, one can obtain a total effect decomposition and the subsequent definition of [NDEs] that are \textbf{analogous} to those in \citetSM{pearl2001direct}.''}}   \\  \hline  \hline 
\multirow{2}[15]{*}{$\mathbb{E}_{P^F}[Y^{a, i=1}]$}   & \tiny{\citeMeth[pg. ~4009, 4016]{lok2016defining}}&  \multirow{4}{2cm}{{Organic DE}} & \tiny{{{``To \textbf{overcome} these problems, this article proposes instead to base mediation analysis on newly defined ``organic'' interventions ($I$) on the mediator.''}}} & \tiny{{``I have shown that the proposed organic direct and indirect effects are identified by the same expressions as developed previously in the literature for [NDEs].''} }      \\  \cline{2-2}\cline{4-5}  
 & \tiny{\citeMeth[pg. ~414, 418]{lok2021causal}} & & \tiny{ {``Organic indirect and direct effects: an intervention-based \textbf{approach} \textbf{avoiding} cross-worlds counterfactuals and assumptions''}} &   \tiny{{``Under the respective conditions usually made and if there are no [recanting witnesses], [NDEs], their randomized counterparts, and their organic counterparts relative to $a = 1$ all lead to the same numerical results in settings where these effects are all well defined.''}}   \\ \hline  \hline 
$\text{ }$ \newline $\mathbb{E}_{P^F}[Y^{g, M}]$ & \tiny{\citeMeth[pg. ~2, 2]{diaz2020causal}} &  {Population
intervention DE}  & \tiny{ {``In an attempt to \textbf{solve} these problems, several researchers have proposed \textbf{methods} that do away with cross-world counterfactual independences.''}} &   \tiny{{``These \textbf{methods} can be divided into two types: identification of bounds and \textbf{alternative} definitions of the (in)direct effect. Here, we take the second \textbf{approach}[$\dots$].''}}   \\ \hline  \hline
$\text{ }$ \newline $\mathbb{E}_{P^F}[Y^{g, G}]$ & \tiny{\citeMeth[pg. ~4, 7]{hejazi2020nonparametric}} & {Stochastic
interventional DE}  & \tiny{ {``Our proposed class of effects are the first to simultaneously \textbf{avoid} the requirement of cross-world counterfactual independencies[$\dots$]and remain identifiable despite intermediate confounding.''}} &   \tiny{{``Decomposition as the sum of direct and indirect effects affords an interpretation \textbf{analogous} to the corresponding standard decomposition of the average treatment effect into the [NDE] and [NIE].''}}   \\ \hline  \hline 
 \hline
\end{tabular}
 \begin{tablenotes}
      \tiny
      \item N(I)DE: Natural (in)direct effect; DE: direct effect; M: mediator; Y: outcome; L: pre-treatment covariate; W: post-treatment covariate.\
      \item We use [brackets] to indicate when we have modified excerpts, for brevity, and refer readers to Tables \ref{tab: Intconcern1} to \ref{tab: IDconcern3}, which provide excerpts in their entirety. 
      \item $Y^{a, G_{a*\mid L}}$ is defined by a regime in which patients are assigned treatment level $a$ and a patient with baseline covariate $L=l$ is assigned a value of the mediator according to the conditional distribution of $M^{a*}$ given  $L=l$. 
      \item $Y^{a, G_{a*\mid L, W}}$ is defined by a regime in which patients are assigned treatment level $a$ and a patient with baseline covariate $L=l$ and post-treatment covariate $W^{a}=w$ is assigned a value of the mediator according to the conditional distribution of $M^{a*}$ given  $L=l$ and $W^{a*}=w$. 
      \item $Y^{a, i=1}$ is defined by a regime in which patients are assigned treatment level $a$ and a patient is assigned a value of some exogenous variable $I=i$ for which (by definition) $M^{a=1, i=1}$ given $L=l$ follows the same distribution as $M^{a=0}$ given $L=l$ and for which $Y^{a=1, i=1}$ given $M^{a=1, i=1}=m$ and $L=l$ follows the same distribution as $Y^{a=1}$ given $M^{a=1}=m$ and $L=l$. 
      \item $\mathbb{E}[Y^{g, M}]$ is defined by a regime in which patients are assigned treatment level according to some arbitrary distribution $g$ (possibly a function of $L$) and each patient is deterministically assigned the value of the mediator $M$ that they would have received under the factual (unknown) regime. 
      \item $\mathbb{E}[Y^{g, G}]$ is defined by a regime in which patients are assigned treatment level according to some arbitrary distribution $g$ (possibly a function of $L$ and/or $P_0^{full}$) and each patient with baseline covariate $L=l$ and assigned treatment level $a$ is then assigned a value of the mediator according to the conditional (factual) distribution of $M$ given  $A=a$ and $L=l$. 
    \end{tablenotes}
\end{threeparttable}

\end{table}
}

\end{landscape}
\restoregeometry
\doublespacing

\subsection{Evidence of \textit{identity slippage} in the applied mediation literature}\label{subsec: case1 app}

Evidently, stochastic mediation parameters constitute ``techniques''/``approaches''/``methods'' that may be substituted as ``alternative'' parameters in the service of a general causal mediation inquiry (see Web Appendix D). In Section \ref{sec: pragideal}, we conjectured that a \textit{pragmatic framework} is more conducive to errors of \textit{identity slippage}: under a \textit{pragmatic framework}, an investigator with a fixed scientific task might be inclined to choose their causal parameter from a set of acceptable approaches based not on its precise causal content, but on its statistical properties. Then, an investigator will have the opportunity to make the error of interpreting their computed quantity not as their originally selected parameter, but as a different parameter that is also in the set of acceptable approaches for their scientific task. In other words, they will have a clear opportunity for \textit{identity slippage}. In contrast, an investigator following an \textit{ideal framework} will not have this same opportunity.

We found evidence that stochastic mediation parameters have primarily been developed and promoted within a \textit{pragmatic framework} for causal inference. We thus expect that there is an elevated frequency of \textit{identity slippage} within the applied literature estimating stochastic mediation parameters. Therefore, we designed and conducted a systematic review of a portion of this literature, specifically the population of applied articles that estimated $\gamma_{RDE}$, the randomized interventional analogue of the natural direct effects. The methods of the review, including inclusion/exclusion criteria, data abstraction, and analytic approach are described in Web Appendix E. 


\subsubsection{Results}\label{subsec: res}

The results of our online literature search are depicted in a PRISMA flow diagram in Figure \ref{fig: consort} of Web Appendix E. In total, the methodological articles developing $\gamma_{RDE}$ (Table \ref{tab: RIAstoch}) were cited 965 times. We screened 572 unique titles, 165 abstracts and the full-text of 86 articles, ultimately including 16 articles in our final review. Eight of these applied articles included an author of a methodological article on  $\gamma_{RDE}$, which we henceforth refer to as \textit{methods-authored} articles.

We were able to identify stated rationales for the $\gamma_{RDE}$ in 14 of the 16 articles (7/8 of \textit{methods-authored} articles). These excerpts are listed in Table \ref{tab: rationales}.

All but 2 of the identified rationales for the included articles, and all 7 of those for the \textit{methods-authored} articles, explicitly cited the identifiability properties of the $\gamma_{RDE}$ as the rationale for its selection. Of those 12 articles, 8 directly compared the identifiability of $\gamma_{RDE}$ to other ``causal mediation approaches'' \citepApp{wang2022does},``mediation methods'' \citepApp{rudolph2021explaining, casey2019unconventional},  ``alternative decompositions'' \citepApp{wodtke2020neighborhood}, or $\gamma_{NDE}$ itself \citepApp{komulainen2019childhood, hanson2019does, rudolph2021helped, rudolph2018mediation}. We provide one example below from \citetMain[pg. ~592]{rudolph2018mediation}, which includes authors on several of the most recent methodological articles on stochastic mediation parameters \citepMeth{rudolph2018robust, rudolph2021transporting, diaz2021nonparametric, diaz2020causal, hejazi2020nonparametric}:

{\small{\textit{``We estimated stochastic direct and indirect effects instead of the more common natural direct and indirect effects, because the stochastic effects do not require the absence of post-treatment confounding of the mediator-outcome relationship, $M^{a^*} \CI Y^{a,m} \mid W$, an assumption that is required for identifying their natural counterparts.''}}}

We therefore concluded that 12/14 of these articles (7/7 of \textit{methods-authored} articles) implicitly followed a \textit{pragmatic framework} to causal inference.

Of the 16 articles included in the review, only 5 indicated that they targeted a stochastic mediation parameter in the abstract. Not a single one of the non-\textit{methods-authored} articles indicated that they targeted a stochastic mediation parameter in the abstract.

\newgeometry{margin=1cm} 
\begin{landscape}
\thispagestyle{empty}

{\renewcommand{\arraystretch}{1}
\begin{table}[!htb]
\begin{threeparttable}

\caption{Stated rationales for stochastic mediation parameters in Applications \label{tab: rationales}}


\begin{tabular}{|m{3cm}|m{2cm}|m{2.5cm}|m{16cm}|}
\toprule
 & \textbf{\emph{Article}} & {\small \textbf{\emph{Parameter in abstract}}} & \textbf{\emph{Stated rationale for stochastic mediation parameter}} \\   \hline
 \multirow{8}{3cm}{{\small \textbf{{No authorship overlap}}}}   
   & \tiny{\citeApp[pg. ~1634]{wang2022mechanisms}} & No & \tiny{{{``This approach allowed us to also evaluate the confounding effect of $[W]$, which was identified as a post-exposure (post [$A$]) confounder[$\dots$].''}}} \\  \cline{2-4} 
   &\tiny{\citeApp[pg. ~1189]{wang2022does}} & No & \tiny{{{``Specifically, we estimated the so-called ‘interventional’ direct and indirect effects, which are estimable under relaxed assumptions compared to previous causal mediation approaches.''}}} \\  \cline{2-4}
   &\tiny{\citeApp{campbell2021placental}} & No & \tiny{{{[None provided]}}} \\  \cline{2-4}
   &\tiny{\citeApp[pg. ~20]{wodtke2020neighborhood}} & No & \tiny{{{``We focus on a decomposition defined in terms of randomized interventions on [$A$] because its components can be identified under more defensible assumptions than those required of other effect decompositions[$\dots$]unlike the components of alternative decompositions[$\dots$these effects ]can be identified in the presence of exposure-induced confounders.''}}} \\  \cline{2-4}
   &\tiny{\citeApp[pg. ~2]{liu2020inflammation}} & No & \tiny{{{``While the “total (causal) effect” of [$A$] on [$Y$] is ill-defined, it is relevant to examine the extent to which [$A$]-related disparities in [$Y$] would be reduced, if the distribution of [$M$] in those with $[A=1]$ were reduced to levels observed in[ those with $A=0$], thereby representing potential `interventional effects'.''}}} \\  \cline{2-4}
   &\tiny{\citeApp[pg. ~e197]{komulainen2019childhood}} & No & \tiny{{{``In the presence of exposure-induced confounding, [NDEs] are not identified, but randomized interventional analogs for [NDEs] can be estimated.''}}} \\  \cline{2-4}
   &\tiny{\citeApp[pg. ~155]{hanson2019does}} & No & \tiny{{{``The [RDE] is based on fewer assumptions and can therefore be identified in settings with time-varying exposures and mediators and when mediator-outcome confounding affected by exposure may be a problem.''}}} \\  \cline{2-4}
   &\tiny{\citeApp[pg. ~631]{gage2013maternal}} & No & \tiny{{{``Effect decomposition can then be estimated using a procedure similar to direct standardization. \citet{geneletti2007identifying} called this a `generated direct effect,' which is similar to Pearl’s [NDE]''}}} \\  \hline \hline
\multirow{8}{3cm}{{\small \textbf{{Authorship overlap}}}}   
   & \tiny{\citeApp[pg. ~337]{rudolph2021helped}} & Yes & \tiny{{{``[NDE and NIE] are common mediation estimands, together adding to the total effect, but are not identified in the presence of such a variable. Consequently, we estimate interventional (also known as stochastic) [DEs], which do not require the absence of post-treatment confounders of the [$M-Y$] relationship for identification but are analogous to [NDEs] in the absence of such variables.''}}} \\  \cline{2-4} 
   & \tiny{\citeApp[pg. ~2095]{rudolph2021explaining}} & No & \tiny{{{``Until recently, the statistical mediation methods available required numerous restrictive assumptions that are unlikely to hold in practice, and may result in significantly biased results.[$\dots$]We used a novel mediation approach that overcomes these limitations by allowing for[$\dots$]post-treatment intermediate outcomes.''}}} \\  \cline{2-4}
   & \tiny{\citeApp[pg. ~822, 824]{goin2020mediation}} & Yes & \tiny{{{``We believed that the assumption of no postexposure confounding was violated in this study[$\dots$]; therefore, we opted to estimate stochastic direct and indirect effects.''; ``By drawing from the distribution of [$M$] rather than assigning [$M$]  based on what we would expect it to be under different [values of $A$], we are able to estimate direct and indirect effects without needing to invoke the controversial `cross-world' assumption and can allow for postexposure confounders.''}}} \\  \cline{2-4}
   & \tiny{\citeApp{ploubidis2019lifelong}} & No & \tiny{{{[None provided]}}} \\  \cline{2-4}
   & \tiny{\citeApp[pg. ~1]{casey2019unconventional}} &  Yes &\tiny{{{``Our analytic approach toward evaluating the mediating effect of [$M$] on the association between [$A$] and [$Y$] allowed for the estimation of both direct and indirect pathways and for inclusion of post-exposure mediator-outcome confounders. The inability to deal with such confounders is a limitation of other mediation methods.''}}} \\  \cline{2-4}
   & \tiny{\citeApp[pg. ~592]{rudolph2018mediation}} & Yes & \tiny{{{``We estimated stochastic direct and indirect effects instead of the more common [NDEs], because the stochastic effects do not require the absence of posttreatment confounding of the [$M-Y$] relationship,[$\dots$]an assumption that is required for identifying their natural counterparts.''}}} \\  \cline{2-4}
   & \tiny{\citeApp[pg. ~882]{maika2017associations}} & Yes & \tiny{{{``Because of this intermediate confounding, we used the effect decomposition method derived[$\dots$]for effect decomposition in the presence of exposure-induced mediator-outcome confounding[$\dots$].''}}} \\  \cline{2-4}
   & \tiny{\citeApp[pg. ~112]{pearce2016early}} & No & \tiny{{{``This approach is therefore suited to situations where there is just one mediating pathway of interest, which is likely to be biased by intermediate confounding.''}}} \\   \hline  \hline 
\bottomrule
 \hline
\end{tabular}
 \begin{tablenotes}
      \tiny
      \item N(I)DE: Natural (in)direct effect; RDE: randomized interventional analogue of the natural direct effect; DE: direct effect; A: treatment; M: mediator; Y: outcome; L: pre-treatment covariate.
      \item We use [brackets] to indicate when we have modified excerpts, for brevity. Table \ref{tab: AppRat} in Web Appendix E provides excerpts in their entirety. 
    \end{tablenotes}
\end{threeparttable}

\end{table}
}

\end{landscape}
\restoregeometry
\doublespacing

We examined the extent to which interpretive claims in the text of the Results and  Discussion/Conclusion sections of the main texts and abstracts of the articles contained indicators that could permit a reader to distinguish whether or not claims were made using $\mathcal{I}_{\gamma_{RDE}}$ or $\mathcal{I}_{\gamma_{NDE}}$. When a claim allowed such disambiguation and correctly indicated $\mathcal{I}_{\gamma_{RDE}}$, we abstracted a modified version of that excerpt.  If no such unambiguous claim could be identified, we abstracted a representative excerpt and we counted this as a case of \textit{identity slippage} for that section of the abstract or main text. If no claims, whatsoever were made about $\gamma_{RDE}$ in a particular section of an article, then this article was excluded from denominators. Due to space constraints, abstracted excerpts for all articles are presented according to section in  Tables  \ref{tab: results} (Results) and  \ref{tab: discussion} (Discussion / Conclusion) in Web Appendix E. 

In the Results section of the abstract, we counted \textit{identity slippage} in 13/15 (8/8 non-\textit{methods-authored}) articles. In the Results section of the main text, we counted \textit{identity slippage} in 9/16 (5/8 non-\textit{methods-authored}) articles. In the Discussion / Conclusion section of the abstract, we counted \textit{identity slippage} in 12/13 (7/7 non-\textit{methods-authored}) articles. In the Discussion / Conclusion section of the main text, we counted \textit{identity slippage} in 11/15 (5/7 non-\textit{methods-authored}) articles. 

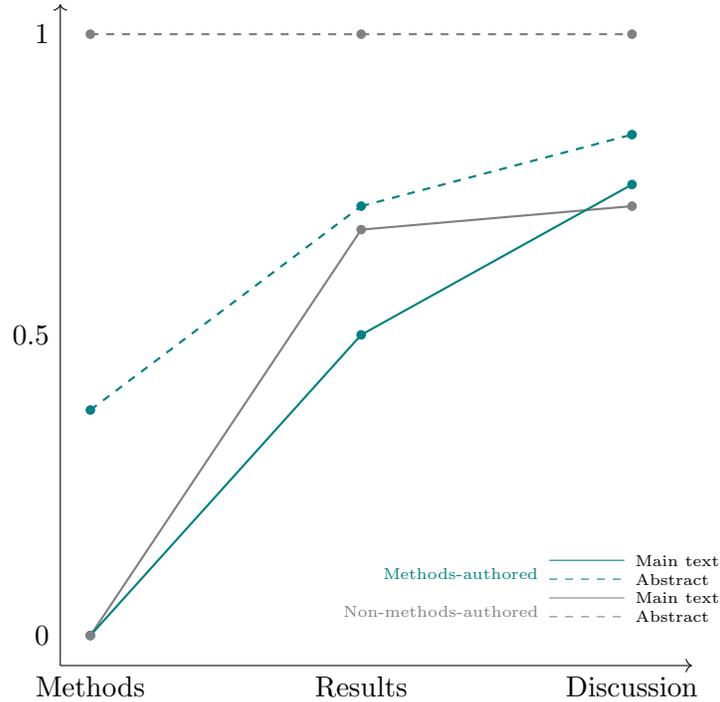
\begin{figure}[!htb]
    \centering
    \begin{tikzpicture}
  \draw[->] (0, -0.4) -- (8.4, -0.4) ;
  \draw[->] (0, -0.4) -- (0, 8.4);

  \draw[-] (0.4, 8- 0) --       (4.  , 8- 0) [thick, color=gray, dashed];
  \draw[-] (4. , 8- 0) --       (7.6 , 8- 0) [thick, color=gray, dashed];
  \draw[-] (0.4, 8- 8*1) --     (4.  , 8- 8*0.325) [thick, color=gray  ];
  \draw[-] (4. , 8- 8*0.325) -- (7.6 , 8- 8*0.286) [thick, color=gray];
  \draw[-] (0.4, 8- 8*0.625) -- (4.  , 8- 8*0.286) [thick, color=teal, dashed];
  \draw[-] (4. , 8- 8*0.286) -- (7.6 , 8- 8*0.167)   [thick, color=teal, dashed];
  \draw[-] (0.4, 8- 8*1) --     (4.  , 8- 8*0.50) [thick, color=teal];
  \draw[-] (4. , 8- 8*0.50) --  (7.6 , 8- 8*0.25)   [thick, color=teal];

  \draw[-] (6.5, 1)    -- (7.5, 1) [color=teal];
  \draw[-] (6.5, 0.75) -- (7.5, 0.75) [color=teal, dashed];
  \draw[-] (6.5, 0.5)  -- (7.5, 0.5) [color=gray];
  \draw[-] (6.5, 0.25) -- (7.5, 0.25) [color=gray, dashed];

  \node (MainMet)         at  (0.4,8-8*1.00)  [label=left:,point, color=teal];
  \node (MainRes)         at  (4.0,8-8*0.50)  [label=left:,point, color=teal];
  \node (MainDis)         at  (7.6,8-8*0.25)  [label=left:,point, color=teal];
  \node (nonMainMet)      at  (0.4,8-8*1)     [label=left:,point, color=gray];
  \node (nonMainRes)      at  (4.0,8-8*0.325) [label=left:,point, color=gray];
  \node (nonMainDis)      at  (7.6,8-8*0.286) [label=left:,point, color=gray];
  \node (AbsMet)          at  (0.4,8-8*0.625) [label=left:,point, color=teal];
  \node (AbsRes)          at  (4.0,8-8*0.286) [label=left:,point, color=teal];
  \node (AbsDis)          at  (7.6,8-8*0.167) [label=left:,point, color=teal];
  \node (nonAbsMet)       at  (0.4,8-8*0)     [label=left:,point, color=gray];
  \node (nonAbsRes)       at  (4.0,8-8*0)     [label=left:,point, color=gray];
  \node (nonAbsDis)       at  (7.6,8-8*0)     [label=left:,point, color=gray];
  
  \node ()       at  (0,0)          [anchor=east]{\small{$0$}};
  \node ()       at  (0,4)          [anchor=east]{\small{$0.5$}};
  \node ()       at  (0,8)          [anchor=east]{\small{$1$}};
  \node ()       at  (0.4,-0.4)     [anchor=north]{\small{$\text{Methods}$}};
  \node ()       at  (4.0,-0.4)     [anchor=north]{\small{$\text{Results}$}};
  \node ()       at  (7.6,-0.4)     [anchor=north]{\small{$\text{Discussion}$}};  
  
 \node ()       at  (6.5,0.825)     [anchor=east, color=teal]{\tiny{$\text{Methods-authored}$}};
 \node ()       at  (6.5,0.325)     [anchor=east, color=gray]{\tiny{$\text{Non-methods-authored}$}};
 \node ()       at  (7.5,1)        [anchor=west, color=black]{\tiny{$\text{Main text}$}};
 \node ()       at  (7.5,0.75)     [anchor=west, color=black]{\tiny{$\text{Abstract}$}};
 \node ()       at  (7.5,0.5)        [anchor=west, color=black]{\tiny{$\text{Main text}$}};
 \node ()       at  (7.5,0.25)     [anchor=west, color=black]{\tiny{$\text{Abstract}$}};
\end{tikzpicture}
    \caption{Prevalence of identity slippage across the major sections of a research article in the applied literature estimating $\gamma_{RDE}$.}
    \label{fig: slippage}
\end{figure}

In Table \ref{tab: examples},  we provide an example excerpt of \textit{identity slippage} from a \textit{methods-authored} article for each of the four sections we considered. Figure \ref{fig: slippage} plots the prevalence of \textit{identity slippage} across these major sections of a research article, stratified by Abstract vs.\ Main text and by \textit{methods-authored} vs. non-\textit{methods-authored} articles. \textit{Identity slippage} in this literature becomes increasingly prevalent across the major sections of a research article (from-Methods-to-Results-to-Discussion/Conclusion), is more severe in abstracts vs.\ the main texts of articles, and is more severe in non-\textit{methods-authored} articles vs.\ \textit{methods-authored} articles.

{\renewcommand{\arraystretch}{1}
\begin{table}[!htb]
\begin{threeparttable}

\caption{Example excerpts of identity slippage in \textit{methods-authored} articles estimating $\gamma_{RDE}$. \label{tab: examples}}

\begin{tabular}{|m{3cm}|m{3cm}|m{10cm}|}
\toprule
    \emph{\textbf{Section}} & \emph{\textbf{Author}} & \emph{\textbf{Excerpt}} \\ \hline
 Abstract - \newline Results    &  \tiny{\citetMain[pg. ~2094]{rudolph2021explaining}}  & {\tiny{{``[$\dots$]the protective indirect path contributed a [$X\%$] reduced risk of [$Y$] comparing [$A=1$ to $A=0$] (explaining [$X\%$] of the total effect).''}}} \\  \hline 
  Main - \newline Results &  \tiny{\citetMain[pg. ~2099]{rudolph2021explaining}}  & {\tiny{{``The indirect effect accounted for much of the overall increased risk of [$Y$]  on [$A=1$ versus $A=0$] in [$L=1$] (contributing a [$X\%$] reduced risk of [$Y$] corresponding to [$X\%$] of the total effect).''}}} \\  \hline 
 Abstract - \newline Discussion   &  \tiny{\citetMain[pg. ~1]{casey2019unconventional}}&  {\tiny{{``We observed a relationship between [$A$] and [$M$], which did not mediate the overall association between [$A$] and [$Y$].''}}}\\  \hline 
 Main - \newline Discussion &  \tiny{\citetMain[pg. ~116]{pearce2016early}}   &  {\tiny{{``Decomposition of the indirect pathway showed that around [$X\%$] was through [$M\dots$].''}}}\\  \hline 
\bottomrule
 \hline

\end{tabular}
 \begin{tablenotes}
      \tiny
      \item A: treatment; M: mediator; Y: outcome; L: pre-treatment covariate; W: post-treatment covariate.
      \item  A \textit{methods-authored} article is one whose set of authors overlaps with the collection of authors on at least one of the 22 articles included in the systematic of the methodological literature on stochastic mediation parameters.\
      \item We use [brackets] to indicate when we have modified excerpts to facilitate examination of interpretive claims, according to procedures described in Web Appendix \ref{subsec: met}. We refer readers to Tables \ref{tab: AppResAb} to \ref{tab: AppDisMain} in Web Appendix \ref{appsubsec: uanbrevExcerpts}, which provide excerpts in their entirety.
    \end{tablenotes}
\end{threeparttable}

\end{table}
}

\section{Concluding Remark} \label{sec: conc}

Routinely, an investigator with a specific scientific question will be unable to identify a causal parameter of interest under realistic models. Nevertheless, a \textit{pragmatic} framework for causal inference will often allow a causal analysis to proceed  -- by offering a menu of acceptable alternatives that may be more compatible with the data on hand. However, this increased permissiveness may come at the cost of interpretational errors.  Our studies of \textit{identity slippage} support this claim. They suggest that statisticians, epidemiologists, and other investigators should apply increased scrutiny to the frameworks collectively and individually deployed in causal parameter design and selection. In particular, an \textit{ideal} framework may protect against harmful interpretational errors and yield analyses that are more likely to remain faithful to an investigator's original scientific question.

In addition, our unique systematic review methodology is a template for further investigation of interpretive errors in practice. Future work might repeat such a review for parameters that are suspected to be developed under a \textit{pragmatic} framework for causal inference, for example, the local average treatment effect, or other principal stratum parameters.



\bibliographystyleMain{unsrtnat-init}

\bibliographyMain{refs.bib}

\clearpage

\clearpage

\begin{appendices}
\appendixpage
\doublespacing
\setcounter{page}{1}
\titlelabel{Web Appendix \thetitle.\quad}

\DoToC
\clearpage
\listoftables
\clearpage

\addtocontents{toc}{\protect\setcounter{tocdepth}{1}}
\section{Contemporary motivating example: dynamic treatment regimes and stochastic interventions} \label{sec: case2}

The case study in section \ref{sec: case1} traced a history of the development of mediation parameters, beginning with the path analysis of \citetSM{wright1920relative} and ending in a suite of stochastic mediation parameters. Evidently, the development of these parameters since \citetSM{baron1986moderator} was driven primarily by identifiability concerns: \citetSM{robins1992identifiability} and \citetSM{pearl2001direct} provided general model-free formulations of mediation parameters that disentangled their definitions from unrealistic parametric assumptions, but at the same time, revealed that nonparametric identification would often necessitate other assumptions that are  contradicted by common causal structures. The stochastic mediation literature proposes parameters that do not suffer from this property, and promote novel estimands that can be understood as alternatives within a \textit{pragmatic} framework to causal inference. We showed that identity slippage is highly prevalent in the resulting applied literature targeting these parameters. 

In this section we present a brief second case study, also related to a causal parameter defined by a stochastic interventions in time-varying treatment settings.

\subsection{Preliminaries}

To facilitate this presentation, we consider a longitudinal study with $k=0,1,2,\ldots,K$ denoting a follow-up interval where $K$ is the end of the follow-up of interest. Let $A_K$ denote the observed treatment during interval $k$, $L_k$ a vector of additional time-varying covariates measured in interval $k$, and $Y\equiv L_{K+1}$ an outcome of interest measured in the last interval. Henceforth, we assume the ordering $(L_0,A_0, \ldots, L_{K}, A_{K}, Y)$, and we let $O_1,\ldots, O_n$ denote a sample of iid observations where $O=(L_0,A_0, \ldots, L_{K}, A_{K}, Y)\sim P$. Following \citetSM{Richardson2013}, we let all counterfactual outcomes be defined via recursive substitution using the structural equations for each of the variables in the preceding ordering. 

We let $g$ denote a treatment regime that specifies how treatment is assigned at each $k=0,1,2,\ldots, K$. Let $L^{g}_k$, $A_k^g$ and $A_k^{g+}$ denote the natural values of covariates, natural treatment variable and the assigned treatment variable at $k$ under $g$, respectively.  Here, $A_k^g$ is the value that treatment variable would take in a hypothetical world in which the treatment strategy $g \equiv \{g_1,\dots, g_{K}\}$ was followed just until the instant prior to assigning treatment in interval $k$, and $A_k^{g+} = g_k(\overline{L}^{g}_k, \overline{A}^{g}_{k}, \overline{A}^{g+}_{k-1})$. Thus, the set $O^g\equiv \{\overline{L}^{g}_{K+1}, \overline{A}^{g}_{K}, \overline{A}^{g+}_{K}\}$ comprise the complete set of relevant variables under regime $g$, whose causal relations may be depicted in a dynamic single world intervention graph (SWIG) of \citetSM{Richardson2013}, denoted $\mathcal{G}(g)$. We also, let $\mathcal{G}$ and $\mathcal{G}^{\mathbf{a^{\dagger}}}$, represent SWIGs, respectively, for the factual variables $O$ and the counterfactuals $O^{\mathbf{a^{\dagger}}}$ that arise under a regime that assigns $\overline{A}^{g+}_{K} =\mathbf{a^{\dagger}}$. Using these SWIGs, we can define the following subsets of $O^g$, $O^{\mathbf{a^{\dagger}}}$, and $O$, where we let $\text{anc}(V^g)$ denote the ancestors of $V^g$ , and $PA_t^+$ denote the parents of $A_t^{g+}$ in $\mathcal{G}(g)$.

\begin{definition}[Z-sets and S-sets] \label{def: Zset}
As in \citetSM{Richardson2013}, we define  $Z^g$, $S^{g}$ and $Z_k^{g}$ as the set of nodes in $\mathcal{G}(g)$ as follows:
\begin{align*}
    Z^g & \equiv \text{anc}(Y^g) \setminus \overline{A}_K^{g+} , \\
    Z_k^{g} &  \equiv Z^g \setminus \{\overline{L}_k^{g}, \overline{A}_k^{g}\}, \\
    S^{g} & \equiv \{L^{g}, A^{g}\} \setminus Z^g, \\
    S_k^{g} & \equiv S^{g} \setminus \{\overline{L}_k^{g}, \overline{A}_k^{g}\}.
\end{align*}
Furthermore, let $\{Z^{\mathbf{a^{\dagger}}}, Z_k^{\mathbf{a^{\dagger}}}, S^{\mathbf{a^{\dagger}}}, S_k^{\mathbf{a^{\dagger}}}\}$ and $\{Z, Z_k, S, S_k\}$ denote the corresponding node-sets in $\mathcal{G}^{\mathbf{a^{\dagger}}}$ and $\mathcal{G}$, respectively.
\end{definition}

We let $q_t^{g}$ denote the probability mass function or density of $A_t^{g+}$ conditional on its parents in $\mathcal{G}(g)$ (which may include elements in $\overline{A}_t^{g}$, as in regimes that depend on the natural value of treatment.

\subsection{Defining and identifying modified treatment plans}

In this second case study, identity slippage occurs not for a mediation parameter, but for an expected outcome under a special class of dynamic treatment regimes that depend on the natural value of treatment, recently referred to as modified treatment plans (MTPs). \citetSM{Robins2004} first defined an intervention on treatment at a time $k$ that depends on the ``the natural value of treatment'' \citepSM{Richardson2013} at $k$. \citetSM{Haneuse2013} later coined these regimes MTPs, providing identifying assumptions for a mean outcome under such a regime in a point treatment setting and showed that it could be identified by the extended g-formula of \citetSM{Robins2004}. Independent of \citetSM{Haneuse2013},  \citetSM{Richardson2013} and \citetSM{Young2014} had jointly derived identifying conditions sufficient for equating the extended g-formula to the outcome mean under a time-varying treatment regime that assigns treatment at each $k$ based on it's natural value at $k$ and possibly also past patient characteristics, $L_k^{g}$, whose factual counterparts are measured in the observed data. We let the $\Psi_g$ denote extended g-formula functional, defined as

\begin{align}
   \Psi_g(P)=  \sum\limits_{\mathbf{a}^+,  \mathbf{s},  \mathbf{z}}
        y\cdot p(y\mid \mathbf{\overline{l}}_K, \mathbf{\overline{a}}_K^+)
        \prod\limits_{j=1}^{K}
        p(l_j, a_j \mid \mathbf{\overline{l}}_{j-1}, \mathbf{\overline{a}}_{j-1}^+)
        \prod\limits_{t=1}^{K}
        q_t^{g}(a_t^+ \mid \mathbf{pa}_t^+).
\end{align}

Of special interest, \citetSM{Richardson2013} and \citetSM{Young2014} noted the following reformulation of $\Psi_g$, which elucidated the positivity conditions necessary for $\Psi_g$ to be in general well-defined,

\begin{align}
   \Psi_g(P)=  & \sum\limits_{\mathbf{a}^+,  \overline{\mathbf{l}}_{K}, y}
        y\cdot p(y\mid \mathbf{\overline{l}}_K, \mathbf{\overline{a}}_K^+)
        \prod\limits_{j=1}^{K}
        p(l_j, \mid \mathbf{\overline{l}}_{j-1}, \mathbf{\overline{a}}_{j-1}^+)
        \prod\limits_{t=1}^{K}
        \tilde{q}_t^{g}(a_t^+ \mid \mathbf{\overline{l}}_{t}, \mathbf{\overline{a}}_{t-1}^+) \label{eq: exgenmain}\\
       = & \mathbb{E}\bigg[Y\prod\limits_{t=1}^{K}\frac{\tilde{q}_t^{g}(A_t \mid \overline{L}_{t}, \overline{A}_{t-1})}{p_t(A_t \mid \overline{L}_{t}, \overline{A}_{t-1})}\bigg] ,
\end{align}
where for each history $(\mathbf{\overline{l}}_{t}, \mathbf{\overline{a}}_{t-1}^+)$, $\tilde{q}_t^{g}\in[0,1]$ are obtained recursively by 
\begin{align*}
    \tilde{q}_t^{g}(a_t^+ \mid \mathbf{\overline{l}}_{t}, \mathbf{\overline{a}}_{t-1}^+) = 
    \Bigg\{
        \prod\limits_{m=1}^{t-1} 
            \tilde{q}_m^{g}(a_t^+ \mid \mathbf{\overline{l}}_{m}, \mathbf{\overline{a}}_{m-1}^+) 
    \Bigg\}^{-1}
    \Bigg\{ 
        \sum\limits_{\overline{a}_t}
        \prod\limits_{m=1}^{t} 
            q_m^{g}(a_m^+ \mid \mathbf{pa}_m^+)p_{\mathbf{a}^+}(a_m \mid \mathbf{pa}_t^+)
    \Bigg\}.
\end{align*}

In particular, \citetSM{Richardson2013} show that in general $\Psi_g(P)$ will identify $\mathbb{E}[Y^g]$ when, for each $m\in\{1,\dots,K\}$, the following conditions hold for all  $\overline{\mathbf{a}}^+_{m}$, $\overline{\mathbf{l}}_{m}$:

\begin{align} \label{eq: seqexmain}
    \textbf{B1.1: } &  Z_m^{\mathbf{a^+}} \CI I(A_m^{\mathbf{a^+}} = a^+_m) \mid \overline{L}_m^{\mathbf{a^+}} = \overline{\mathbf{l}}_ m, \overline{A}_{m-1}^{\mathbf{a^+}} = \overline{\mathbf{a}}^+_{m-1}, \\
    \textbf{B1.2: } & \tilde{q}_m^{g}(a_m^+ \mid \mathbf{\overline{l}}_{m}, \mathbf{\overline{a}}_{m-1}^+) >0  \implies  p_m(a_m^+ \mid \mathbf{\overline{l}}_{m}, \mathbf{\overline{a}}_{m-1}^+) >0 .
\end{align}

The assumptions of \textbf{B1.1} are known as sequential exchangeability for treatment $A$ with respect to the ancestors of $Y$ under regime $g$, $Z$. Indeed, when a regime $g$ is considered that depends on the natural values of treatment, then $Z_m$ will in general include future treatments $\underline{A}_{m+1}$, so that the presence of unmeasured common causes of treatments over time will contradict this condition. Then, a general exchangeability condition that will be sufficient for any regime that might possibly depend on the natural value of treatment is the following:

\begin{align} \label{eq: seqex2}
    \textbf{B1.3: } &  \{\underline{L}_{m+1}^{\mathbf{a^+}}, \underline{A}_{m+1}^{\mathbf{a^+}}\} \CI I(A_m^{\mathbf{a^+}} = a^+_m) \mid \overline{L}_m^{\mathbf{a^+}} = \overline{\mathbf{l}}_ m, \overline{A}_{m-1}^{\mathbf{a^+}} = \overline{\mathbf{a}}^+_{m-1}.
\end{align}

When, regime $g$ does not depend on the natural value of treatment, and is simply a deterministic function of a patient's past assigned treatment and covariate history, one can simply take $\tilde{q}_t^{g}(a_t^+ \mid \mathbf{\overline{l}}_{t}, \mathbf{\overline{a}}_{t-1}^+) = I(a_t^+ = g_t(\mathbf{\overline{l}}_{t}, \mathbf{\overline{a}}_{t-1}^+))$. Under a stochastic regime, $\tilde{q}_t^{g}$ may follow immediately by definition. For example, the longitudinal incremental propensity score interventions of \citetSM{Kennedy2019} define a class of regimes such that, for each $t\in\{1,\dots,K\}$, and $\beta \in (0, \infty)$ 
$$\tilde{q}_t^{g}(1 \mid \mathbf{\overline{l}}_{t}, \mathbf{\overline{a}}_{t-1}^+) = \frac{\beta p(1 \mid \mathbf{\overline{l}}_{t}, \mathbf{\overline{a}}_{t-1}^+)}{\beta  p(1 \mid \mathbf{\overline{l}}_{t}, \mathbf{\overline{a}}_{t-1}^+) + 1 - p(1 \mid \mathbf{\overline{l}}_{t}, \mathbf{\overline{a}}_{t-1}^+)}.$$ As none of these regimes depends on the natural value of treatment, a condition that is strictly weaker than \textbf{B1.3} will be sufficient for g-formula identification:

\begin{align} \label{eq: seqex3}
    \textbf{B1.4: } &  \underline{L}_{m+1}^{\mathbf{a^+}} \CI I(A_m^{\mathbf{a^+}} = a^+_m) \mid \overline{L}_m^{\mathbf{a^+}} = \overline{\mathbf{l}}_ m, \overline{A}_{m-1}^{\mathbf{a^+}} = \overline{\mathbf{a}}^+_{m-1}.
\end{align}

We now can consider two nested models: $\mathcal{M}_1$, defined by the set of laws that satisfies assumptions \textbf{B1.2} and \textbf{B1.4}; and $\mathcal{M}_2\subset \mathcal{M}_1$, defined by the set of laws that satisfies assumptions \textbf{B1.2} and \textbf{B1.3}. 

We may also consider a set of causal parameters of interest. Let $\gamma_{MTP}$ be a first causal parameter of interest, representing the expected potential outcome under a regime $g_1$ that depends on the natural value at each time-point, and a patient's history of assigned treatment and covariates. Let $P_{A_t \mid \overline{\mathbf{l}}_t, \overline{\mathbf{a}}_{t-1}}$, a parameter of $P$, denote the probability mass function for treatment $A_t$ conditional on past treatment and covariate history $\{\overline{L}_t = \overline{\mathbf{l}}_t, \overline{A}_{t-1}= \overline{\mathbf{a}}_{t-1}\}$. Similarly, let $P^{g_1}_{A_t \mid \overline{\mathbf{l}}_t, \overline{\mathbf{a}}_{t-1}}$, a parameter of $P^{full}$ denote the probability mass function for the natural value of treatment $A^{g_1}_t$ conditional on past assigned treatment and covariate history under regime $g_1$. Then let $\phi_t$ denote some functional of $P_{A_t \mid \overline{\mathbf{l}}_t, \overline{\mathbf{a}}_{t-1}}$ so that $\phi_t(P_{A_t \mid \overline{\mathbf{l}}_t, \overline{\mathbf{a}}_{t-1}})$ is a function that maps values of $\{a_t, \overline{\mathbf{l}}_t, \overline{\mathbf{a}}_{t-1}\}$ to some number in $[0,1]$ and is a valid conditional probability mass function or density function for a variable with the domain of $A_t$. 

Under $\mathcal{M}_2$, $\gamma_{MTP}$ is identified by $\Psi_{g}(P)$ and suppose that in particular it is identified by a $\Psi_{g}(P)$ in which 

$$\tilde{q}_t^{g_1}(a_t^+ \mid \mathbf{\overline{l}}_{t}, \mathbf{\overline{a}}_{t-1}^+) = \phi_t(P_{A_t \mid \overline{\mathbf{l}}_t, \overline{\mathbf{a}}_{t-1}})(a_t^+ \mid \mathbf{\overline{l}}_{t}, \mathbf{\overline{a}}_{t-1}^+).$$ 


Accordingly, we can write the following structural equation for $A_t^{g_1+}$ as

\begin{align}  
    A^{g_1+}_t = g_{1,t}(A^{g_1}_{t}, \overline{L}^{g_1}_t, \overline{A}^{g_1+}_{t-1}) \label{eq: SE- LMTP}.
\end{align}

Let $\gamma_{MTP-SI^{g_1}}$ be a second causal parameter of interest representing the expected potential outcome under a stochastic regime $g_2$ that depends only on a patient's history of assigned treatment and covariates (and not their natural values of treatment), wherein, at each time point, treatment is assigned according to a random draw from the $\phi_t(P^{g_1}_{A_t \mid \overline{\mathbf{l}}_t, \overline{\mathbf{a}}_{t-1}})$ consistent with their realized history of $\{\overline{L}^{g_2}_t, \overline{A}^{g_2+}_{t-1}\}$. If treatments were binary at each time point, $A_t^{g_2+}$ might be assigned according to the following structural equation:

\begin{align}  
    A^{g_2+}_t = I\Big(\delta_t^{g_2} < \phi_t(P^{g_1}_{A_t \mid \overline{\mathbf{l}}_t, \overline{\mathbf{a}}_{t-1}})(1 \mid \overline{L}^{g_2}_{t}, \overline{A}^{g_2+}_{t-1})\Big) \label{eq: SE- LMTP-SIg1},
\end{align}
where $\delta_t^{g_2}$ is (as in \eqref{eq: SE- RDE}) an investigator-controlled randomizer distributed uniformly over the unit interval that is by definition jointly independent of all other variables.

Finally, let $\gamma_{LMTP-SI}$ be a third parameter representing the expected potential outcome under a stochastic regime $g_3$ nearly identical to $g_2$ of $\gamma_{MTP-SI^{g_1}}$, except treatment is assigned according to a random draw from the $\phi_t(P_{A_t \mid \overline{\mathbf{l}}_t, \overline{\mathbf{a}}_{t-1}})$. If treatments were binary at each time point, $A_t^{g_3+}$ might be assigned according to the following structural equation:

\begin{align}  
    A^{g_3+}_t = I\Big((\delta_t^{g_3} < \phi_t(P_{A_t \mid \overline{\mathbf{l}}_t, \overline{\mathbf{a}}_{t-1}})(1 \mid \overline{L}^{g_3}_{t}, \overline{A}^{g_3+}_{t-1})\Big) \label{eq: SE- LMTP-SI}.
\end{align}

$\gamma_{MTP-SI}$ was first introduced in \citetSM{Young2014} and \citetSM{Richardson2013}, and then was later discussed in \citetSM{diaz2021mtpJASA}. $\gamma_{MTP-SI^{g_1}}$ was introduced in a pre-print version of \citetSM{diaz2021mtpJASA} (see \citetSM{diaz2020non}), but was not considered further in later versions. To reiterate,  $\gamma_{MTP-SI^{g_1}}$ and $\gamma_{MTP-SI}$ are both parameters defined by stochastic regimes that do not depend on the natural value of treatment. However, whereas the assigned treatment distribution for $\gamma_{LMTP-SI^{g_1}}$ is defined with respect to the distribution of the natural value of treatment under $g_1$, that of $\gamma_{LMTP-SI}$ is defined with respect to the distribution of treatment in the factual data.

Under $\mathcal{M}_2$, $\Psi_{g}(P) = \gamma_{MTP} = \gamma_{MTP-SI^{g_1}}=\gamma_{MTP-SI}$, that is, all three parameters are identified by $\Psi_{g}(P)$. This identification result of course follows by design. Under $\mathcal{M}_2$, the expected outcome under any dynamic treatment regime is identified by a dynamic g-formula of \citetSM{Richardson2013}, and so any set of parameters whose intervention densities $\tilde{q}_t^{g}$ agree under the model, will be identified by the same particular dynamic g-formula. It follows then that the causal parameters themselves will also agree at all laws in the model. Regimes $g_1$ and $g_3$ agree on $\tilde{q}_t^{g}$ immediately by definition, and $g_2$ and $g_3$ agree on $\tilde{q}_t^{g}$ because  $P^{g_1}_{A_t \mid \overline{\mathbf{l}}_t, \overline{\mathbf{a}}_{t-1}} = P_{A_t \mid \overline{\mathbf{l}}_t, \overline{\mathbf{a}}_{t-1}}$ under $\mathcal{M}_2$. 

In contrast, under  $\mathcal{M}_1$,  $\Psi_{g}(P)$ identifies $\gamma_{MTP-SI}$ (see \citetSM{Richardson2013} and \citetSM{Young2014}). Furthermore, $\gamma_{MTP} = \gamma_{MTP-SI^{g_1}}$ agree under all laws in $\mathcal{M}_1$. However, $\gamma_{MTP-SI^{g_1}}$ will not in general equal $\gamma_{MTP-SI}$ precisely because there may still be unmeasured common causes of treatment over time under $\gamma_{MTP-SI}$:  $P^{g_1}_{A_t \mid \overline{\mathbf{l}}_t, \overline{\mathbf{a}}_{t-1}}$ may not equal $P_{A_t \mid \overline{\mathbf{l}}_t, \overline{\mathbf{a}}_{t-1}}$. \citetSM{diaz2020non} claim that $\gamma_{MTP-SI^{g_1}}$ is identified under a model compatible with such unmeasured common causes. However, it follows here that this claim is erroneous  (and was indeed later corrected in \citetSM{diaz2021mtpJASA}): neither $\gamma_{MTP}$ nor $\gamma_{MTP-SI^{g_1}}$ are identified by $\Psi_{g}(P)$ under  $\mathcal{M}_1$.

\subsection{Identity slippage for the effects of modified treatment plans}

Suppose an investigator is originally interested in the effect of a longitudinal MTP, $\gamma_{MTP}$, for example the effect of shifting by a fixed amount the daily arterial partial-pressure-of-oxygen-to-fraction-of-inspired-oxygen (P/F) ratio for mechanically-ventilated patients in intensive care wards, as in \citetSM{diaz2021mtpJASA}. Suppose however they are uncomfortable with model $\mathcal{M}_2$, perhaps because they cannot rule out the presence of ICU care providers that might systematically manage patients' P/F ratio more or less aggressively over time. Such a variable would be a common cause of P/F ratio over time that is not measured in the observed data (and thus would contradict $\mathcal{M}_2$), but may not be associated with survival among patients with the same treatment history (and thus would not contradict $\mathcal{M}_1$). Thus, suppose this investigator instead assumes model $\mathcal{M}_1$ and nominally targets $\gamma_{MTP-SI}$, identified by $\Psi_{g}(P)$ and consistently estimated by $\hat{\psi}_{g}$. However, the investigator draws claims from $\mathcal{I}_{MTP}(\hat{\psi}_{g})$. This would be a case of identity slippage, which might be facilitated by the known concordance between $\gamma_{MTP}$ and $\gamma_{MTP-SI^{g_1}}$ under $\mathcal{M}_1$, and/or beliefs in the conceptual proximity of $\gamma_{MTP-SI}$ to $\gamma_{MTP-SI^{g_1}}$ and $\gamma_{MTP}$

As in Case Study 1 (of stochastic mediation parameters), the evidence from the small but growing literature on MTP parameters suggests they are being developed within a pragmatic framework for causal inference. \citetSM{diaz2021mtpJASA} motivate these parameters by noting both \textit{identifiability} and \textit{interpretability concerns}.  As a case of the former,  \citetSM{diaz2021mtpJASA} writes:  

{\small{\textit{``LMTPs also have the advantage that they can be designed to satisfy the positivity assumption required for causal inference. ''}}}

As cases of the latter (analogous to the approach of \citetSM{vansteelandt2020assumption}), \citetSM{diaz2021mtpJASA} write:

{\small{\textit{``the LMTP functional [$\Psi_{g}$] can still be interpreted as a causal effect even under the standard identifiability assumptions required for dynamic regimes [, those of $\mathcal{M}_1$.]''}}}

In explicitly naming $\gamma_{MTP-SI}$ the ``longitudinal modified treatment policy stochastic intervention'',  \citetSM{diaz2021mtpJASA} also make explicit conceptual linkages between $\gamma_{MTP}$ and $\gamma_{MTP-SI}$. 

To our knowledge, there are of yet no analyses targeting $\gamma_{MTP-SI}$ in applied journals. However, if investigators become increasingly concerned with unmeasured confounding of sequential treatments, as many are for unmeasured confounding of sequential survival status in inference for causal hazard ratios \citepSM{Robins1986, hernan2004structural, hernan2010hazards}, then increased vigilance for identity slippage in this setting will be needed.

\clearpage\doublespacing

\section{Additional historical examples of identity slippage}



\subsection*{2-sample T-test vs.\ Mann-Whitney U-test.}
Let $\gamma_2$ be an indicator that the mean of some variable $X$ is equal across two groups defined by $A$ and let $\gamma_1$ be an indicator that the median of some variable $X$ is equal across two such groups.  Suppose a student investigator considers two models:  $\mathcal{M}_2$, defined by the assumption that $X$ is distributed normally conditional on both $A=a$ and $A=a'\neq a$, with common variance $\sigma^2$, and; $\mathcal{M}_1$, defined by the assumption that $X$ is distributed according to some common distribution conditional on both $A=a$ and $A=a'\neq a$ with common variance $\sigma^2$, so that $\mathcal{M}_2 \subset \mathcal{M}_1$. Suppose this student investigator is not comfortable assuming normality and so they instead assume the larger $\mathcal{M}_1$. Having learned that the T-test is not ``valid'' unless normality holds, they run a Mann-Whitney U-test which is ``robust'' to such normality violations. In the discussion section of their paper, however, the investigators make claims about differences in the expected values of $X$ between those with and without $A=a$. This is a case of identity slippage. The investigator had assumed the correct model and chosen a procedure that is consistent for a parameter that may be of some interest (the median-difference) under that model, but then proceeded to interpret the results of that procedure as if it were consistent for a parameter only identified under the stricter model (that further assumed that the common distributions were in fact normal).

\clearpage\doublespacing
\section{Defining randomized interventional analogues}

\begin{figure}[!h] 
\centering
\begin{tikzpicture}
\begin{scope}[every node/.style={thick,draw=none}]

    \node (W) at (2,2) {$W$};
    \node (A) at (0,0) {$A$};
    \node (M) at (2,0) {$M$};
    \node (Y) at (4,0){$Y$};

\end{scope}

\begin{scope}[>={Stealth},
              every node/.style={fill=white,circle},
              every edge/.style={draw=black}]

\path [->] (A) edge (W);
\path [->] (L) edge (M);
\path [->] (L) edge (Y);
\path [->] (A) edge (M);
\path [->] (M) edge (Y);
\path [->] (A) edge[bend right] (Y);

\end{scope}
\end{tikzpicture}

\caption{Simple directed acyclic graph (DAG) with treatment-induced mediator-outcome confounding and no baseline confounding}
\label{fig: RIADAG}
\end{figure}
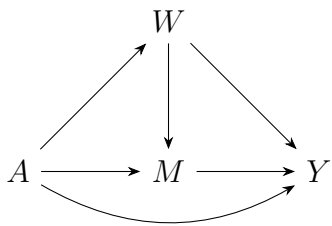

To fix ideas, consider a variable set $V \equiv \{A, W, M, Y\}$ generated recursively according to the directed acyclic graph in Figure \ref{fig: RIADAG}, representing the following structural equations that might arise in the context of an ideal randomized controlled trial for treatment $A$ with full compliance:

\begin{align*}
    A = & f_A(\epsilon_A) \\
    W = & f_W(A, \epsilon_W) \\
    M = & f_M(A, W, \epsilon_M)\\
    Y = & f_Y(A, W, M, \epsilon_Y).
\end{align*}

$\{f_A, f_W, f_M, f_Y\}$ and $\{\epsilon_A, \epsilon_W, \epsilon_M, \epsilon_Y\}$ are structural functions and error terms whose form and joint distribution, respectively, are not restricted by definitions, but may be restricted by assumptions (possibly enforced by experimental design). All counterfactuals defined by an intervention setting $V_j$ to $v_j$ are then defined by recursive substitution using these structural functions. For illustrative purposes,  note that the counterfactual $M^{a^*}$, the mediator under an intervention setting $A$ to $a^*$, may equivalently be expressed by composing the structural functions $f_M$ and $f_W$ as

\begin{align}
   M^{a^*} = f^{a^*}_M(\epsilon_M, \epsilon_W) \coloneqq f_M(a^*,  \epsilon_M, f_W(a^*, \epsilon_W)). \label{eq: SE- NDE} 
\end{align}

Now we define the randomized interventional analogues of the natural direct and indirect effects  
$\gamma^{RDE}$ and $\gamma^{RIE}$ as

\begin{align}  
    \gamma^{RDE}(P^F)= \mathbb{E}_{P^F}[Y^{a=1, G_{a=0}} - Y^{a=0, G_{a=0}}] \label{RDEapp}\\
    \gamma^{RIE}(P^F)= \mathbb{E}_{P^F}[Y^{a=1, G_{a=1}}- Y^{a=1, G_{a=0}}]. \label{RIEapp}
\end{align}

The parameters $\gamma^{RDE}$ and $\gamma^{RIE}$ have also variously been called ``generated (in)direct effects'' \citepSM{geneletti2007identifying}, ``stochastic (in)direct effects'' \citepSM{rudolph2018robust, rudolph2021transporting}, ``interventional direct effects'' \citepSM{vansteelandt2017interventional, mittinty2019effect, benkeser2021nonparametric, loh2020heterogeneous, diaz2021nonparametric}, or even simply ``natural (in)direct effects'' \citepSM{didelez2012direct}. 


The randomized interventional analogues (RIA) of natural direct and indirect effects, are defined by counterfactuals of the type $Y^{a, G_{a^*}}$. Here, $G_{a^*}$ indicates a variable following the counterfactual distribution of the mediator $M$ under an intervention that sets $A$ to $a^*$. Then, $Y^{a, G_{a^*}}$ is the potential outcome that would occur under an intervention that sets treatment $A$ to $a$ and sets $M$ to a value randomly drawn from that distribution. In the case of a binary mediator, this could be achieved through specification of the following structural equation for $G_{a^*}$:

\begin{align}  
    G_{a^*} = I(\delta < \pi^{a*}), \label{eq: SE- RDE}
\end{align}

where $\delta$ is an investigator-controlled randomizer distributed uniformly over the unit interval that is by definition jointly independent of all other variables, and $\pi^{a*} \coloneqq  P(M^{a*}=1)$ is assumed known to the hypothetical investigator who would in principle implement the regime.


Evidently, \eqref{eq: SE- RDE} is a \textit{known} function of an \textit{investigator-controlled} term $\delta$. In contrast, \eqref{eq: SE- NDE} is an \textit{unknown} function ($f^{a^*}_M$) of \textit{unobserved} terms $(\epsilon_M, \epsilon_W)$. This contrast highlights the fundamental definitional difference between the parameters $\gamma^{NDE}$ and $\gamma^{RDE}$, which results in the previously-highlighted practical differences in implementability and identifiability. 

There also exist alternatively defined randomized interventional analogues of natural direct effects ($\gamma^{RDE-W}$), given by a stochastic intervention on the \textit{conditional} distribution of $M^{a^*}$ given $W^{a^*}$,

\begin{align}  
    \gamma^{RDE-W}(P^F) = \mathbb{E}_{P^F}[Y^{a=1, G_{a=0 \mid W}} - Y^{a=0, G_{a=0\mid W}}]. \label{RDE-Wapp}
\end{align}

Similarly, these stochastic mediation parameters are defined by counterfactuals of the type $Y^{a, G_{a^*\mid W}}$. Here, $G_{a^*\mid W}$ indicates a variable following the counterfactual distribution of the mediator $M$ conditional on $W$ under an intervention that sets $A$ to $a^*$. When $M$ is binary, we can let $\pi_W^{a*}(w)$ denote that particular distribution for $W^{a^*}=w$,  $P(M^{a*}=1 \mid W^{a^*}=w)$.  Then, $Y^{a, G_{a^*\mid W}}$ is the potential outcome that would occur under the following regime: first, treatment $A$ is set to $a$ and subsequently the value of $W^a$ is observed (say, $w$) ; second, $M$ is set to a value randomly drawn from the distribution $\pi_W^{a*}(w)$. As before, in the case of a binary mediator this could be achieved for example through specification of the following structural equation for $G_{a^* \mid W}$:

\begin{align}  
    G_{a^*\mid W} = I(\delta < \pi_W^{a*}(W^a)), \label{eq: SE- RDE-W}
\end{align}

where again  $\delta$ is an investigator controlled randomizer, distributed as in \eqref{eq: SE- RDE}, and $\pi_W^{a*}$ is assumed known to the hypothetical investigator who would in principle implement the regime. While $G_{a^*\mid W}$ is indeed assigned as a function of a counterfactual variable, it is precisely the variable that would be observed under this single regime. Since \eqref{eq: SE- RDE-W} is a \textit{known} function of an \textit{investigator-controlled} term $\delta$ and an \textit{investigator-observed} variable $W^a$,  $\gamma^{RDE-W}$ shares identifiability properties with $\gamma^{RDE}$: both are identified without parametric or cross-world assumptions and both are in principle implementable without special knowledge of $f_M$. The parameter $\gamma^{RDE-W}$, and its indirect counterpart, have also been called ``natural (in)direct effects'' \citepSM{zheng2012causal, zheng2017longitudinal} when $W$ is possibly caused by treatment $A$, or are simply treated as particular randomized interventional analogues $\gamma^{RDE}$ ( $\gamma^{RIE}$) whenever $W$ is assumed a non-descendent of $A$, and thus $W$ is subsumed into the vector of baseline ``pre-treatment'' covariates.

\subsubsection{Identifying randomized interventional analogues}

Now, suppose an investigator defines $\mathcal{M}_1$ via the following assumptions:

\begin{align}
    \textbf{A1.1: }& (Y^{a, m}, M^a) \CI A  \ \forall \ m,a \label{eq: A1.1}\\
    \textbf{A1.2: }& Y^{a, m} \CI M^a \mid W, A=a, \ \forall \ m,a \label{eq: A1.2}\\
    \textbf{A1.3: }&  P_0(A=a )>0 \text{, and } \\
    & \text{if } P_0(W=w, A=a)>0, \text{ then } P_0(M=m \mid W=w, A=a)>0  \ \forall \ m, w,  a. \nonumber
\end{align}

It is well known that $\gamma^{NDE}$ is not identified under $\mathcal{M}_1$. In contrast, both $\gamma^{RDE}$ and $\gamma^{RDE-W}$ are identified by $\Psi_{RDE}(P)$ and $\Psi_{RDE-W}(P)$, respectively:

\begin{align}
    \Psi_{RDE}(P) = \int_{m}\int_{w}\mathbb{E}[Y \mid m, w, a] f(m \mid a^*)f(w \mid a)d\mu(w,m); \label{eq: psi RDEapp}
\end{align}

\begin{align}
    \Psi_{RDE-W}(P) = \int_{m}\int_{w}\mathbb{E}[Y \mid m, w, a] f(m \mid w, a^*)f(w \mid a)d\mu(w,m). \label{eq: psi RDE-Wapp}
\end{align}

An investigator might additionally consider the following assumptions:

\begin{align}
    \textbf{A2.1: }& (Y^{a, m}, M^{a^*}) \CI A \mid W  \ \forall \ m,a, a^* \label{eq: A2.1}\\
    \textbf{A2.2: }& Y^{a, m} \CI M^{a^*} \mid W, A, \ \forall \ m,a, a^* \label{eq: A2.2}.
\end{align}

Let $\mathcal{M}_2$ denote the intersection of $\mathcal{M}_1$ and the set of laws $P^{full}$ that satisfy assumptions \eqref{eq: A2.1} and \eqref{eq: A2.2}. Under $\mathcal{M}_2$, which is a strict submodel of $\mathcal{M}_1$, both $\gamma^{NDE}$ and $\gamma^{RDE-W}$ are identified by $\Psi_{RDE-W}$.

\clearpage\doublespacing
\addtocontents{toc}{\protect\setcounter{tocdepth}{2}}
\section{Methodological review of stochastic mediation parameters}

\subsection{Bibliography for methodological review of stochastic mediation parameters} \label{appsubsec: BibMeth}

    \bibliographystyleMeth{unsrtnat}
    \bibliographyMeth{refs}

\subsection{Unabbreviated excerpts documenting Interpretability and Identifiability concerns}

\thispagestyle{empty}

{\renewcommand{\arraystretch}{1}
\begin{table}[!htb]
\begin{threeparttable}

\caption{Example excerpts for Interpretability concerns: ``\textbf{Approach}''. \label{tab: Intconcern1}}

\begin{tabular}{|m{3cm}|m{13cm}|}
\toprule
     \emph{\textbf{Author}} & \emph{\textbf{Excerpt}} \\ \hline
 \tiny{\citetMeth[pg. ~205]{geneletti2007identifying}}  & {\tiny{{``This \textbf{approach} is similar to Pearl’s natural direct effect''}}} \\  \hline 
 \tiny{\citetMeth[pg. ~30]{zheng2012causal}}  & {\tiny{{``As an alternative, we proposed to adopt the stochastic interventions (SI) \textbf{approach} of \citetMeth{didelez2012direct}, where the mediator is considered an intervention variable, onto which a given distribution is enforced.''}}} \\  \hline 
 \tiny{\citetMeth[pg. ~300]{vanderweele2014effect}}  & {\tiny{{``In this article, we describe three alternative \textbf{approaches} to effect decomposition that give quantities that can be interpreted as direct and indirect effects and that can be identified from data even in the presence of an exposure-induced mediator-outcome confounder''}}} \\  \hline 
 \tiny{\citetMeth[pg. ~126]{vanderweele2015explanation}}  & {\tiny{{``In the following section we will discuss alternative \textbf{approaches} to natural direct and indirect effects that can be used in the presence of exposure-induced mediator–outcome confounders even though natural direct and indirect effects themselves are not identified.''}}} \\  \hline
 \tiny{\citetMeth[pg. ~4010]{lok2016defining}}  & {\tiny{{``However, the common \textbf{approaches} to causal mediation analysis all rely on the existence of all $Y^{a,m}$.''}}} \\  \hline
 \tiny{\citetMeth[pg. ~918]{vanderweele2017mediation}}  & {\tiny{{``In this paper we propose an \textbf{approach} to pathway analysis that can be used in settings with time varying exposures and mediators.''}}} \\  \hline
 \tiny{\citetMeth[pg. ~1]{lin2017interventional}}  & {\tiny{{``In this paper, the authors propose a method, defining a randomly interventional analogue of PSE (rPSE), as an alternative \textbf{approach} for mechanism investigation.''}}} \\  \hline
 \tiny{\citetMeth[pg. ~3]{rudolph2018robust}}  & {\tiny{{``In this paper we build on the \textbf{approach} described by \citetMeth{vanderweele2017mediation} in which they defined the stochastic distribution on M as: [$\dots$]''}}} \\  \hline
 \tiny{\citetMeth[pg. ~3]{hejazi2020nonparametric}}  & {\tiny{{``In the present work, we outline a general framework encompassing many prior causal mediation analysis \textbf{approaches}, including the natural (in)direct effects, their interventional effect counterparts, and the stochastic (in)direct effects.''}}} \\  \hline
 \tiny{\citetMeth[pg. ~414]{lok2021causal}}  & {\tiny{{``Organic indirect and direct effects: an intervention-based \textbf{approach} avoiding cross-worlds counterfactuals and assumptions''}}} \\  \hline
 \tiny{\citetMeth[pg. ~631]{diaz2021nonparametric}}  & {\tiny{{``To overcome this problem while retaining the path decomposition employed by the natural direct and indirect effects, we adopt an \textbf{approach} previously outlined by \citetMeth{petersen2006estimation, van2008direct, zheng2012causal, rudolph2018robust}, defining direct and indirect effects using stochastic interventions on the mediator.''}}} \\  \hline
 \tiny{\citetMeth[pg. ~172]{benkeser2021nonparametric}}  & {\tiny{{``A debate in this literature has emerged pertaining to the reliance of methodology on cross-world independence assumptions that are fundamentally untestable even in randomized controlled experiments. One \textbf{approach} to this problem is to utilize methods that attempt to estimate bounds on effects (\citetSM{robins2010alternative}, \citetSM{tchetgen2014bounds}, among others). A second \textbf{approach} considers seeking alternative definitions of mediation parameters that do not require such cross-world assumptions (\citetMeth{vanderweele2014effect}, \citetMeth{rudolph2018robust}, among others).''}}} \\  \hline
 \tiny{\citetMeth[pg. ~17]{zheng2017longitudinal}}  & {\tiny{{``In this paper, we proposed to random interventions \textbf{approach} to formulate parameters of interest in longitudinal mediation analysis with time varying mediator and exposures.''}}} \\  \hline
 \tiny{\citetMeth[pg. ~2]{diaz2020causal}}  & {\tiny{{``In an attempt to solve these problems, several researchers have proposed methods that do away with cross-world counterfactual independences. These methods can be divided into two types: identification of bounds \citepSM{robins2010alternative, tchetgen2014bounds, miles2015partial} and alternative definitions of the (in)direct effect \citepSM{petersen2006estimation, van2008direct, vansteelandt2012natural, vanderweele2014effect}. Here, we take the second \textbf{approach}, defining the (in)direct effect in terms of a decomposition of the total effect of a stochastic intervention on the population exposure.''}}} \\  \hline
 \tiny{\citetMeth[pg. ~5085]{mittinty2019effect}}  & {\tiny{{``One \textbf{approach} was proposed by \citetSM{vanderweele2014effect}.''}}} \\  \hline
\bottomrule
 \hline

\end{tabular}
 \begin{tablenotes}
      \tiny
      \item  
    \end{tablenotes}
\end{threeparttable}

\end{table}
}

{\renewcommand{\arraystretch}{1}
\begin{table}[!htb]
\begin{threeparttable}

\caption{Example excerpts for Interpretability concerns: ``\textbf{Alternative}''. \label{tab: Intconcern2}}

\begin{tabular}{|m{3cm}|m{13cm}|}
\toprule
     \emph{\textbf{Author}} & \emph{\textbf{Excerpt}} \\ \hline
 \tiny{\citetMeth[pg. ~7]{van2008direct}}  & {\tiny{{``As a side note, we point out that \citetSM{robins2003semantics} provides yet another \textbf{alternative} definition of a natural direct effect that does not rely on an additional identifying assumption.''}}} \\  \hline 
 \tiny{\citetMeth[pg. ~7]{zheng2012causal}}  & {\tiny{{``In this respect, the parameters defined here aim to provide \textbf{alternative} formulations to questions that arise in mediation analysis in the current survival setting`''}}} \\  \hline 
 \tiny{\citetMeth[pg. ~300]{vanderweele2014effect}}  & {\tiny{{``In this article, we describe three \textbf{alternative} approaches to effect decomposition that give quantities that can be interpreted as direct and indirect effects and that can be identified from data even in the presence of an exposure-induced mediator-outcome confounder.''}}} \\  \hline 
 \tiny{\citetMeth[pg. ~126]{vanderweele2015explanation}}  & {\tiny{{``In the following section we will discuss \textbf{alternative} approaches to natural direct and indirect effects that can be used in the presence of exposure-induced mediator–outcome confounders even though natural direct and indirect effects themselves are not identified.''}}} \\  \hline 
 \tiny{\citetMeth[pg. ~4017]{lok2016defining}}  & {\tiny{{``In future work, I will show that in contrast to natural direct and indirect effects, organic direct and indirect effects can be extended to provide an identification result for the case where there are post-treatment mediator-outcome confounders. This will provide another \textbf{alternative} to the three quantities described in \citetMeth{vanderweele2014effect}.''}}} \\  \hline 
 \tiny{\citetMeth[pg. ~927]{vanderweele2017mediation}}  & {\tiny{{``As yet another \textbf{alternative}, though one that we argue is not suitable for mediation analysis, we could have, at each time $t$, fixed the mediator $M_t$ for that time $t$ to a random draw from the mediator distribution under a particular exposure history up to that point in time $t$.''}}} \\  \hline 
 \tiny{\citetMeth[pg. ~3]{lin2017interventional}}  & {\tiny{{``Therefore, \textbf{alternative} definitions for NDE and NIE, i. e. randomly interventional analogues of natural direct effect (rNDE) and of natural indirect effects (rNIE), have been proposed for settings with time-varying confounders.''}}} \\  \hline 
 \tiny{\citetMeth[pg. ~1]{rudolph2018robust}}  & {\tiny{{``We build on recent work that identified \textbf{alternative} estimands that do not require this assumption and propose a flexible and double robust targeted minimum loss-based estimator for stochastic direct and indirect effects.''}}} \\  \hline 
 \tiny{\citetMeth[pg. ~258]{vansteelandt2017interventional}}  & {\tiny{{``These concerns all originate from the fact that natural (in)direct effects are defined in terms of so-called cross-world counterfactuals that are unobservable, even from experimental data; they call for \textbf{alternative} effect measures that are less remote from the observed data.''}}} \\  \hline 
 \tiny{\citetMeth[pg. ~3]{hejazi2020nonparametric}}  & {\tiny{{``In this vein, our work formulates \textbf{alternative} (in)direct effect estimands that retain the flexibility of the (in)direct effects of \citetMeth{diaz2020causal};''}}} \\  \hline 
 \tiny{\citetMeth[pg. ~628]{diaz2021nonparametric}}  & {\tiny{{``The first class of methods aim to identify and estimate bounds on the natural (in)direct effects in the presence of intermediate confounders (e.g., \citetSM{robins2010alternative, tchetgen2014bounds, miles2015partial}), whereas the second class is concerned with \textbf{alternative} definitions of the (in)direct effects;''}}} \\  \hline 
\tiny{\citetMeth[pg. ~172]{benkeser2021nonparametric}}  & {\tiny{{``A second approach considers seeking \textbf{alternative} definitions of mediation parameters that do not require such cross-world assumptions (\citetMeth{vanderweele2014effect}, \citetMeth{rudolph2018robust}, among others).''}}} \\  \hline 
\tiny{\citetMeth[pg. ~2]{zheng2017longitudinal}}  & {\tiny{{``As an \textbf{alternative}, random (stochastic) interventions (RI) based formulation to causal mediation was proposed in \citetMeth{didelez2012direct}''}}} \\  \hline 
\tiny{\citetMeth[pg. ~2]{diaz2020causal}}  & {\tiny{{``These methods can be divided into two types: identification of bounds \citepSM{robins2010alternative, tchetgen2014bounds, miles2015partial} and \textbf{alternative} definitions of the (in)direct effect \citepSM{petersen2006estimation, van2008direct, vansteelandt2012natural, vanderweele2014effect}.''}}} \\  
\bottomrule
 \hline

\end{tabular}
 \begin{tablenotes}
      \tiny
      \item  
    \end{tablenotes}
\end{threeparttable}

\end{table}
}

{\renewcommand{\arraystretch}{1}
\begin{table}[!htb]
\begin{threeparttable}

\caption{Example excerpts for Interpretability concerns: ``\textbf{Analog}''. \label{tab: Intconcern3}}

\begin{tabular}{|m{3cm}|m{13cm}|}
\toprule
     \emph{\textbf{Author}} & \emph{\textbf{Excerpt}} \\ \hline
 \tiny{\citetMeth[pg. ~0]{zheng2012causal}}  & {\tiny{{``The natural direct and indirect effects can be defined \textbf{analogously} to the ideas in \citetSM{pearl2001direct}.''}}} \\  \hline 
 \tiny{\citetMeth[pg. ~300]{vanderweele2014effect}}  & {\tiny{{``These are not the natural direct and indirect effects considered earlier but are instead \textbf{analogs} arising from not fixing the mediator for each person to the level it would have been under a particular exposure, but rather to a level that is randomly chosen from the distribution of the mediator among all those with a particular exposure.''}}} \\  \hline 
 \tiny{\citetMeth[pg. ~135]{vanderweele2015explanation}}  & {\tiny{{``Although these natural direct and indirect effects are not identified, certain \textbf{analogues} of these effects based on randomized interventions on the mediator will be identified even in the presence of such exposure-induced mediator outcome confounding.''}}} \\  \hline 
 \tiny{\citetMeth[pg. ~157]{chen2016mediation}}  & {\tiny{{``Instead we consider randomized interventional \textbf{analogs} of these effects, following the work of \citetMeth{vanderweele2017mediation}.''}}} \\  \hline 
 \tiny{\citetMeth[pg. ~4011]{lok2016defining}}  & {\tiny{{``\textbf{Analogously} to natural direct and indirect effects, this article focuses on interventions $I$ that cause the mediator under treatment $A = 1$ and intervention $I$, $M_{1,I=1}$, to have the same distribution as $M_0$, given the pre-treatment common causes $C$ of mediator and outcome.''}}} \\  \hline 
 \tiny{\citetMeth[pg. ~917]{vanderweele2017mediation}}  & {\tiny{{``However, we define a randomized interventional \textbf{analogue} of natural direct and indirect effects that are identified in this setting.''}}} \\  \hline 
 \tiny{\citetMeth[pg. ~8]{lin2017interventional}}  & {\tiny{{``Rather than the standard definitions based on cross-world counterfactual outcomes, our method used definitions of randomly interventional \textbf{analogues}, which is not exactly how the mechanisms perform in nature. ''}}} \\  \hline 
 \tiny{\citetMeth[pg. ~7]{hejazi2020nonparametric}}  & {\tiny{{``Decomposition as the sum of direct and indirect effects affords an interpretation \textbf{analogous} to the corresponding standard decomposition of the average treatment effect into the natural direct and indirect effects.''}}} \\  \hline 
 \tiny{\citetMeth[pg. ~18]{zheng2017longitudinal}}  & {\tiny{{``Under the proposed formulation, one can obtain a total effect decomposition and the subsequent definition of natural direct and indirect effects that are \textbf{analogous} to those in \citetSM{pearl2001direct}.}''}} \\  \hline 
 \tiny{\citetMeth[pg. ~1]{diaz2020causal}}  & {\tiny{{``We present an \textbf{analogous} decomposition of the population intervention effect, defined through stochastic interventions on the exposure.''}}} \\  \hline 
 \tiny{\citetMeth[pg. ~5086]{mittinty2019effect}}  & {\tiny{{``Apart from their joint mediators and path-specific approaches, \citetMeth{vanderweele2014effect} introduced so-called randomised interventional \textbf{analogue} (in)direct effects.''}}} \\  \hline 
\bottomrule
 \hline

\end{tabular}
 \begin{tablenotes}
      \tiny
      \item  
    \end{tablenotes}
\end{threeparttable}

\end{table}
}

{\renewcommand{\arraystretch}{1}
\begin{table}[!htb]
\begin{threeparttable}

\caption{Example excerpts for Interpretability concerns: ``\textbf{Natural direct effect}''. \label{tab: Intconcern4}}

\begin{tabular}{|m{3cm}|m{13cm}|}
\toprule
     \emph{\textbf{Author}} & \emph{\textbf{Excerpt}} \\ \hline
 \tiny{\citetMeth[pg. ~281]{petersen2006estimation}}  & {\tiny{{``Similarly, according to equation (3), the \textbf{natural direct effect} is simply an average of the controlled direct effect at each level of the intermediate weighted with respect to the distribution of the intermediate variable in the unexposed.''}}} \\  \hline 
 \tiny{\citetMeth[pg. ~24]{van2008direct}}  & {\tiny{{``A \textbf{natural direct effect} is then simply defined as a summary measure of the controlled direct effect, specifically a weighted average, where weighting is determined by distribution of intermediate given baseline covariates in absence of exposure.''}}} \\  \hline 
 \tiny{\citetMeth[pg. ~6]{didelez2012direct}}  & {\tiny{{``The \textbf{natural direct effect} can be formulated as a special case of a standardized direct effect$[\dots]$''}}} \\  \hline 
 \tiny{\citetMeth[pg. ~7]{zheng2012causal}}  & {\tiny{{``We refer to the difference $P(T(1,Z(g_1,1)) > t_0)-P(T(1,Z(g_1,0)) > t_0)$ as the \textbf{natural indirect effect} (NIE) and the difference $P(T(1,Z(g_1,0))>t_0)-P(T(0,Z(g_0,0))>t_0)$ as the \textbf{natural direct effect} (NDE).''}}} \\  \hline 
 \tiny{\citetMeth[pg. ~5]{zheng2017longitudinal}}  & {\tiny{{``To contrast the effects of two exposure regimens on an outcome at time $\tau$, the corresponding\textbf{ natural indirect effect} is defined as $\mathbb{E}[Y_{\tau}(1, \overline{\Gamma}^1)] - \mathbb{E}[Y_{\tau}(1, \overline{\Gamma}^0)]$ and the \textbf{natural direct effect} is $\mathbb{E}[Y_{\tau}(1, \overline{\Gamma}^0)] - \mathbb{E}[Y_{\tau}(0, \overline{\Gamma}^0)]$.''}}} \\  \hline 
\bottomrule
 \hline

\end{tabular}
 \begin{tablenotes}
      \tiny
      \item  
    \end{tablenotes}
\end{threeparttable}

\end{table}
}

{\renewcommand{\arraystretch}{1}
\begin{table}[!htb]
\hspace*{-2cm}
\begin{threeparttable}

\caption{Example excerpts for Interpretability concerns: concordance between the stochastic mediation parameter and $\gamma_{NDE}$. \label{tab: Intconcern5}}

\begin{tabular}{|m{3cm}|m{16cm}|}
\toprule
     \emph{\textbf{Author}} & \emph{\textbf{Excerpt}} \\ \hline
 \tiny{\citetMeth[pg. ~280]{petersen2006estimation}}  & {\tiny{{``We note that, even when our direct effect assumption (7) fails to hold, equation (3) still estimates an interesting causal parameter: a summary of the direct effect of the exposure in the population with the intermediate controlled at its mean counterfactual level in the absence of exposure.''}}} \\  \hline 
 \tiny{\citetMeth[pg. ~3]{van2008direct}}  & {\tiny{{``The article focuses on modelling and estimation of this natural direct effect parameter, which happens to agree with the conventional natural direct effect parameter $E(Y_{a,Z_0} - Y_{0,Z_0})$ under any of the identifying assumptions of \citetSM{robins1992identifiability}, \citetSM{van2004estimation} or \citetSM{pearl2001direct}, but does not rely on any of these additional assumptions.''}}} \\  \hline 
 \tiny{\citetMeth[pg. ~3]{zheng2012causal}}  & {\tiny{{``Importantly, however, one should note that even though these SI-based parameters and their non-SI-based counterparts in appendix A4.2 all identify to the same statistical parameters, they are formally different causal parameters defined under different formulations (but aim to answer the same type of mediation questions)''}}} \\  \hline 
 \tiny{\citetMeth[pg. ~304]{vanderweele2014effect}}  & {\tiny{{``These expressions reduce to the mediation formulae (1) and (2) when L does not confound the association between $M$ and $Y$, conditional on covariates $C$.''}}} \\  \hline 
 \tiny{\citetMeth[pg. ~183]{vanderweele2015explanation}}  & {\tiny{{``One way to approach the interpretation of the estimators for direct and indirect effects then, if we can control for exposure–outcome, mediator–outcome, and exposure–mediator confounding, is that, without further assumptions, we can still interpret them as the interventional analogues of natural direct and indirect effects. If we are further willing to make the cross-world independence assumption, then we can also interpret these estimates as estimates of the natural direct and indirect effects themselves.''}}} \\  \hline 
 \tiny{\citetMeth[pg. ~4016]{lok2016defining}}  & {\tiny{{``I have shown that the proposed organic direct and indirect effects are identified by the same expressions as developed previously in the literature for natural direct and indirect effects.''}}} \\  \hline 
 \tiny{\citetMeth[pg. ~934]{vanderweele2017mediation}}  & {\tiny{{``When identified, these interventional direct and indirect effects do reduce to the natural direct and indirect effects where there is no mediator–outcome confounder affected by exposure (e.g. when there are no time varying confounders) but the interventional analogues can be estimated in a broader range of settings even when natural direct and indirect effects are not identified with the data.''}}} \\  \hline 
 \tiny{\citetMeth[pg. ~6]{lin2017interventional}}  & {\tiny{{``When the time-varying confounders are not affected by exposure, i. e. all $L_1$, $L_2$, and $M_2$ are all empty, $r_{\phi}(a, a', a'', a''')$ reduces to $\sum E[Y \mid v, a, m_1]Pr(m_1\mid v, a')Pr(v)$, which is the expression of the standard mediation parameter $E[Y(a,M_1(a'))]$.''}}} \\  \hline 
 \tiny{\citetMeth[pg. ~2]{rudolph2018robust}}  & {\tiny{{``The SDE and SIE coincide with the NDE and NIE in the absence of intermediate confounders.''}}} \\  \hline 
 \tiny{\citetMeth[pg. ~260]{vansteelandt2017interventional}}  & {\tiny{{``Under these assumptions, these effects reduce to expressions (1) and (2) obtained for average direct and indirect interventional effects, but with $L$ empty. It thus follows that in single mediator models without post-treatment confounding, natural (in)direct effects obtained under assumption (iv) can also be interpreted as interventional (in)direct effects (even when that assumption is violated).''}}} \\  \hline 
 \tiny{\citetMeth[pg. ~7]{hejazi2020nonparametric}}  & {\tiny{{``By generalizing the effect definitions of \citetMeth{diaz2020causal}, our proposed (in)direct effects include, as special cases, the natural (in)direct effects (under a static intervention on binary A and no intermediate confounders $L$).''}}} \\  \hline 
 \tiny{\citetMeth[pg. ~418]{lok2021causal}}  & {\tiny{{``Under the respective conditions usually made and if there are no post–treatment common causes of mediator and outcome, natural direct and indirect effects, their randomized counterparts, and their organic counterparts relative to $a = 1$ all lead to the same numerical results in settings where these effects are all well defined.''}}} \\  \hline 
 \tiny{\citetMeth[pg. ~628]{diaz2021nonparametric}}  & {\tiny{{``We base our approach on a proposal first outlined by \citetMeth{petersen2006estimation} and \citetMeth{van2008direct}, who argue that in the absence of cross-world assumptions, the standard identification formula for the natural direct effect may be interpreted as a weighted average of controlled direct effects, with weights given by a counterfactual distribution on the mediators. In subsequent work, \citetMeth{vanderweele2014effect} and \citetMeth{vansteelandt2017interventional} showed that the direct effect parameter of \citetMeth{petersen2006estimation} corresponds to a decomposition of a total causal effect defined in terms of a non- deterministic intervention on the mediator, and extended the methods to the setting of intermediate confounders, calling the methods interventional effects.''}}} \\  \hline 
 \tiny{\citetMeth[pg. ~5095]{mittinty2019effect}}  & {\tiny{{``These expressions reduce to mediation formulae in (1) and (2) when $L$ does not confound the association between $M_1$, $M_2$, and $Y$ conditional on covariates $C$.''}}} \\  \hline 
 \tiny{\citetMeth[pg. ~2]{loh2020heterogeneous}}  & {\tiny{{``Furthermore, the interpretation of interventional effects remains meaningful even when the exposure cannot be manipulated at the individual level, and the natural and interventional effects may coincide empirically.''}}} \\  \hline 
\bottomrule
 \hline

\end{tabular}
 \begin{tablenotes}
      \tiny
      \item  
    \end{tablenotes}
\end{threeparttable}

\end{table}
}

{\renewcommand{\arraystretch}{1}
\begin{table}[!htb]
\hspace*{-2cm}
\begin{threeparttable}

\caption{Example excerpts for Identifiability concerns: identifiability of the stochastic mediation parameter when $\gamma_{NDE}$ is not identified. \label{tab: IDconcern1}}
\vspace*{-2cm}

\begin{tabular}{|m{3cm}|m{16cm}|}
\toprule
     \emph{\textbf{Author}} & \emph{\textbf{Excerpt}} \\ \hline
 \tiny{\citetMeth[pg. ~280]{petersen2006estimation}}  & {\tiny{{``We note that, even when our direct effect assumption (7) fails to hold, equation (3) still estimates an interesting causal parameter: a summary of the direct effect of the exposure in the population with the intermediate controlled at its mean counterfactual level in the absence of exposure.''}}} \\  \hline 
 \tiny{\citetMeth[pg. ~204]{geneletti2007identifying}}  & {\tiny{{`Thus, in the DT framework that I propose from Section 3 onwards, neither the assumption of \citetSM{pearl2001direct} nor of \citetMeth{petersen2006estimation} is necessary; nor are any additional counterfactual assumptions (see \citetSM{robins2003semantics}) necessary for the identification of direct and indirect effects provided that the appropriate conditional independences hold.''}}} \\  \hline 
 \tiny{\citetMeth[pg. ~24]{van2008direct}}  & {\tiny{{``We note that these alternative definitions no longer depend on untestable assumptions regarding unmeasured confounding, nor are any additional assumptions necessary to make the natural direct effect identifiable.''}}} \\  \hline 
 \tiny{\citetMeth[pg. ~3]{zheng2012causal}}  & {\tiny{{``Consequently, the identifiability conditions of the resulting parameters would impose restrictions on the event indicators — conditions that we find too strong for the purpose of a survival study (this causal formulation is elaborated in detail in appendix A4.2, the resulting statistical parameters are the same as those in the main text).''}}} \\  \hline 
 \tiny{\citetMeth[pg. ~303]{vanderweele2014effect}}  & {\tiny{{``Although natural direct and indirect effects with $M$ alone as the mediator of interest are not identified, alternative effects that randomly set $M$ to a value chosen from the distribution of a particular exposure level can be identified.''}}} \\  \hline 
 \tiny{\citetMeth[pg. ~135]{vanderweele2015explanation}}  & {\tiny{{``Although these natural direct and indirect effects are not identified, certain analogues of these effects based on randomized interventions on the mediator will be identified even in the presence of such exposure-induced mediator outcome confounding.''}}} \\  \hline 
 \tiny{\citetMeth[pg. ~157]{chen2016mediation}}  & {\tiny{{``Unfortunately, in settings like Figure 1A in which there is a variable $L(1)$ that is affected by the exposure and confounds the relationship between the mediator and the outcome, natural direct and indirect effects are not identified.''}}} \\  \hline 
 \tiny{\citetMeth[pg. ~4017]{lok2016defining}}  & {\tiny{{``The assumptions are weaker than for natural direct and indirect effects.''}}} \\  \hline 
 \tiny{\citetMeth[pg. ~923]{vanderweele2017mediation}}  & {\tiny{{``Although these interventional direct and indirect effects defined here are not identical to natural direct and indirect effects, they are in some sense the best we may be able to do as the natural direct and indirect effects themselves will not be identified when a mediator–outcome confounder is affected by the exposure''}}} \\  \hline 
 \tiny{\citetMeth[pg. ~9]{lin2017interventional}}  & {\tiny{{``Our approach, based on the definition of randomly interventional analogues, requires only the first two assumptions described above to identify rPSEs non-parametrically.''}}} \\  \hline 
 \tiny{\citetMeth[pg. ~1]{rudolph2018robust}}  & {\tiny{{``We build on recent work that identified alternative estimands that do not require this assumption and propose a flexible and double robust targeted minimum loss-based estimator for stochastic direct and indirect effects.''}}} \\  \hline 
 \tiny{\citetMeth[pg. ~260]{vansteelandt2017interventional}}  & {\tiny{{``It thus follows that in single mediator models without post-treatment confounding, natural (in)direct effects obtained under assumption (iv) can also be interpreted as interventional (in)direct effects (even when that assumption is violated).''}}} \\  \hline 
 \tiny{\citetMeth[pg. ~4]{hejazi2020nonparametric}}  & {\tiny{{``Our proposed class of effects are the first to simultaneously avoid the requirement of cross-world counterfactual independencies; leverage stochastic interventions to be applicable to binary, categorical, and continuous-valued exposures; and remain identifiable despite intermediate confounding.''}}} \\  \hline 
 \tiny{\citetMeth[pg. ~199]{rudolph2021transporting}}  & {\tiny{{``Stochastic direct and indirect effects are similar to natural direct and indirect effects but do not require the natural direct and indirect effects identifying assumption of no measured or unmeasured post-treatment confounder of the mediator-outcome relationship.''}}} \\  \hline
 \tiny{\citetMeth[pg. ~414]{lok2021causal}}  & {\tiny{{``Organic indirect and direct effects: an intervention-based approach avoiding cross-worlds counterfactuals and assumptions''}}} \\  \hline 
 \tiny{\citetMeth[pg. ~627]{diaz2021nonparametric}}  & {\tiny{{``Interventional effects for mediation analysis were proposed as a solution to the lack of identifiability of natural (in)direct effects in the presence of a mediator-outcome confounder affected by exposure.''}}} \\  \hline 
 \tiny{\citetMeth[pg. ~172]{benkeser2021nonparametric}}  & {\tiny{{``A debate in this literature has emerged pertaining to the reliance of methodology on cross-world independence assumptions that are fundamentally untestable even in randomized controlled experiments $[\dots]$. A second {approach} considers seeking alternative definitions of mediation parameters that do not require such cross-world assumptions (\citetMeth{vanderweele2014effect}, \citetMeth{rudolph2018robust}, among others).''}}} \\  \hline
 \tiny{\citetMeth[pg. ~2]{zheng2017longitudinal}}  & {\tiny{{``The corresponding natural direct effect and indirect effect parameters have different interpretation than those under the formulations in \citetSM{robins1992identifiability} and \citetSM{pearl2001direct}, but their identifiability is at hand even in the presence of exposure-induced mediator-outcome confounder (e.g. \citetMeth{vanderweele2014effect}).''}}} \\  \hline 
 \tiny{\citetMeth[pg. ~2]{loh2020heterogeneous}}  & {\tiny{{``This distinction allows identification of interventional indirect effects in the context of multiple mediators under empirically verifiable assumptions that could be ensured in a randomized trial.''}}} \\  \hline 
 \tiny{\citetMeth[pg. ~2]{diaz2020causal}}  & {\tiny{{``In an attempt to solve these problems, several researchers have proposed methods that do away with cross-world counterfactual independences.''}}} \\  \hline 
 \tiny{\citetMeth[pg. ~5086]{mittinty2019effect}}  & {\tiny{{``These can be identified under much weaker conditions than natural (in)direct effects, but sum to a total interventional causal effect, not the TCE.''}}} \\  \hline 
  
\bottomrule
 \hline

\end{tabular}
 \begin{tablenotes}
      \tiny
      \item  
    \end{tablenotes}
\end{threeparttable}

\end{table}
}

{\renewcommand{\arraystretch}{1}
\begin{table}[!htb]
\begin{threeparttable}

\caption{Example excerpts for Identifiability concerns: ``\textbf{Method}''. \label{tab: IDconcern2}}

\begin{tabular}{|m{3cm}|m{13cm}|}
\toprule
     \emph{\textbf{Author}} & \emph{\textbf{Excerpt}} \\ \hline
 \tiny{\citetMeth[pg. ~300]{vanderweele2014effect}}  & {\tiny{{``As such, the \textbf{methods} presented in this article may help to address one of the major limitations of the causal mediation analysis literature.''}}} \\  \hline 
 \tiny{\citetMeth[pg. ~4014]{lok2016defining}}  & {\tiny{{``In order to directly apply the \textbf{method} described in this article, we therefore re-code $A = 0$ if a person was treated with AZT, and $A = 1$ if a person was not treated with AZT''}}} \\  \hline 
 \tiny{\citetMeth[pg. ~933]{vanderweele2017mediation}}  & {\tiny{{``In this paper we have considered \textbf{methods} for time varying exposures and mediators.''}}} \\  \hline 
 \tiny{\citetMeth[pg. ~1]{lin2017interventional}}  & {\tiny{{``This \textbf{method} is valid under assumptions of no unmeasured confounding and allows settings with mediators dependent on each other, interaction, and mediator-outcome confounders which are affected by exposure.''}}} \\  \hline 
 \tiny{\citetMeth[pg. ~1]{rudolph2018robust}}  & {\tiny{{``The proposed \textbf{method} intervenes stochastically on the mediator using a distribution which conditions on baseline covariates and marginalizes over the intermediate confounder.''}}} \\  \hline 
 \tiny{\citetMeth[pg. ~629]{diaz2021nonparametric}}  & {\tiny{{``To the best of our knowledge, this is the first \textbf{method} with all these properties.''}}} \\  \hline 
 \tiny{\citetMeth[pg. ~2]{diaz2020causal}}  & {\tiny{{``In an attempt to solve these problems, several researchers have proposed \textbf{methods} that do away with cross-world counterfactual independences.''}}} \\  \hline 
 \tiny{\citetMeth[pg. ~5098]{mittinty2019effect}}  & {\tiny{{``\textbf{Methods} for the decomposition of the total causal effect and the circumstances under which each is appropriate''}}} \\  \hline 
\bottomrule
 \hline

\end{tabular}
 \begin{tablenotes}
      \tiny
      \item  
    \end{tablenotes}
\end{threeparttable}

\end{table}
}

{\renewcommand{\arraystretch}{1}
\begin{table}[!htb]
\hspace*{-2cm}
\begin{threeparttable}

\caption{Example excerpts for Identifiability concerns: miscellaneous terms. \label{tab: IDconcern3}}

\begin{tabular}{|m{3cm}|m{3cm}|m{12cm}|}
\toprule
 \emph{\textbf{Term}} &    \emph{\textbf{Author}} & \emph{\textbf{Excerpt}} \\ \hline
\multirow{3}{*}{\textbf{{``solution''}} } & \tiny{\citetMeth[pg. ~7]{hejazi2020nonparametric}}  & {\tiny{{``Removal of the directed path from $L$ to $Z$ in the interventional distribution \textbf{resolves} this complication, analogously to \textbf{solutions} presented by those authors for the identification of simpler path-specific effects.''}}} \\  \cline{2-3}
 & \tiny{\citetMeth[pg. ~627]{diaz2021nonparametric}}  & {\tiny{{``Interventional effects for mediation analysis were proposed as a \textbf{solution} to the lack of identifiability of natural (in)direct effects in the presence of a mediator-outcome confounder affected by exposure.''}}} \\  \cline{2-3} 
& \tiny{\citetMeth[pg. ~8]{diaz2020causal}}  & {\tiny{{``\citetMeth{vanderweele2014effect} proposed a \textbf{solution} to this problem which involves a stochastic intervention on the mediator $Z$.''}}} \\  \hline 
\multirow{1}{*}{\textbf{{``relax''}} } & \tiny{\citetMeth[pg. ~3]{rudolph2018robust}}  & {\tiny{{``There has been recent work to \textbf{relax} the assumption of no intermediate confounder, $M_{a^*} \CI Y^{a,m} \mid W$, by using a stochastic intervention on $M$.''}}} \\  \hline 
\multirow{4}{*}{\textbf{{``overcome''}}} \textbf{}& \tiny{\citetMeth[pg. ~164]{vanderweele2015explanation}}  & {\tiny{{``We will, however, instead be able to employ randomized interventional analogues to natural direct and indirect effects, as we did in Chapter 5, to help \textbf{overcome} this difficulty.''}}} \\  \cline{2-3} 
& \tiny{\citetMeth[pg. ~4009]{lok2016defining}}  & {\tiny{{``To \textbf{overcome} these problems, this article proposes instead to base mediation analysis on newly defined “organic” interventions ($I$) on the mediator.''}}} \\  \cline{2-3} 
& \tiny{\citetMeth[pg. ~258]{vansteelandt2017interventional}}  & {\tiny{{``We will adapt this proposal to \textbf{overcome} this, and then extend it to the case of multiple mediators.''}}} \\  \cline{2-3}
& \tiny{\citetMeth[pg. ~631]{diaz2021nonparametric}}  & {\tiny{{``To \textbf{overcome} this problem while retaining the path decomposition employed by the natural direct and indirect effects, we adopt an {approach} previously outlined by \citetMeth{petersen2006estimation, van2008direct, zheng2012causal, rudolph2018robust}, defining direct and indirect effects using stochastic interventions on the mediator.''}}} \\  \hline 
\multirow{4}{*}{\textbf{{``avoid''}}} & \tiny{\citetMeth[pg. ~305]{vanderweele2014effect}}  & {\tiny{{``In contrast to the previous two approaches, approach 3 \textbf{avoids} assumptions of unconfoundedness with respect to smoking.''}}} \\  \cline{2-3} 
& \tiny{\citetMeth[pg. ~261]{vansteelandt2017interventional}}  & {\tiny{{``The latter is chosen independently of $M_1$, so as to \textbf{avoid} assumptions on the joint distribution of the counterfactuals $M_1^a$ and $M_2^{a^*}$ corresponding to different exposure levels.''}}} \\  \cline{2-3} 
& \tiny{\citetMeth[pg. ~4]{hejazi2020nonparametric}}  & {\tiny{{``Our proposed class of effects are the first to simultaneously \textbf{avoid} the requirement of cross-world counterfactual independencies; leverage stochastic interventions to be applicable to binary, categorical, and continuous-valued exposures; and remain identifiable despite intermediate confounding.''}}} \\  \cline{2-3} 
& \tiny{\citetMeth[pg. ~414]{lok2021causal}}  & {\tiny{{``Organic indirect and direct effects: an intervention-based approach \textbf{avoiding} cross-worlds counterfactuals and assumptions''}}} \\  \hline 
\multirow{4}{*}{\textbf{{``allow''}}} & \tiny{\citetMeth[pg. ~918]{vansteelandt2017interventional}}  & {\tiny{{``The approach that we develop draws on both Robins’s g-formula and Pearl’s mediation formula but unites these in a single framework that \textbf{allows} us to assess mediation with time varying exposures and mediators in the presence of time varying confounders.''}}} \\  \cline{2-3} 
& \tiny{\citetMeth[pg. ~1]{lin2017interventional}}  & {\tiny{{``This method is valid under assumptions of no unmeasured confounding and \textbf{allows} settings with mediators dependent on each other, interaction, and mediator-outcome confounders which are affected by exposure.''}}} \\  \cline{2-3} 
& \tiny{\citetMeth[pg. ~1]{hejazi2020nonparametric}}  & {\tiny{{``Our (in)direct effects are identifiable without a restrictive assumption on cross-world counterfactual independencies, \textbf{allowing} for substantive conclusions drawn from them to be validated in randomized controlled trials.''}}} \\  \cline{2-3} 
& \tiny{\citetMeth[pg. ~2]{loh2020heterogeneous}}  & {\tiny{{``This distinction \textbf{allows} identification of interventional indirect effects in the context of multiple mediators under empirically verifiable assumptions that could be ensured in a randomized trial.''}}} \\  \hline 
\multirow{1}{*}{\textbf{{``technique''}}} & \tiny{\citetMeth[pg. ~1]{lin2017interventional}}  & {\tiny{{``Mediation analysis is a \textbf{technique} to decompose the total effect of an exposure on an outcome into a direct effect (the effect not through a mediator) and an indirect effect (the effect through a mediator).''}}} \\  \hline 
\bottomrule
 \hline

\end{tabular}
 \begin{tablenotes}
      \tiny
      \item  
    \end{tablenotes}
\end{threeparttable}

\end{table}
}

\clearpage

\clearpage\doublespacing
\section{Supplementary materials for systematic review of the applied literature} \label{appsec: SysRev}

\subsection{Systematic review methods} \label{subsec: met}
We queried a commonly used online literature database (Google Scholar, on August 11, 2022) to return and de-duplicate the titles of every single article that had cited at least one of the methodological papers developing theory for $\gamma_{RDE}$ (listed in Table \ref{tab: RIAstoch}, with the exception of \citetSM{vanderweele2015explanation}, a compendium of theory and practical tools related to investigations of causal mediation and interaction).  This de-duplicated list of titles, abstracts, and full-texts were successively screened according to a set of exclusion criteria listed in Table \ref{tab: excrit}. Notably, we excluded articles that focused on disparities and estimating the extent to which a factual difference between two groups would remain under an intervention that assigned a variable $M$ to both the groups according to its distribution among one of the groups. Clearly this parameter is distinct from  $\gamma_{RDE}$, despite the fact that they are often discussed as conceptually-related (see, for example \citetSM{vansteelandt2017interventional}). The articles that remained after applying the exclusion criteria constituted our analytic sample of applied articles.

{\renewcommand{\arraystretch}{1}
\begin{table}[!htb]

\begin{threeparttable}

\caption{Exclusion criteria for systematic review of the applied stochastic mediation literature \label{tab: excrit}}

\begin{tabular}{|p{10cm}|}
\toprule
    \emph{\textbf{Exclusion criterion}} \\ \hline
    Contribution is primarily methodological \\  \hline 
    Did not target a randomized interventional analogue parameter \\  \hline 
    Multiple mediators$^*$ \\  \hline 
    Sequential treatment - mediator setting \\  \hline 
    Disparities-based$^{**}$  \\  \hline 
\bottomrule
 \hline
\end{tabular}
 \begin{tablenotes}
      \tiny
      \item $*$ Articles could be included if they treated multiple mediators as a single multi-dimensional intermediate variable
      \item $**$ Disparities-based parameters considered estimands of the form $\mathbb{E}[Y^{G'_{a^* \mid L}} \mid A=a] - \mathbb{E}[Y^{G'_{a^* \mid L}} \mid A=a^*]$, where $G'_{a^* \mid l}$ is a distribution of the mediator $M$ \textit{among} individuals with $A=a^*$ and $L=l$, and $Y^{G'_{a^* \mid L}}$ is a counterfactual defined simply as the outcome under and intervention that assigns the mediator as a random draw from this distribution. 
    \end{tablenotes}
\end{threeparttable}

\end{table}
}

For each article in the our analytic sample, we abstracted the following data: an indicator of overlap in authorship with the set of authors for the methodological papers in Table \ref{tab: RIAstoch}; an indicator of whether or not the abstract explicitly stated that the article targets a stochastic mediation parameter; excerpt(s) of text representing the stated rationale for targeting $\gamma_{RDE}$; excerpts of interpretive claims $\mathfrak{i}$ for $\gamma_{RDE}$ in the results section of the abstract; claims $\mathfrak{i}$ in the discussion/conclusion section of the abstract; claims $\mathfrak{i}$ in the results section of the main text; and claims $\mathfrak{i}$ in the discussion/conclusion section of the main text.

To facilitate analysis and presentation of the excerpts, we modified the text of each excerpt by replacing each article's $$\{\text{treatment, mediator, outcome, baseline covariates, post-treatment covariates}\}$$ with the symbols $\{A, M, Y, L, W\}$ and replaced any numbers $$\{\text{number 1, number 2, number 3,}\dots \}$$ with the symbols $\{X, X', X'',\dots\}$. For each stated rationale, we classified whether or not the authors implicitly followed a \textit{pragmatic} or \textit{ideal} framework to causal inference by determining whether their stated rationale for selecting $\gamma_{RDE}$ was based on its correspondence with a \textit{scientific task} or based on its statistical properties. For each interpretive claim $\mathfrak{i}$, we determined if the claim had any features that would permit a reader to disambiguate its membership to the set $\mathcal{I}_{RDE}(\hat{\gamma}_{RDE})$ versus the set $\mathcal{I}_{NDE}(\hat{\gamma}_{RDE})$. We analysed the data marginally and also conditional on authorship overlap, because we hypothesized that results might qualitatively differ based on this feature - particularly we hypothesized that articles with authors who also authored methodological papers on $\gamma_{RDE}$ would exhibit lower rates of identity slippage than those articles without such an author.

\subsection{Bibliography for review of applied articles estimating $\gamma_{RDE}$} \label{appsubsec: BibApp}
 \begin{figure}[!htb]
    \centering
\begin{tikzpicture}
 \node [font=\small, anchor=west] (0)  at (4,10.5)  {  \textbf{Total Citing Articles:} 965};
    \node [font=\tiny, anchor=west] (1)  at (4.5,10)  {  $\bullet$ \citeMeth{geneletti2007identifying}: 102};
    \node [font=\tiny, anchor=west] (2)  at (4.5,10-0.5*1)   {  $\bullet$ \citeMeth{didelez2012direct}: 152};
    \node [font=\tiny, anchor=west] (3)  at (4.5,10-0.5*2)   {  $\bullet$ \citeMeth{vanderweele2014effect}: 275};
    \node [font=\tiny, anchor=west] (4)  at (4.5,10-0.5*3)   {  $\bullet$ \citeMeth{chen2016mediation}: 7};
    \node [font=\tiny, anchor=west] (5)  at (4.5,10-0.5*4)   {  $\bullet$ \citeMeth{vanderweele2017mediation}: 169};
    \node [font=\tiny, anchor=west] (6)  at (4.5,10-0.5*5)   {  $\bullet$ \citeMeth{lin2017interventional}: 25};
    \node [font=\tiny, anchor=west] (7)  at (4.5,10-0.5*6)   {  $\bullet$ \citeMeth{rudolph2018robust}: 30};
    \node [font=\tiny, anchor=west] (8)  at (4.5,10-0.5*7)   {  $\bullet$ \citeMeth{rudolph2021transporting}: 4};
    \node [font=\tiny, anchor=west] (9)  at (4.5,10-0.5*8)   {  $\bullet$ \citeMeth{vansteelandt2017interventional}: 168};
    \node [font=\tiny, anchor=west] (10) at (4.5,10-0.5*9)  {  $\bullet$ \citeMeth{mittinty2019effect}: 5};
    \node [font=\tiny, anchor=west] (11) at (4.5,10-0.5*10)  {  $\bullet$ \citeMeth{benkeser2021nonparametric}:8};
    \node [font=\tiny, anchor=west] (12) at (4.5,10-0.5*11) {  $\bullet$ \citeMeth{loh2020heterogeneous}: 1};
    \node [font=\tiny, anchor=west] (13) at (4.5,10-0.5*12) {  $\bullet$ \citeMeth{diaz2021nonparametric}: 19};
\node[draw=black, thick,fit={(0) (1) (2) (3)  (4) (5) (6) (7) (8) (9) (10) (11) (12) (13)} , inner sep=.1cm]   (box1) {};   

    \node [font=\small  , anchor=west] (total) at (4,1) {  \textbf{Total (unique) for title review:} 572};
\node[draw=black, thick,fit={(total)} , inner sep=.1cm]   (box2) {};   

    \node [font=\small  , anchor=west] (dups) at (14,2.5) {  \textbf{Duplicates:} 393};
\node[draw=black, thick,fit={(dups)} , inner sep=.1cm]   (box3) {};

    \node [font=\small  , anchor=west] (titex) at (14,-0.5) {  \textbf{Not applications:} 407};
\node[draw=black, thick,fit={(titex)} , inner sep=.1cm]   (box4) {};

    \node [font=\small  , anchor=west] (total) at (4,-2) {  \textbf{Included in abstract review:} 165};
\node[draw=black, thick,fit={(total)} , inner sep=.1cm]   (box5) {};

    \node [font=\small  , anchor=west] (absrev0) at (14,-2) {  \textbf{Excluded because:}};
    \node [font=\tiny, anchor=west] (absrev1)  at (14.5,-2-0.5*1)   {  Primarily methodological: 21};
    \node [font=\tiny, anchor=west] (absrev2)  at (14.5,-2-0.5*2)   {  Not related to stochastic mediation: 11};
    \node [font=\tiny, anchor=west] (absrev3)  at (14.5,-2-0.5*3)   {  Not single time-point: 12};
    \node [font=\tiny, anchor=west] (absrev4)  at (14.5,-2-0.5*4)   {  Not single mediator: 8};
    \node [font=\tiny, anchor=west] (absrev5)  at (14.5,-2-0.5*5)   {  Disparities-based: 27};
\node[draw=black, thick,fit={(absrev0) (absrev1) (absrev2) (absrev3)  (absrev4) (absrev5)} , inner sep=.1cm]   (box6) {}; 

    \node [font=\small  , anchor=west] (total) at (4,-5) {  \textbf{Included in full-text review:} 86};
\node[draw=black, thick,fit={(total)} , inner sep=.1cm]   (box7) {};

    \node [font=\small, anchor=west](txtrev0) at (14,-6) {  \textbf{Excluded because:}};
    \node [font=\tiny, anchor=west] (txtrev1)  at (14.5,-6-0.5*1)   {  Not related to stochastic mediation: 50};
    \node [font=\tiny, anchor=west] (txtrev2)  at (14.5,-6-0.5*2)   {  Not single time-point: 9};
    \node [font=\tiny, anchor=west] (txtrev3)  at (14.5,-6-0.5*3)   {  Not single mediator: 9};
    \node [font=\tiny, anchor=west] (txtrev4)  at (14.5,-6-0.5*4)   {  Disparities-based: 2};
\node[draw=black, thick,fit={(txtrev0) (txtrev1) (txtrev2) (txtrev3)  (txtrev4)} , inner sep=.1cm]   (box8) {};

    \node [font=\small  , anchor=west] (total) at (4,-8) {  \textbf{Included in systematic review:} 16};
    \node [font=\small, anchor=west] (total1)  at (4.5,-8-0.5*1)   {Authorship overlap with a methodological paper: 8};
\node[draw=black, thick,fit={(total) (total1)} , inner sep=.1cm]   (box9) {};

     \coordinate [right=1cm of box1.south west, anchor=south ] (pt1);   
     \coordinate [right=1cm of box2.north west, anchor=north ] (pt2); 
     \coordinate [right=1cm of box2.south west, anchor=south ] (pt3); 
     \coordinate [right=1cm of box5.north west, anchor=north ] (pt4);
     \coordinate [right=1cm of box5.south west, anchor=south ] (pt5);
     \coordinate [right=1cm of box7.north west, anchor=north ] (pt6);
     \coordinate [right=1cm of box7.south west, anchor=south ] (pt7);
     \coordinate [right=1cm of box9.north west, anchor=north ] (pt8);
    \draw[->] [black, thick] (pt1) -- (pt2);
\draw [thick, black, ->] (pt1) -- ++(0, -1cm) |- (box3.west) ;
\draw [thick, black, ->] (pt3) -- ++(0, -1cm) |- (box4.west) ;
\draw [thick, black, ->] (pt5) -- ++(0, -1cm) |- (box6.west) ;
\draw [thick, black, ->] (pt7) -- ++(0, -1cm) |- (box8.west) ;
    \draw[->] [black, thick] (pt3) -- (pt4);
    \draw[->] [black, thick] (pt5) -- (pt6);
    \draw[->] [black, thick] (pt7) -- (pt8);

\end{tikzpicture}
    \caption{Diagram depicting selection process for articles included in the systematic text review. }
    \label{fig: consort}
\end{figure}
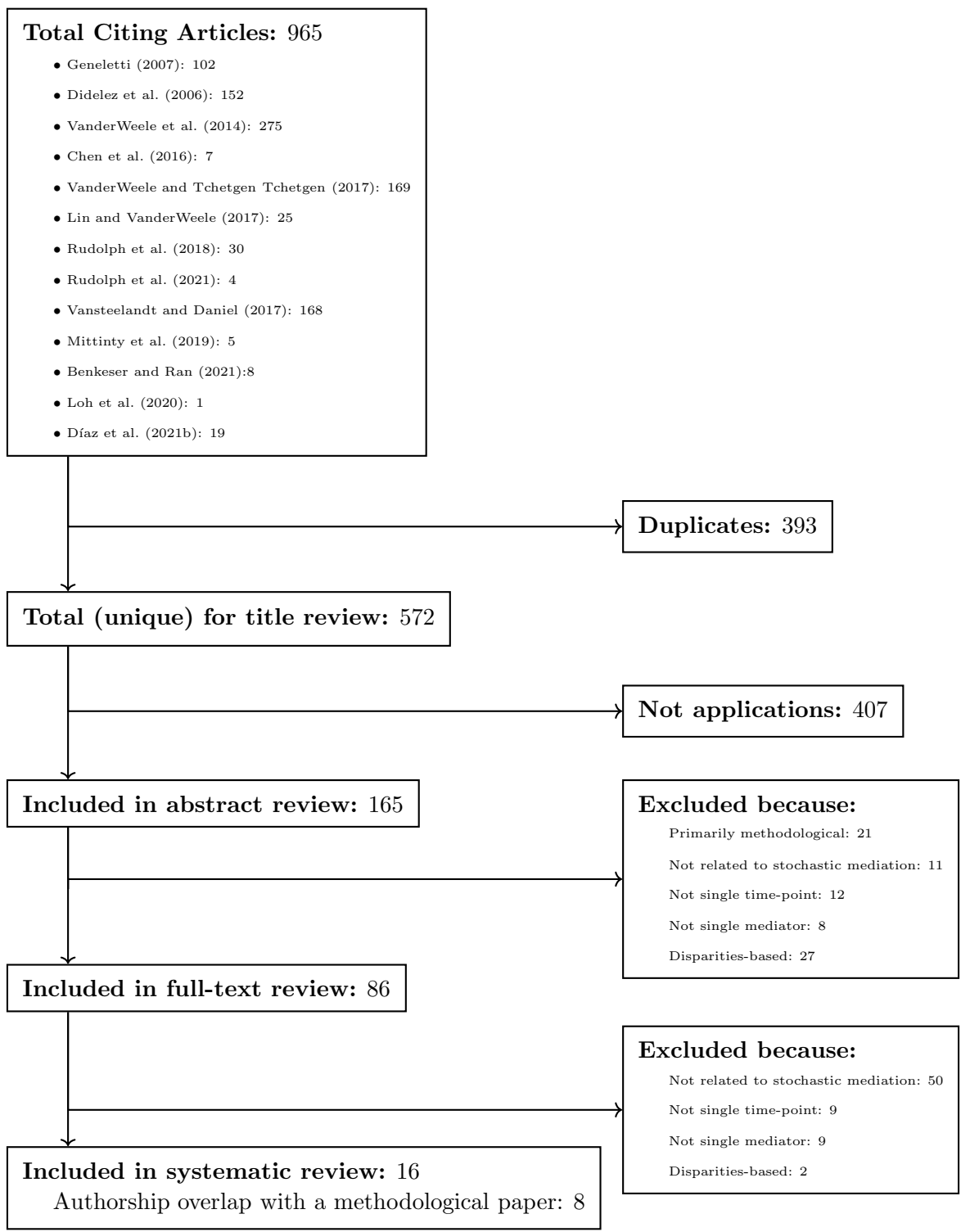\clearpage

    \bibliographystyleApp{unsrtnat}
    \bibliographyApp{refs}

\subsection{Interpretation excerpts from Applications of $\gamma_{RDE}$} \label{appsubsec: Excerpts}
    \newgeometry{margin=1cm} 
\begin{landscape}
\thispagestyle{empty}

{\renewcommand{\arraystretch}{1}
\begin{table}[!htb]
\begin{threeparttable}

\caption{Interpretation excerpts from Results sections of Applications \label{tab: results}}


\begin{tabular}{|m{2.5cm}|m{2cm}|m{10cm}|m{10cm}|}
\toprule
\multirow{2}{*}{} & \multirow{2}{*}{\textbf{\emph{Article}} } & \multicolumn{2}{|c|}{\textbf{\emph{Excerpts}}}   \\   \cline{3-4} 
& & \multicolumn{1}{|c|}{\textbf{{Abstract}}} & \multicolumn{1}{|c|}{\textbf{{Main text}}} \\ \hline
 \multirow{8}{2.5cm}{{\small \textbf{{No authorship overlap}}}}   
   & \tiny{\citeApp[pg. ~1630, 1634]{wang2022mechanisms}} & \tiny{{{``[$X\%$] excess [$Y$] risk from [$A$] was mediated by [$M$].''}}} & \tiny{{{``[$X\%$] of this effect was mediated through [$M$]: the estimated indirect effect was [$X\%$]. The remaining [$X\%$] was unexplained: the estimated direct effect not mediated through [$M$] was [$X\%$].''}}} \\  \cline{2-4} 
   &\tiny{\citeApp[pg. ~1188, 1191]{wang2022does}} & \tiny{{{``[$X\%$] of the total effect was mediated via [$M$] (indirect effect: $X'$[$\dots$]). }}} & \tiny{{{``[$X\%$ to $X'\%$] of the total effects of [$A$] on hearing loss were explained by [$M\dots$].''}}} \\  \cline{2-4}
   &\tiny{\citeApp[pg. ~1, 8]{campbell2021placental}} & \tiny{{{``[$M$] explained [$X\%$] of the association between [$Y$] and [$A$].''}}} & \tiny{{{``[$X\%$] of the association between $A$ and $Y$ was attributable to differences in [$M\dots$].''}}} \\  \cline{2-4}
   &\tiny{\citeApp[pg. ~1, 31]{wodtke2020neighborhood}} & \tiny{{{``We find no evidence that [$A$] effects are mediated by or interact with [$M$].''}}} & \tiny{{{``[$\dots$]estimates of the [RDE] indicate that [$A\dots$], would still reduce [$Y$] by [$X$] even after an intervention to fix the [$M$] to that observed [among $A=0$].''}}} \\  \cline{2-4}
   &\tiny{\citeApp[pg. ~1, 4]{liu2020inflammation}} & \tiny{{{``[Those with $A=1$] had [$X$ higher $Y$], of which [$X\%$] was mediated via [$M$]''}}} & \tiny{{{``Children with [$A=1$] had [$X$ higher $Y$] compared to those with [$A=0$], of which [$X\%$] was mediated via [$M\dots$].''}}} \\  \cline{2-4}
   &\tiny{\citeApp[pg. ~e195, e198]{komulainen2019childhood}} & \tiny{{{``There was no evidence for indirect effects from [$A$] to [$Y$] through [$M$] after controlling for time-dependent confounding by [$W$] (indirect effect = [$X\dots$]).''}}} & \tiny{{{``The analog for the direct effect on [$Y$] indicated a corresponding difference of [$X$] in [$Y$]. Approximately [$X\%$] of the associations of [$A$] with [$Y$] were estimated to be mediated through [$M\dots$].''}}} \\  \cline{2-4}
   &\tiny{\citeApp[pg. ~153, 157]{hanson2019does}} & \tiny{{{``A mediation model including [$W$] further showed that [$X\%$] of the association between [$A$] and [$Y$] was attributable to a weak indirect effect through [$M$].''}}} & \tiny{{{``The randomized analogue of the indirect effect was similar but not statistically significant in the adjusted model[$\dots$].''}}} \\  \cline{2-4}
   &\tiny{\citeApp[pg. ~615, 627]{gage2013maternal}} & \tiny{{{``[$Y$] declines significantly (by a factor of [$X$]) through the direct effect of [$A$]. The indirect effect of [$A$] among [$L=1$] is small but significant in three cohorts.''}}} & \tiny{{{``Further decomposition of the change of [$Y$] into direct (independent of [$M$]) and indirect (associated with changes in [$M$]) effects of [$A$] indicates a strong direct effect in the primary subpopulation, which reduces [$Y\dots$].''}}} \\  \hline \hline
\multirow{8}{2.5cm}{{\small \textbf{{Authorship overlap}}}}   
   & \tiny{\citeApp[pg. ~336, 340]{rudolph2021helped}} & \tiny{{{``The majority ([$X\%$]) of the total negative long-term effects could be explained by indirect effects through the mediators considered.''}}} & \tiny{{{``[$\dots$]nearly all of the effect of [$A$] on [$Y$] operates through the mediators considered, explaining [$X\%$] of the total interventional effect[$\dots$].''}}} \\  \cline{2-4} 
   & \tiny{\citeApp[pg. ~2094, 2099]{rudolph2021explaining}} & \tiny{{{``[$\dots$]the protective indirect path contributed a [$X\%$] reduced risk of [$Y$] comparing [$A=1$ to $A=0$] (explaining [$X\%$] of the total effect).''}}} & \tiny{{{``The indirect effect accounted for much of the overall increased risk of [$Y$]  on [$A=1$ versus $A=0$] in [$L=1$] (contributing a [$X\%$] reduced risk of [$Y$] corresponding to [$X\%$] of the total effect).''}}} \\  \cline{2-4}
   & \tiny{\citeApp[pg. ~820, 827]{goin2020mediation}} & \tiny{{{``The largest indirect effects for the association between [$A$] and [$Y$] were observed for [$M$] (stochastic indirect effect = [$X$]).''}}} & \tiny{{{``The largest stochastic indirect effects were observed for [$M$] (stochastic indirect effect = [$X\dots$]).''}}} \\  \cline{2-4}
   & \tiny{\citeApp[pg. ~41, 46]{ploubidis2019lifelong}} & \tiny{{{``We found that [$A$] influence [$Y$] either directly or indirectly 65 years later.''}}} & \tiny{{{``[$A$] had both direct and indirect -via [$M$] -effects on [$Y$].''}}} \\  \cline{2-4}
   & \tiny{\citeApp[pg. ~1, 6]{casey2019unconventional}} &  \tiny{{{``We found no relationship between [$A$] and [$Y$] and no mediation effect either overall or when stratifying by [$L$].''}}} &\tiny{{{``The direct effect of [$A$] on [$Y$] appeared protective for [$L=1\dots$].''}}} \\  \cline{2-4}
   & \tiny{\citeApp[pg. ~590, 594]{rudolph2018mediation}} & \tiny{{{``[$M$] partially mediated the effect of [$A$] on [$Y$] (e.g., stochastic indirect effect: [$X\%\dots$]).''}}} & \tiny{{{``Using the above data-dependent stochastic interventions, we estimated the direct and indirect effects of [$A$] on [$Y\dots$] [$A$] led to a [$X\%$] increased probability of [$Y$] through [$M\dots$].''}}} \\  \cline{2-4}
   & \tiny{\citeApp[pg. ~884]{maika2017associations}} & \tiny{{{[None provided]}}} & \tiny{{{``From[$\dots$]the intervention analog approach, the indirect effect of [$A$] mediated through [$M$] was smaller[$\dots$].''}}} \\  \cline{2-4}
   & \tiny{\citeApp[pg. ~108, 115]{pearce2016early}} & \tiny{{{``Around two-thirds of this elevated risk was ‘direct’ and the majority of the remainder was mediated by [$M$].''}}} & \tiny{{{``[$\dots$]the small indirect pathway from [$A$] to [$Y$] through [$M$] was not attenuated after adjustment for intermediate confounding.''}}} \\   \hline  \hline 
\bottomrule
 \hline
\end{tabular}
 \begin{tablenotes}
      \tiny
      \item  A: treatment; M: mediator; Y: outcome; L: pre-treatment covariate; W: post-treatment covariate.
      \item We use [brackets] to indicate when we have modified excerpts to facilitate examination of interpretive claims, according to procedures described in Web Appendix \ref{subsec: met}. We refer readers to Tables \ref{tab: AppResAb} and \ref{tab: AppResMain} in Web Appendix \ref{appsubsec: uanbrevExcerpts}, which provide excerpts in their entirety.
    \end{tablenotes}
\end{threeparttable}

\end{table}
}

\end{landscape}
\restoregeometry

    \newgeometry{margin=1cm} 
\begin{landscape}
\thispagestyle{empty}

{\renewcommand{\arraystretch}{1}
\begin{table}[!htb]
\begin{threeparttable}

\caption{Interpretation excerpts from Discussion / Conclusion sections of Applications \label{tab: discussion}}


\begin{tabular}{|m{2.5cm}|m{2cm}|m{10cm}|m{10cm}|}
\toprule
\multirow{2}{*}{} & \multirow{2}{*}{\textbf{\emph{Article}} } & \multicolumn{2}{|c|}{\textbf{\emph{Excerpts}}}   \\   \cline{3-4} 
& & \multicolumn{1}{|c|}{\textbf{{Abstract}}} & \multicolumn{1}{|c|}{\textbf{{Main text}}} \\ \hline
 \multirow{8}{2.5cm}{{\small \textbf{{No authorship overlap}}}}   
   & \tiny{\citeApp[pg. ~1630, 1636]{wang2022mechanisms}} & \tiny{{{``Evidence of [$M$] mediating the effect of [$A$] on [$Y$] supports this proposed mechanism.''}}} & \tiny{{{``[$\dots$]estimating interventional effects allowed us to evaluate how much [$Y$] risk could be reduced with a clinical intervention that would reduce [$M$]. [$M$] mediated approximately [$X\%$] of the excess [$Y$] risk for [having $A=1$].''}}} \\  \cline{2-4} 
   &\tiny{\citeApp[pg. ~1188]{wang2022does}} & \tiny{{{``[$M$] may play an important mediating role in the [$Y$] reductions associated with [$A\dots$].''}}} & \tiny{{[None provided]}} \\  \cline{2-4}
   &\tiny{\citeApp[pg. ~1, 12]{campbell2021placental}} & \tiny{{{``Our findings indicate substantial [$M$] heterogeneity in [$A$] that predicts previously [$Y$] differences.''}}} & \tiny{{{``A substantial proportion of the [effect of $A$] in [$Y$] was mediated by [$M$].''}}} \\  \cline{2-4}
   &\tiny{\citeApp[pg. ~36]{wodtke2020neighborhood}} & \tiny{{{[None provided]}}} & \tiny{{{``[$\dots$]we find no evidence that [$A$] effects are mediated by [$M\dots$].''}}} \\  \cline{2-4}
   &\tiny{\citeApp[pg. ~1, 5]{liu2020inflammation}} & \tiny{{{``[$M$] mediated associations between [$A$] and [$Y$]''}}} & \tiny{{{``We found that the association of [$A$] with [$Y$] was mediated by [$M\dots$].''}}} \\  \cline{2-4}
   &\tiny{\citeApp[pg. ~e195, e200]{komulainen2019childhood}} & \tiny{{{``After accounting for [$W$], [$M$] explain[s] little of the associations.''}}} & \tiny{{{``Findings from the causal mediation analysis suggest that this association is not explained by [$M$] once [$L,W$] are accounted for.''}}} \\  \cline{2-4}
   &\tiny{\citeApp[pg. ~153, 159]{hanson2019does}} & \tiny{{{``This study indicates an association between [$A$] and [$Y$] that is partially ascribed to [$M$].''}}} & \tiny{{{``Our results imply that [$M$] may be a mechanism that explains the association between [$A$] and [$Y$].''}}} \\  \cline{2-4}
   &\tiny{\citeApp[pg. ~615, 631]{gage2013maternal}} & \tiny{{{``Overall, our results are consistent with the view that the decrease in [$Y$] is not mediated by [$M$].''}}} & \tiny{{{``The methods of effect decomposition used here are based on statistical decision theory \citep{geneletti2007identifying}, as opposed to the more common `counterfactual' approaches[$\dots$]. By standardizing [$M$] (within [$A=1$]), the main effect of the association of [$A$] and [$Y$] is eliminated, and the regressions can estimate the direct effect of [$A$] on [$Y$].''}}} \\  \hline \hline
\multirow{8}{2.5cm}{{\small \textbf{{Authorship overlap}}}}   
   & \tiny{\citeApp[pg. ~336, 344]{rudolph2021helped}} & \tiny{{{``This evidence suggests that, even though [$A$] had the desired effects on [$M$], these “positives” ultimately negatively impacted [$Y$].''}}} & \tiny{{{``Using[$\dots$]interventional direct and indirect effects, we concluded that most of the overall harmful effects operated through [$M$].''}}} \\  \cline{2-4} 
   & \tiny{\citeApp[pg. ~2094, 2099]{rudolph2021explaining}} & \tiny{{{``A novel approach to mediation analysis shows that much of the difference in [$A$] effectiveness ([$A=1$] versus [$A=0$]) on [$Y$] appears to be explained by [$M$].''}}} & \tiny{{{``Using mediation analysis, we found that the pathway from [$A$] through [$M$] contributed to much of the increased risk of [$Y$].''}}} \\  \cline{2-4}
   & \tiny{\citeApp[pg. ~820, 828]{goin2020mediation}} & \tiny{{{``[$A$] was associated with [$Y$], and this association was partially mediated by [$M$].''}}} & \tiny{{{``Our findings suggest that [$A$] increases risk of [$Y$], with the strongest mediation associations being seen for [$M\dots$].''}}} \\  \cline{2-4}
   & \tiny{\citeApp[pg. ~54]{ploubidis2019lifelong}} & \tiny{{{``[None provided]''}}} & \tiny{{{``[$\dots A$] had only an indirect association with [$Y$].''}}} \\  \cline{2-4}
   & \tiny{\citeApp[pg. ~1, 7]{casey2019unconventional}} &  \tiny{{{``We observed a relationship between [$A$] and [$M$], which did not mediate the overall association between [$A$] and [$Y$].''}}} & \tiny{{{``Our study indicates that [$M$] did not act as a mediator of the relationship between [$A$] and [$Y$].''}}} \\  \cline{2-4}
   & \tiny{\citeApp[pg. ~590, 597]{rudolph2018mediation}} & \tiny{{{``[$M$] played little role in mediating the effect of [$A$] on [$Y$].''}}} & \tiny{{{``In summary, we found evidence that [$M$] weakly mediated the relationship between [$A$] and [$Y$].''}}} \\  \cline{2-4}
   & \tiny{\citeApp[pg. ~886]{maika2017associations}} & \tiny{{{[None provided]}}} & \tiny{{{``From effect decomposition, we found that [$A$] had a bigger direct effect on [$Y$] than via its mediated effect through [$M$].[$\dots$]Finally, in the intervention analog approach, effect decomposition is estimated as an analog of a sequentially randomized mediator based on the exposure level, which effectively removes [$W$].''}}} \\  \cline{2-4}
   & \tiny{\citeApp[pg. ~108, 116]{pearce2016early}} & \tiny{{{``Early interventions [on $M$] also hold potential for reducing inequalities in [$Y$].''}}} & \tiny{{{``Decomposition of the indirect pathway showed that around [$X\%$] was through [$M\dots$].''}}} \\   \hline  \hline 
\bottomrule
 \hline
\end{tabular}
 \begin{tablenotes}
      \tiny
      \item  A: treatment; M: mediator; Y: outcome; L: pre-treatment covariate; W: post-treatment covariate.
      \item We use [brackets] to indicate when we have modified excerpts to facilitate examination of interpretive claims, according to procedures described in Web Appendix \ref{subsec: met}. We refer readers to Tables \ref{tab: AppDisAb} and \ref{tab: AppDisMain} in Web Appendix \ref{appsubsec: uanbrevExcerpts}, which provide excerpts in their entirety.
    \end{tablenotes}
\end{threeparttable}

\end{table}
}

\end{landscape}
\restoregeometry

\subsection{Unabbreviated excerpts from Applications of $\gamma_{RDE}$}  \label{appsubsec: uanbrevExcerpts}

    \thispagestyle{empty}

{\renewcommand{\arraystretch}{1}
\begin{table}[!htb]
\begin{threeparttable}

\caption{Application excerpts: Rationales. \label{tab: AppRat}}

\begin{tabular}{|m{3cm}|m{13cm}|}
\toprule
     \emph{\textbf{Author}} & \emph{\textbf{Excerpt}} \\ \hline   
   \tiny{\citeApp[pg. ~1634]{wang2022mechanisms}} & \tiny{{{``This approach allowed us to also evaluate the confounding effect of adiposity, measured by body mass index (BMI), which was identified as a post-exposure (post H. pylori) confounder for the association between GERD and Barrett’s esophagus.''}}} \\  \hline
   \tiny{\citeApp[pg. ~1189]{wang2022does}} & \tiny{{{``Specifically, we estimated the so-called ‘interventional’ direct and indirect effects, which are estimable under relaxed assumptions compared to previous causal mediation approaches.''}}} \\  \hline
   \tiny{\citeApp{campbell2021placental}} & \tiny{{{[None provided]}}}  \\  \hline
   \tiny{\citeApp[pg. ~20]{wodtke2020neighborhood}} & \tiny{{{``We focus on a decomposition defined in terms of randomized interventions on school quality because its components can be identified under more defensible assumptions than those required of other effect decompositions. In particular, unlike the components of alternative decompositions, all of the effects outlined previously can be identified in the presence of exposure-induced confounders.''}}}  \\  \hline
   \tiny{\citeApp[pg. ~2]{liu2020inflammation}} & \tiny{{{``While the ``total (causal) effect'' of obesity on any outcome is ill-defined, it is relevant to examine the extent to which obesity-related disparities in retinal microvascular parameters would be reduced, if the distribution of inflammatory markers in those with obesity were reduced to levels observed in the non-obese, thereby representing potential `interventional effects'. ''}}}  \\  \hline
   \tiny{\citeApp[pg. ~e197]{komulainen2019childhood}} &  \tiny{{{``In the presence of exposure-induced confounding, natural direct and indirect effects are not identified, but randomized interventional analogs for natural direct and indirect effects can be estimated.''}}} \\  \hline
   \tiny{\citeApp[pg. ~155]{hanson2019does}} & \tiny{{{``The NIE$^R$ is based on fewer assumptions and can therefore be identified in settings with time-varying exposures and mediators and when mediator-outcome confounding affected by exposure may be a problem. ''}}}  \\  \hline
   \tiny{\citeApp[pg. ~631]{gage2013maternal}} & \tiny{{{``Effect decomposition can then be estimated using a procedure similar to direct standardization. \citetSM{geneletti2007identifying} called this a ``generated direct effect,'' which is similar to Pearl’s ``natural direct effect''.''}}}  \\  \hline
   \tiny{\citeApp[pg. ~337]{rudolph2021helped}} & \tiny{{{``Natural direct and indirect effects are common mediation estimands, together adding to the total effect, but are not identified in the presence of such a variable. Consequently, we estimate interventional (also known as stochastic) direct and indirect effects, which do not require the absence of posttreatment confounders of the mediator–outcome relationship for identification but are analogous to natural direct and indirect effects in the absence of such variables.''}}}  \\  \hline
   \tiny{\citeApp[pg. ~2095]{rudolph2021explaining}} & \tiny{{{``Until recently, the statistical mediation methods available required numerous restrictive assumptions that are unlikely to hold in practice, and may result in significantly biased results. The most common of these is the parametric structural equation modeling approach popularized by Baron \& Kenny in which the indirect effect (i.e. path through the mediator) is calculated as the product of two linear regression coefficients representing the treatment on the mediator and the mediator on the outcome. This approach is appealing for its simplicity, but its assumptions are often incompatible with real-world data analysis [e.g. needing to assume: (i) no intermediate variables exist (post-treatment variables that affect mediators and outcomes, e.g. medication initiation), (ii) a linear relationship between the mediator and outcome and (iii) that mediators are unrelated to each other]. We used a novel mediation approach that overcomes these limitations by allowing for multiple, inter-related mediating variables in addition to including baseline covariates, treatment and post-treatment intermediate outcomes.''}}}  \\  \hline
   \tiny{\citeApp[pg. ~822, 824]{goin2020mediation}} & \tiny{{{``We believed that the assumption of no postexposure confounding was violated in this study (see Figure 1 and Web Appendix 2 for additional discussion); therefore, we opted to estimate stochastic direct and indirect effects.''; ``By drawing from the distribution of mediator values rather than assigning a mediator value based on what we would expect it to be under different exposure values, we are able to estimate direct and indirect effects without needing to invoke the controversial “cross-world” assumption and can allow for postexposure confounders.''}}}  \\  \hline
   \tiny{\citeApp{ploubidis2019lifelong}} & \tiny{{{[None provided]}}}  \\  \hline
   \tiny{\citeApp[pg. ~1]{casey2019unconventional}} & \tiny{{{``Our analytic approach toward evaluating the mediating effect of antenatal depression and anxiety on the association between UNGD activity and adverse birth outcomes allowed for the estimation of both direct and indirect pathways and for inclusion of post-exposure mediator-outcome confounders. The inability to deal with such confounders is a limitation of other mediation methods.''}}}  \\  \hline
   \tiny{\citeApp[pg. ~592]{rudolph2018mediation}} & \tiny{{{``We estimated stochastic direct and indirect effects instead of the more common natural direct and indirect effects because the stochastic effects do not require the absence of posttreatment confounding of the mediator–outcome relationship, $M^{a^*} \CI Y^{a,m} \mid W$, an assumption that is required for identifying their natural counterparts.''}}}  \\  \hline
   \tiny{\citeApp[pg. ~882]{maika2017associations}} & \tiny{{{``Because of this intermediate confounding, we used the effect decomposition method derived by \citetSM{vanderweele2014effect}, who introduced 3 approaches for effect decomposition in the presence of exposure-induced mediator-outcome confounding: 1) joint mediators, 2) path-specific, and 3) intervention analog.''}}}  \\  \hline
   \tiny{\citeApp[pg. ~112]{pearce2016early}} & \tiny{{{``This approach is therefore suited to situations where there is just one mediating pathway of interest, which is likely to be biased by intermediate confounding.''}}}  \\  \hline \hline
\end{tabular}
 \begin{tablenotes}
      \tiny
      \item  
    \end{tablenotes}
\end{threeparttable}

\end{table}
}

{\renewcommand{\arraystretch}{1}
\begin{table}[!htb]
\begin{threeparttable}

\caption{Application excerpts: Results (Abstract). \label{tab: AppResAb}}

\begin{tabular}{|m{3cm}|m{13cm}|}
\toprule
     \emph{\textbf{Author}} & \emph{\textbf{Excerpt}} \\ \hline
    \tiny{\citeApp[pg. ~1630]{wang2022mechanisms}} & \tiny{{{``For men, 5 of the 15 per 1,000 excess Barrett’s esophagus risk from being seronegative were mediated by GERD.''}}}  \\  \hline 
   \tiny{\citeApp[pg. ~1188]{wang2022does}} &  \tiny{{{``40\% of the total effect was mediated via GlycA (indirect effect: 0.8 dB HL, 95\% CI 0.1–1.4).''}}} \\ \hline
   \tiny{\citeApp[pg. ~1]{campbell2021placental}} &  \tiny{{{``Cellular composition explained 35.1\% of the association between preeclampsia and FLT1 overexpression.''}}}  \\  \hline
   \tiny{\citeApp[pg. ~1]{wodtke2020neighborhood}} &  \tiny{{{``But contrary to expectations, we find no evidence that neighborhood effects are mediated by or interact with school quality.''}}}  \\  \hline
   \tiny{\citeApp[pg. ~1]{liu2020inflammation}} &  \tiny{{{``Children with overweight or obesity had 0.25 to 0.35 SD wider venular calibre, of which 19 to 25\% was mediated via GlycA.''}}}  \\  \hline
   \tiny{\citeApp[pg. ~e195]{komulainen2019childhood}} &  \tiny{{{``There was no evidence for indirect effects from childhood environments to left ventricular outcomes through adult health behaviors after controlling for time-dependent confounding by the adult socioeconomic position (indirect effect b= -0.30, 95\% CI= -1.22, 0.63 for left ventricular mass; b= -0.04, 95\% CI= -0.18, 0.11 for E/e’ ratio).''}}}  \\  \hline
   \tiny{\citeApp[pg. ~153]{hanson2019does}} &  \tiny{{{``A mediation model including workplace social support, IL-6 and diabetes further showed that 10\% of the association between social support and diabetes over the three repeat examinations (total effect $\beta$ = 0.08, CI 0.01–0.15) was attributable to a weak indirect effect through IL-6 ($\beta$ = 0.01, CI 0.00–0.02).''}}}  \\  \hline
   \tiny{\citeApp[pg. ~615]{gage2013maternal}} &  \tiny{{{``Mortality declines significantly (by a factor of 0.40–0.96) through the direct effect of education. The indirect effect of education among normal births is small but significant in three cohorts.''}}}  \\  \hline 
   \tiny{\citeApp[pg. ~336]{rudolph2021helped}} &  \tiny{{{``The majority (between 69\% and 90\%) of the total negative long-term effects could be explained by indirect effects through the mediators considered.''}}}  \\  \hline
   \tiny{\citeApp[pg. ~2094]{rudolph2021explaining}} &  \tiny{{{``For the homeless subgroup, the protective indirect path contributed a 3.4 percentage point reduced risk of relapse [95\% confidence interval (CI) = -12.0, 5.3] comparing XR-NTX to BUP-NX (explaining 21\% of the total effect).''}}}  \\  \hline
   \tiny{\citeApp[pg. ~820]{goin2020mediation}} &  \tiny{{{``The largest indirect effects for the association between violence and preterm birth were observed for infection (stochastic indirect effect = 0.04, 95\% CI: 0.00, 0.08) and substance use (stochastic indirect effect = 0.04, 95\% CI: 0.01, 0.06).''}}}  \\  \hline
   \tiny{\citeApp[pg. ~41]{ploubidis2019lifelong}} &  \tiny{{{``We found that early life circumstances influence behaviour either directly or indirectly 65 years later.''}}}  \\  \hline
   \tiny{\citeApp[pg. ~1]{casey2019unconventional}} &   \tiny{{{``We found no relationship between antenatal anxiety or depression and adverse birth outcomes and no mediation effect either overall or when stratifying by Medical Assistance.''}}}  \\  \hline
   \tiny{\citeApp[pg. ~590]{rudolph2018mediation}} &  \tiny{{{``Having friends who use drugs and involvement in after-school sports or clubs partially mediated the effect of housing voucher receipt on adolescent substance use (e.g., stochastic indirect effect 0.45\% [95\% confidence interval: 0.12\%, 0.79\%] for having friends who use drugs and 0.04\% [95\% confidence interval: -0.02\%, 0.10\%] for involvement in after-school sports or clubs mediating the relationship between housing voucher receipt and marijuana use among boys). ''}}} \\  \hline
   \tiny{\citeApp{maika2017associations}} &  \tiny{{{[None provided]}}}  \\  \hline
   \tiny{\citeApp[pg. ~108]{pearce2016early}} &  \tiny{{{``Around two-thirds of this elevated risk was ‘direct’ and the majority of the remainder was mediated by early cognitive ability and not self-regulation.''}}}  \\   \hline
   \bottomrule
 \hline
\end{tabular}
 \begin{tablenotes}
      \tiny
      \item  
    \end{tablenotes}
\end{threeparttable}

\end{table}
}

{\renewcommand{\arraystretch}{1}
\begin{table}[!htb]
\begin{threeparttable}

\caption{Application excerpts: Results (Main text). \label{tab: AppResMain}}

\begin{tabular}{|m{3cm}|m{13cm}|}
\toprule
     \emph{\textbf{Author}} & \emph{\textbf{Excerpt}} \\ \hline
   \tiny{\citeApp[pg. ~1634]{wang2022mechanisms}} & \tiny{{{``32\% of this effect was mediated through GERD: The estimated indirect effect was 5 (95\% CI, -4-14) per 1,000. The remaining 68\% was unexplained: The estimated direct effect not mediated through GERD was 10 (95\% CI, 0–20) per 1,000.''}}} \\  \hline 
   \tiny{\citeApp[pg. ~1191]{wang2022does}}  & \tiny{{{``11\% to 40\% of the total effects of overweight/obesity on hearing loss were explained by GlycA (indirect effect: overweight 0.1 dB HL 95\% CI -0.1 to 0.4; obesity 0.8 95\% CI 0.1-1.4).''}}} \\ \hline
   \tiny{\citeApp[pg. ~8]{campbell2021placental}}  & \tiny{{{``35.1\% (95\% CI [25.5\%, 46.3\%]) of the association between preeclampsia and FLT1 expression was attributable to differences in placental cell composition between preeclampsia cases and controls (Figure 6).''}}} \\  \hline
   \tiny{\citeApp[pg. ~31]{wodtke2020neighborhood}}  & \tiny{{{``For example, estimates of the RNDE indicate that exposure to a disadvantaged neighborhood at the 80th percentile of the treatment distribution, rather than an advantaged neighborhood at the 20th percentile, would still reduce performance on math and reading assessments by about 0.13 and 0.16 standard deviations, respectively, even after an intervention to fix the school quality distribution to that observed in advantaged neighborhoods.''}}} \\  \hline
   \tiny{\citeApp[pg. ~4]{liu2020inflammation}}  & \tiny{{{``Children with overweight and obesity had 0.25 SD and 0.35 SD wider venular calibre (respectively) compared to those with normal-weight BMI, of which 19 to 25\% was mediated via GlycA (indirect effect: overweight 0.05 SD, 95\% CI 0.0 to 0.10; obese 0.09 SD, 95\% CI 0.01 to 0.18).''}}} \\  \hline
   \tiny{\citeApp[pg. ~e198]{komulainen2019childhood}}  & \tiny{{{``The analog for the direct effect on the LV diastolic function indicated a corresponding difference of 0.14 in the E/e' ratio (95\% CI= -0.29, 0.01).Approximately 23\% and 21\% of the associations of childhood environment with the LV mass and E/e' ratio were estimated to be mediated through adult health behaviors (b= -0.30, 95\% CI, -1.22, 0.63 for LV mass; b= -0.04, 95\% CI= -0.18, 0.11 for E/e' ratio).''}}} \\  \hline
   \tiny{\citeApp[pg. ~157]{hanson2019does}}  & \tiny{{{``The randomized analogue of the indirect effect was similar but not statistically significant in the adjusted model ($\beta$ = 0.01, CI -0.00 to 0.02, p = 0.12).''}}} \\  \hline
   \tiny{\citeApp[pg. ~627]{gage2013maternal}}  & \tiny{{{``Further decomposition of the change of infant mortality into direct (independent of birth weight) and indirect (associated with changes in birth weight) effects of maternal education level indicates a strong direct effect in the primary subpopulation, which reduces infant mortality (Table 4).''}}} \\  \hline 
   \tiny{\citeApp[pg. ~340]{rudolph2021helped}}  & \tiny{{{``For example, nearly all of the effect of voucher receipt on increased risk of any DSM-IV disorder in adolescence operates through the mediators considered (Figure 2), explaining 90\% of the total interventional effect (Equation 1).''}}} \\  \hline
   \tiny{\citeApp[pg. ~2099]{rudolph2021explaining}}  & \tiny{{{``The indirect effect accounted for much of the overall increased risk of relapse on XR-NTX versus BUP-NX in non-homeless patients (contributing a 3.4 percentage point reduced risk of relapse, 95\% CL = –12.0, 5.3, corresponding to 21\% of the total effect).''}}} \\  \hline
   \tiny{\citeApp[pg. ~827]{goin2020mediation}} & \tiny{{{``The largest stochastic indirect effects were observed for infection (stochastic indirect effect = 0.04, 95\% CI: 0.00, 0.08) and substance use (stochastic indirect effect = 0.04, 95\% CI: 0.01, 0.06) (Figure 3, panel A).''}}} \\  \hline
   \tiny{\citeApp[pg. ~46]{ploubidis2019lifelong}}  & \tiny{{{``Early life SEP had both direct and indirect—via later life SEP—effects on smoking.''}}} \\  \hline
   \tiny{\citeApp[pg. ~6]{casey2019unconventional}}  & \tiny{{{``The direct effect of the highest quartile of UNGD activity on term birth weight appeared protective for mothers who did not receive Medical Assistance and harmful for mothers who did receive Medical Assistance, but confidence intervals were wide (Fig. 4 and Table A.8).''}}} \\  \hline
   \tiny{\citeApp[pg. ~594]{rudolph2018mediation}}  & \tiny{{{``Using the above data-dependent stochastic interventions, we estimated the direct and indirect effects of voucher receipt on each of our three outcomes, shown in Figures 2–4.[$\dots$]For girls, receiving a voucher led to a 0.42\% (95\% CI: 0.11\% , 0.73\% ) increased probability of cigarette use through having friends who used drugs, a 0.37\% (95\% CI: - 0.15\% , 0.90\% ) increased probability of marijuana use through having friends who use drugs and a 0.10\% (95\% CI: - 0.03\% , 0.23\% ) increased probability of problematic drug use through having friends who use drugs.''}}} \\  \hline
   \tiny{\citeApp[pg. ~884]{maika2017associations}} & \tiny{{{``From both the path-specific approach and the intervention analog approach, the indirect effect of poverty at age <7 years mediated through poverty at age 7–14 years was smaller ($\beta$ = -0.003, 95\% CI: -0.03, 0.02).''}}} \\  \hline
   \tiny{\citeApp[pg. ~115]{pearce2016early}}  & \tiny{{{``Section B of Table 4 shows that the small indirect pathway from SED to academic achievement through self-regulation was not attenuated after adjustment for intermediate confounding.''}}} \\   \hline  
   \bottomrule
 \hline

\end{tabular}
 \begin{tablenotes}
      \tiny
      \item  
    \end{tablenotes}
\end{threeparttable}

\end{table}
}

{\renewcommand{\arraystretch}{1}
\begin{table}[!htb]
\begin{threeparttable}

\caption{Application excerpts: Discussion / Conclusion (Abstract). \label{tab: AppDisAb}}

\begin{tabular}{|m{3cm}|m{13cm}|}
\toprule
     \emph{\textbf{Author}} & \emph{\textbf{Excerpt}} \\ \hline
   \tiny{\citeApp[pg. ~1630]{wang2022mechanisms}} & \tiny{{{``Evidence of GERD mediating the effect of H. pylori on Barrett’s esophagus risk among men supports this proposed mechanism.''}}}  \\  \hline 
   \tiny{\citeApp[pg. ~1188]{wang2022does}} & \tiny{{{``Inflammation may play an important mediating role in the modest hearing reductions associated with obesity, particularly in children.''}}}  \\  \hline
   \tiny{\citeApp[pg. ~1]{campbell2021placental}} &  \tiny{{{``Our findings indicate substantial placental cellular heterogeneity in preeclampsia that predicts previously observed bulk gene expression differences.''}}}  \\  \hline
   \tiny{\citeApp{wodtke2020neighborhood}} &  \tiny{{{[None provided]}}}  \\  \hline
   \tiny{\citeApp[pg. ~1]{liu2020inflammation}} &  \tiny{{{``Inflammation mediated associations between obesity and retinal venules, but not arterioles from mid-childhood, with higher mediation effects observed in adults.''}}}  \\  \hline
   \tiny{\citeApp[pg. ~e195]{komulainen2019childhood}} &  \tiny{{{``After accounting for socioeconomic positions, adult health behaviors explain little of the associations.''}}}  \\  \hline
   \tiny{\citeApp[pg. ~153]{hanson2019does}} &  \tiny{{{``This study indicates an association between poor workplace support and diabetes that is partially ascribed to an inflammatory response.''}}}  \\   \hline
   \tiny{\citeApp[pg. ~615]{gage2013maternal}} &  \tiny{{{``Overall, our results are consistent with the view that the decrease in infant death by socioeconomic level is not mediated by improved birth weight.''}}}  \\  \hline 
   \tiny{\citeApp[pg. ~336]{rudolph2021helped}} &  \tiny{{{``This evidence suggests that, even though the intervention had the desired effects on neighborhood poverty and the school environment, these `positives' ultimately negatively impacted the long-term mental health and behaviors of boys.''}}}  \\  \hline 
   \tiny{\citeApp[pg. ~2094]{rudolph2021explaining}} &  \tiny{{{``A novel approach to mediation analysis shows that much of the difference in medication effectiveness (extended-release naltrexone versus buprenorphine–naloxone) on opioid relapse among non-homeless adults with opioid use disorder appears to be explained by mediators of adherence, illicit opioid use, depressive symptoms and pain.''}}}  \\  \hline
   \tiny{\citeApp[pg. ~820]{goin2020mediation}} &  \tiny{{{``Firearm violence was associated with risk of preterm delivery, and this association was partially mediated by infection and substance use.''}}}  \\  \hline
   \tiny{\citeApp[pg. ~54]{ploubidis2019lifelong}} &  \tiny{{{[None provided]}}}  \\  \hline
   \tiny{\citeApp[pg. ~1]{casey2019unconventional}} &   \tiny{{{``We observed a relationship between UNGD activity and antenatal anxiety and depression, which did not mediate the overall association between UNGD activity and adverse birth outcomes.''}}}  \\  \hline
   \tiny{\citeApp[pg. ~590]{rudolph2018mediation}} &  \tiny{{{``Measured school- and peer-environment variables played little role in mediating the effect of housing voucher receipt on subsequent adolescent substance use.''}}}  \\  \hline
   \tiny{\citeApp{maika2017associations}} &  \tiny{{{[None provided]}}}  \\  \hline
   \tiny{\citeApp[pg. ~108]{pearce2016early}} &  \tiny{{{``Early interventions to improve cognitive ability (rather than self-regulation) also hold potential for reducing inequalities in children's academic outcomes.''}}}  \\   \hline
   \bottomrule
 \hline

\end{tabular}
 \begin{tablenotes}
      \tiny
      \item  
    \end{tablenotes}
\end{threeparttable}

\end{table}
}

{\renewcommand{\arraystretch}{1}
\begin{table}[!htb]
\begin{threeparttable}

\caption{Application excerpts: Discussion / Conclusion (Main text). \label{tab: AppDisMain}}

\begin{tabular}{|m{3cm}|m{13cm}|}
\toprule
     \emph{\textbf{Author}} & \emph{\textbf{Excerpt}} \\ \hline
   \tiny{\citeApp[pg. ~1636]{wang2022mechanisms}}  & \tiny{{{``Moreover, estimating interventional effects allowed us to evaluate how much Barrett’s esophagus risk could be reduced with a clinical intervention that would reduce occurrence of daily GERD symptoms. GERD mediated approximately 32\% of the excess Barrett’s esophagus risk for being H. pylori seronegative.''}}} \\  \hline 
   \tiny{\citeApp{wang2022does}}  & \tiny{{{[None provided]}}} \\  \hline
   \tiny{\citeApp[pg. ~12]{campbell2021placental}}  & \tiny{{{``A substantial proportion of the overexpression of the FLT1 in preeclampsia was mediated by placental cell composition.''}}} \\  \hline
   \tiny{\citeApp[pg. ~36]{wodtke2020neighborhood}} & \tiny{{{``At the same time, however, we find no evidence that neighborhood effects are mediated by school quality because differences in the socioeconomic composition of neighborhoods do not appear to be strongly linked with differences in school quality.''}}} \\  \hline
   \tiny{\citeApp[pg. ~5]{liu2020inflammation}} & \tiny{{{``We found that the association of obesity with venular calibre was mediated by inflammation in both age groups.''}}} \\  \hline
   \tiny{\citeApp[pg. ~e200]{komulainen2019childhood}} & \tiny{{{``Findings from the causal mediation analysis suggest that this association is not explained by ideal cardiovascular health behaviors in adulthood once age, sex, and adult SEP are accounted for.''}}} \\  \hline
   \tiny{\citeApp[pg. ~159]{hanson2019does}}  & \tiny{{{``Our results imply that inflammation may be a mechanism that explains the association be- tween social relationships and diabetes.''}}} \\   \hline
   \tiny{\citeApp[pg. ~631]{gage2013maternal}}  & \tiny{{{``The methods of effect decomposition used here are based on statistical decision theory, as opposed to the more common “counterfactual” approaches qualitatively applied to infant mortality by Hernández-Diaz et al. (2008). Here, we model the educational level–specific birth weight density as a mixture of two Gaussian distributions with subpopulation birth weight–specific mortality curves as second-degree polynomials of standardized birth weight. By standardizing birth weight (within specific educational levels and subpopulations), the main effect of the association of education and birth weight is eliminated, and the regressions can estimate the direct effect of education on infant mortality versus any indirect effect as interaction terms of education and birth weight on infant mortality.''}}} \\  \hline 
   \tiny{\citeApp[pg. ~344]{rudolph2021helped}}  & \tiny{{{``Using a novel, robust, and efficient nonparametric estimator of interventional direct and indirect effects, we concluded that most of the overall harmful effects operated through objectively measured aspects of the neighborhood and school environments as well as the instability of these environments.''}}} \\  \hline 
   \tiny{\citeApp[pg. ~2099]{rudolph2021explaining}} & \tiny{{{``Using mediation analysis, we found that the pathway from medication assignment (and subsequent initiation) through mediators of adherence, early illicit opioid use, depressive symptoms and pain contributed to much of the increased risk of relapse on XR-NTX versus BUP-NX for non-homeless participants, but did not substantially contribute to the reduced risk of relapse among homeless participants.''}}} \\  \hline
   \tiny{\citeApp[pg. ~828]{goin2020mediation}}  & \tiny{{{``Our findings suggest that firearm violence increases risk of spontaneous preterm birth, with the strongest mediation associations being seen for infection and substance use.''}}} \\  \hline
   \tiny{\citeApp[pg. ~54]{ploubidis2019lifelong}}  & \tiny{{{``A homogeneous pattern emerged for all types of physical activity in men, where later life SEP was positively associated with mild, moderate and vigorous physical activity and early life SEP had only an indirect association with all types of activity.''}}} \\  \hline
   \tiny{\citeApp[pg. ~7]{casey2019unconventional}}  & \tiny{{{``Our study indicates that maternal mental health, measured as antenatal anxiety or depression, did not act as a mediator of the relationship between UNGD activity and adverse birth outcomes.''}}} \\  \hline
   \tiny{\citeApp[pg. ~597]{rudolph2018mediation}}  & \tiny{{{``In summary, we found evidence that aspects of the peer environment weakly mediated the relationship between Section 8 voucher receipt and subsequent move out of public housing and substance use among adolescents.''}}} \\  \hline
   \tiny{\citeApp[pg. ~886]{maika2017associations}}  & \tiny{{{``Finally, in the intervention analog approach, effect decomposition is estimated as an analog of a sequentially randomized mediator based on the exposure level, which effectively removes L.[$\dots$]From effect decomposition, we found that poverty at age <7 years had a bigger direct effect on cognitive function than via its mediated effect through poverty at age 7–14 years.''}}} \\  \hline
   \tiny{\citeApp[pg. ~116]{pearce2016early}} & \tiny{{{``Decomposition of the indirect pathway showed that around 80-90\% was through cognitive ability rather than self-regulation, in part reflecting the weaker association between self-regulation and both the exposure (maternal education) and the outcome (academic achievement).''}}} \\   \hline   \bottomrule
 \hline

\end{tabular}
 \begin{tablenotes}
      \tiny
      \item  
    \end{tablenotes}
\end{threeparttable}

\end{table}
}

\clearpage

\clearpage\doublespacing

\bibliographystyleSM{unsrtnat}
\bibliographySM{refs}

\clearpage

\end{appendices}

\end{document}